\documentclass[aps,twocolumn,prd,superscriptaddress]{revtex4-1}

\usepackage[utf8]{inputenc}

\usepackage{mathtools}
\usepackage{amsfonts}
\usepackage{mathrsfs}
\usepackage{bbm}
\usepackage{slashed}

\usepackage{graphicx}
\usepackage[dvipsnames]{xcolor}
\usepackage{array}

\usepackage{placeins}

\usepackage{xspace}
\usepackage{siunitx}
\usepackage{hyperref}
\usepackage[nameinlink]{cleveref}

\usepackage{xifthen}
\usepackage{dsfont}
\usepackage{appendix}
\usepackage{enumitem}
\usepackage{booktabs}
\usepackage{units}

\setkeys{Gin}{width=0.48\textwidth}

\graphicspath{
	{./figures/}
}

\renewcommand{\Im}{\rm Im}

\DeclareMathOperator{\sign}{sgn}
\DeclareMathOperator{\logG}{ln\Gamma}

\newcommand{\imag}{\text{i}}
\newcommand{\hps}{\hat{p}^{2}}
\newcommand{\tinytext}[1]{\text{\tiny{#1}}}


\newcommand{\gettitle}{Reconstructing the gluon}

\hypersetup{
	colorlinks,
	linkcolor={red!75!black},
	citecolor={blue!75!black},
	urlcolor={blue!75!black},
	pdftitle={\gettitle},
	pdfauthor={Cyrol, Pawlowski, Rothkopf, Wink},
	pdfkeywords={Spectral function} {Analytic continuation}
	{Correlations functions} {Gluon} 
	{Bayesian inference}
	{Real time} {Infrared},
	bookmarksopen=true,
	bookmarksopenlevel=2,
	bookmarksnumbered=true
}

\begin{document}

\title{\gettitle}

\author{Anton~K.~Cyrol\,} 
\affiliation{Institute for Theoretical Physics, Universit\"at
  Heidelberg, Philosophenweg 12, D-69120 Germany}

\author{Jan~M.~Pawlowski\,} 
\affiliation{Institute for Theoretical Physics, Universit\"at
  Heidelberg, Philosophenweg 12, D-69120 Germany}
\affiliation{ExtreMe Matter Institute EMMI, GSI, Planckstr. 1,
  D-64291 Darmstadt, Germany}

\author{Alexander~Rothkopf\,}
\affiliation{Institute for Theoretical Physics, Universit\"at
  Heidelberg, Philosophenweg 12, D-69120 Germany}
\affiliation{Faculty of Science and Technology, University of Stavanger, NO-4036 Stavanger, Norway}

\author{Nicolas~Wink\,} 
\affiliation{Institute for Theoretical Physics, Universit\"at
  Heidelberg, Philosophenweg 12, D-69120 Germany}

\begin{abstract}
  We reconstruct the gluon spectral function in Landau gauge QCD from
  numerical data for the gluon propagator. The reconstruction relies
  on two novel ingredients: Firstly we derive analytically the low
  frequency asymptotics of the spectral function. Secondly we
  construct a functional basis from a careful consideration of the
  analytic properties of the gluon propagator in Landau gauge. This
  allows us to reliably capture the non-perturbative regime of the
  gluon spectrum. We also compare different reconstruction methods and
  discuss the respective systematic errors.
\end{abstract}

\maketitle

\section{Introduction}

Real-time correlation functions play a pivotal
r$\hat{\rm o}$le for the theoretical understanding of heavy-ion
collisions and the hadron spectrum.  Their direct numerical
computation in the strongly correlated regime of QCD is hampered by
the fact that non-perturbative real-time methods are still in their
infancies. Even though impressive results have been obtained in models
and even in Yang-Mills theory, still more work is required on the way towards a
full real-time approach to QCD.

In turn, non-perturbative Euclidean first principles methods such as
Euclidean functional approaches and lattice simulations have been
extensively used to obtain numerical results for QCD correlation
functions.  When analytically continuing these to the real-time regime
or equivalently reconstructing their spectral function by means of
solving an inverse integral transformation, one encounters large
systematic uncertainties as in the case of single particle spectral
functions
\cite{Haas:2013hpa,Qin:2013ufa,Dudal:2013yva,Christiansen:2014ypa,%
  Rothkopf:2016luz,Ilgenfritz:2017kkp,Fischer:2017kbq,Silva:2017mds}
or the energy momentum tensor (EMT) \cite{Astrakhantsev:2017nrs,
  Pasztor:2018yae, Mages:2015rea, Astrakhantsev:2015jta,Meyer:2007dy,
  Meyer:2007ic}, to name another pertinent correlator.  Typically, the
uncertainty even grows larger at small frequency. This is particularly
harmful for the computation of transport coefficients which are
related to the vanishing frequency limit of correlation functions of
the EMT.

This problem of a large systematic uncertainty at low frequencies can be partially circumvented
by a diagrammatic representation of the correlation functions of the
EMT in terms of loops of real-time correlation functions of quarks and
gluons, as discussed in \cite{Haas:2013hpa,Christiansen:2014ypa}. There the
spectral function of the EMT has been computed from the
single-particle spectral function of the gluon. The gluon spectral function itself,
in the above mentioned paper has been reconstructed via a Bayesian 
spectral reconstruction method \cite{Hobson:1998bz}, which is a variant of
the Maximum Entropy Method (MEM) \cite{Jarrell:1996rrw,Asakawa:2000tr}.
This approach also extends to general real-time computations of correlation functions on
the basis of real-time single particle spectral functions of quarks
and gluons.  

A further reduction of the systematic error comes from prior
information about the properties of the single particle spectral
functions used as input. Often, such prior information is available
for the high frequency asymptotics of the spectral function.  This is
the off-shell limit with the Minkowski four momentum $p^2\to\infty$.

In the present work we argue that the low frequency asymptotics
is determined by the infrared (IR) limit in the Euclidean domain
using only rather general assumptions.
This leaves us with a well-constrained spectral
function, which allows for a qualitatively enhanced spectral
reconstruction. We apply this argument to the spectral reconstruction
of the single particle gluon spectral function. The results
presented here provide the starting point for the
computation of transport coefficients in the spirit of
the work presented in \cite{Haas:2013hpa,Christiansen:2014ypa}, as
well as direct real-time computation of thermodynamical properties and
the QCD hadron spectrum.

This paper is organized as follows: In \Cref{sec:anspec} we derive
a general relation between the low frequency behavior of
bosonic spectral functions and the infrared (IR) behavior of the
corresponding Euclidean correlator.  In \Cref{sec:Analytical} we
summarize known properties of the gluon spectral function, 
its normalization and its asymptotics in the ultraviolet
regime (UV). Then we turn to the low frequency behavior of the gluon
spectral function in \Cref{sec:lowfreq}. Both the analytic structure of the
scaling scenario and several realizations of the decoupling scenario
are discussed. In \Cref{sec:SpecFunc}, we reconstruct the 
gluon spectral function with a novel reconstruction method
from numerical data from \cite{Cyrol:2016tym}. We conclude
in \Cref{sec:conclusion} and provide a comparison of different
reconstruction approaches in \Cref{app:der-analytic}.

\section{Low frequency asymptotics of spectral functions}
\label{sec:anspec}

In general, a spectral function can be obtained from analytic
continuation of its Euclidean propagator or from the inverse integral
transformation via the K{\"a}ll{\'e}n–Lehmann spectral representation.
In this section we first introduce some basic definitions
and then derive a novel general relation,
\labelcref{eq:key_relation}, between the low frequency behavior of the
spectral function and the infrared behavior of the Euclidean
propagator.  The relation is illustrated at a Breit-Wigner example
before it is applied to the gluon spectral function in
\Cref{sec:lowfreq}.

Throughout this section we assume that the propagator
admits the K{\"a}ll{\'e}n–Lehmann spectral representation 
\begin{align}\label{eq:specrep}
  G(p_0)&=
  \int_{0}^{\infty} \frac{\mathrm{d}\lambda}{\pi} \frac{\lambda\,
    \rho(\lambda)}{\lambda^2+p_0^2}\,.
\end{align}
The existence of a spectral representation has strong consequences for
the analytic structure of the corresponding propagator, i.e. all
non-analyticities are constrained to the $\mathrm{Re}\, p_0 = 0$
line. More details can be found in \Cref{app:der-analytic}.  In
\labelcref{eq:specrep} and in the following we have suppressed the
momentum-dependence $\mathbf{p}$ of the spectral function and the
propagator. Note that all arguments about $p_0=0$ apply equally to
$p^2=0$ at vanishing temperature. In \labelcref{eq:specrep} the restriction
to positive frequencies in the integral follows from the antisymmetry
of the spectral function
\begin{align} \label{eq:antisym}
\rho(\omega) = - \rho(-\omega)\, .
\end{align}
Equivalently, the spectral function can be obtained from the Euclidean
propagator by means of analytic continuation
\begin{align} \label{eq:spec_from_prop}
\rho(\omega) = 2\, \Im\ G(-\imag (\omega + \imag 0^+))\, ,
\end{align}
i.e. from the discontinuity of the propagator. The low frequency
behavior can directly be derived from \labelcref{eq:specrep}, which is done
in the following. We take a derivative of the spectral representation
\labelcref{eq:specrep} with respect to the Euclidean frequency
\begin{align} \label{eq:der_rel_integral}
  \partial_{p_0} G(p_0) = -\int_{-\infty}^{\infty} \frac{\mathrm{d}\lambda}{\pi} \lambda\, p_0\,
  \frac{  \rho(\lambda)}{(\lambda^2+p_0^2)^2} \,.
\end{align}
We now take the limit $p_0\to 0$ in \labelcref{eq:der_rel_integral} in
order to access the low frequency behavior. With the derivative of
the Poisson kernel representation of the delta function
\begin{align}
	\delta(x)=\lim_{\varepsilon\to 0}\frac{1}{\pi}\frac{\varepsilon}{\varepsilon^2+x^2}
\end{align}
one obtains the simple relation
\begin{align} \label{eq:key_relation}
	\lim_{p_0\to 0^+} \partial_{p_0} G(p_0)
	=
	- \frac{1}{2} \lim_{\omega\to 0^+} \partial_{\omega} \rho(\omega)
	\, . 
\end{align}
Equation \labelcref{eq:key_relation} encodes the asymptotic behavior of the spectral
function for small frequencies. 

The low frequency behavior of spectral functions may also have an
additional peculiarity at finite temperature, the transport peak. In
case it is present, or in general at finite temperature, the limits of vanishing frequency $\omega\to 0$
and momenta $|\mathbf{p}|\to 0$ do not commute anymore, for a more
careful discussion on this issue see
e.g. \cite{Pawlowski:2017gxj}. Nevertheless, the analysis performed in
this section holds, as all equations are viewed at fixed a
$\mathbf{p}$.

As an example for the low frequency asymptotics we take a single pair
of complex conjugated poles on the second Riemann sheet, i.e. a
Breit-Wigner.  The propagator is parametrized by
\begin{align} \label{eq:prop_BW}
	G^{(\tinytext{BW})}(p_0) = \frac{A}{(p_0+\Gamma)^2 + M^2}
	\, ,
\end{align}
where $A$ is a suitably chosen normalization, $\Gamma$ is proportional
to the width and $M$ is the mass.  Expanding the derivative of
\labelcref{eq:prop_BW} yields
\begin{align} \label{eq:lowmom_BW}
	\partial_{p_0} G^{(\tinytext{BW})}(p_0) = -\frac{2 A \Gamma}{(M^2 + \Gamma^2)^2} + \mathcal{O}(p_0)
	\, .
\end{align}
Using \labelcref{eq:key_relation}, we obtain the small frequency behavior of the spectral function
\begin{align} \label{eq:lowfreq_BW}
	\rho^{(\tinytext{BW})}(\omega) = \frac{4 A \Gamma \omega}{(M^2 + \Gamma^2)^2} + \mathcal{O}(\omega^2)
	\, ,
\end{align}
which is exactly the low frequency behavior of the full spectral function,
\begin{align} \label{eq:spec_BW}
	\rho^{(\tinytext{BW})}(\omega) = \frac{4 A \Gamma \omega}{4 \Gamma^2 \omega^2 + (M^2 + \Gamma^2 - \omega^2)^2}
	\, .
\end{align}

We emphasize that this derivation is based on the assumption of
sufficient smoothness of the spectral function at low frequencies. A more
careful derivation of \labelcref{eq:key_relation}, discussing the assumptions
and other subtleties, such as modified spectral representations, 
is provided in \Cref{app:der-analytic}.

\section{Known analytic properties of the gluon spectral function}
\label{sec:Analytical}

Throughout this work we assume the existence of a spectral
representation for the gluon propagator.  This
entails that the Euclidean gluon propagator $G_A(p)$ with Euclidean
momenta $p=(p_0,\vec p)$ can be written in terms of a gluon spectral
function $\rho_A(\lambda,\vec p)$ as
\begin{align}\label{eq:specrep-gluon}
G_A(p_0)= \int_0^\infty \frac{d\lambda}{\pi}\frac{\rho_A(\lambda)}{\lambda^2
+p_0^2}+ \sum_{j \in \text{poles}} \frac{R_j}{p_0^2 + M_j^2}\,.
\end{align}
analogously to \labelcref{eq:specrep}, where $M_j\in \mathbb{C}$ are the position of poles with
$\mathrm{Im}\, M_j \neq 0$ and the $R_j$ the corresponding residues.
As there we have suppressed the
spatial momentum dependence in \labelcref{eq:specrep-gluon}. In
\labelcref{eq:specrep-gluon} we also allowed for additional poles for the
sake of maximal generality. In most cases these poles will be
discarded. 

The existence of a spectral representation of the gluon is implicitly
underlying various relations and properties used in covariant
approaches to QCD. In the present context this is most apparent -but
by now means restricted to- for the pinch technique, see e.g.\
\cite{Aguilar:2006gr}. Further works implicitly using gluon spectral 
representation can e.g.\ be found in \cite{Qin:2010pc,Su:2014rma}. 

While the low frequency properties discussed in the preceding section
apply to any bosonic spectral function, we now turn to the gluon spectral
function as a specific example. Their normalization and high frequency
asymptotics are well-known properties of QCD, which we briefly discuss
next. They are exploited in the reconstruction of gluon spectral
functions from Euclidean data in \Cref{sec:SpecFunc}.

The normalization of the gluon spectral function in Landau gauge
follows from the Oehme-Zimmermann superconvergence
relation \cite{oehme1980gauge, Oehme:1990kd},
\begin{align} \label{eq:OehmeZimmermann} \int_{0}^{\infty}
  \mathrm{d}\lambda\ \lambda\, \rho_{\tinytext{A}}(\lambda) = 0\, .
\end{align}
\labelcref{eq:OehmeZimmermann} entails that $ \rho_{\tinytext{A}}(\lambda)$
has positive {\it and} negative values in contrast to spectral functions of
physical (asymptotic) states, e.g. hadronic spectral functions. Then,
the total spectral weight is finite and is typically normalized to
one. Its conversation is related to unitarity. When reconstructing the
gluon spectral function, \labelcref{eq:OehmeZimmermann} serves as a highly
non-trivial consistency check. Alternatively one may simply enforce it
within the reconstruction method as part of the prior information.

Let us further consider the high frequency behavior of the spectral
function. The asymptotic off-shell propagator can be determined in
perturbation theory for arbitrary $p_0\in\mathbb{C}$
\cite{Oehme:1994hf}. In pure glue theory the only scale is the
dynamical QCD scale $\Lambda_\tinytext{QCD}$. Accordingly, its
asymptotic regimes are define with the limits of the dimensional
momenta and frequencies,
\begin{align}\label{eq:asymptotics} 
  \hat p^2= \frac{p^2 }{\Lambda_\tinytext{QCD}^2}\,,\qquad
  \qquad \hat \omega= \frac{\omega}{\Lambda_\tinytext{QCD}}\,.
\end{align}
With the dimensionless momentum $\hat p$ the dimensionless propagator
$\hat G_{\tinytext{A}}= \Lambda_\tinytext{QCD}^2 G_{\tinytext{A}} $
reads
\begin{align}\label{eq:GA-UV}
\lim_{\hat p^2\to\infty}  \hat G_{\tinytext{A}}^{(\tinytext{UV})}(p) = \frac{Z_\tinytext{UV}}{ \hat p^2 
\log(\hat{p}^2)^\gamma} +\text{sub-leading}\, ,
\end{align}
where $Z_\tinytext{UV}$ is an overall dimensionless normalization and 
\begin{align}\label{eq:gluonandim}
\gamma=\frac{13}{22}\,,
\end{align} 
is the one-loop anomalous dimension of the gluon.  Using
\labelcref{eq:spec_from_prop} one can easily obtain the asymptotic behavior
of the dimensionless spectral function
$\hat \rho_\tinytext{A} = \Lambda^2_\tinytext{A}\rho_{\tinytext{A}}$
with
\begin{align} \label{eq:spectral_UV} \lim_{\omega\to\infty}
  \Lambda_\tinytext{QCD}^2
  \rho_{\tinytext{A}}^{(\tinytext{UV})}(\omega) =
  -\frac{Z_\tinytext{UV}}{\hat{\omega}^2
    \log(\hat{\omega}^2)^{1+\gamma}}+\text{sub-leading}\, .
\end{align}
One key aspect is the leading order negativity of the spectral
function at high frequencies. As a direct consequence, the gluon
admits positivity violation in Landau gauge and cannot be an
asymptotic state of the theory. As a consequence, the spectral
function cannot be interpreted in the usual probabilistic sense
anymore. Further details can be found in e.g. \cite{Oehme:1994hf,
  Alkofer:2000wg, Cornwall:2013zra}.

\section{Low frequency properties of the gluon spectral function}
\label{sec:lowfreq}

\begin{figure}
	\hspace*{0.15cm}
	\includegraphics[width=0.2\textwidth]{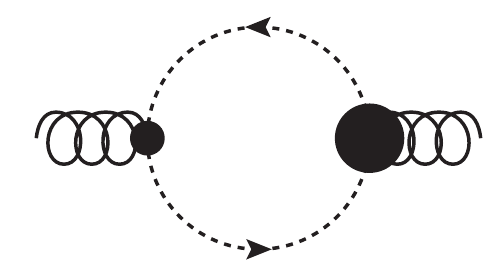}
	\hspace*{0.35cm}
	\includegraphics[width=0.2\textwidth]{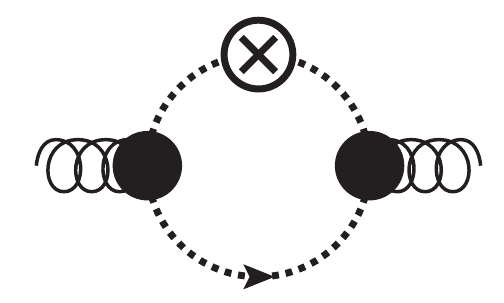}
	\caption{ Ghost loops in the gluon propagator DSE (left), \cite{Fischer:2008uz} and
          FRG (right), \cite{Cyrol:2016tym}.  Internal dashed (wiggly) lines
          represent fully dressed ghost (gluon) propagators.  Thin
          (thick) blobs depict undressed (dressed) vertices.  The
          crossed circle in the FRG diagram denotes the regulator
          insertion.  The massless ghost propagator causes these
          diagrams to yield logarithms as given in
          \labelcref{eq:prop_decoupling}.  The dressing of ghost-gluon vertex
          cannot change this qualitative behavior since it is
          constant in the infrared, see
          \Cref{fig:IrTrivialGhostGluonVertex}.  }
	\label{fig:GluonPropLogs}
\end{figure}

In the present section we evaluate the novel frequency relation of
\labelcref{eq:key_relation} for the gluon spectral function. We show that
for the scaling solution in the Landau gauge, derived from the
Kugo-Ojima criterion \cite{Kugo:1979gm}, the low frequency asymptotics
is negative. For the decoupling solution found on the lattice and in
various analytic approaches the situation is less clear. However, for
the expansion schemes used in the literature we also find negative
spectral functions.

The derivation of \labelcref{eq:key_relation} has been quite general and
holds for a large class of operators. The application of
\labelcref{eq:key_relation} only requires the knowledge of the asymptotic
infrared (IR) behavior of the theory at hand.  Despite the tremendous
progress in understanding the IR sector of Yang Mills theory and QCD, 
we still lack a comprehensive picture. 
Various approaches have been put forward to predict the
analytic IR behavior of the Euclidean gluon propagator, which we use
to determine the small frequency behavior of the gluon spectral
function in the following. In general the Landau gauge gluon
propagator $\hat G_\tinytext{A}= \Lambda^2_\tinytext{QCD}G_{\tinytext{A}}$ 
in the deep IR can be parametrized by
\begin{align} \label{eq:prop_gen}
\hat G_{\tinytext{A}}(p_0)=Z_\tinytext{IR} \, x^{-1+2\kappa}
	\, ,
\end{align}
with the scaling coefficient $\kappa$ and an overall dimensionless IR
normalization $Z_\tinytext{IR}$ and 
\begin{align}\label{eq:x}
	x = \hps+\gamma_G \left( m^2_{\text{\tiny{gap}}} + z_G \,\hps \ln \hps\right)
\end{align}
with $z_G>0\,$. The remainder of this section concerns the structure
of \labelcref{eq:prop_gen}, additionally all equations are understood in
the deep IR limit, i.e. $\hps,|\hat{\omega}|\ll1\,$.

The parameter $\gamma_G\in[0,1]$ is related to the Gribov ambiguity,
together with an appropriate definition of $m_\tinytext{gap}^2$.  The
lower limit $\gamma_G\to 0$ recovers the \textit{scaling} solution,
while the upper limit $\gamma_G\to 1$ can be considered as
implementing the maximal breaking of global BRST symmetry. 
In the following we call the set of solutions with $\gamma_G > 0$ \textit{decoupling}. 
Despite their differences in terms of scaling both solutions have in common
that their $p^2$-derivative diverges in the infrared, 
\begin{align}\label{eq:GAlimit} 
\lim_{p^2\to 0}|\partial_{p^2} G_A(p^2) |\to \infty\,. 
\end{align}
For the scaling solution it follows with $0<\kappa<1$ and
$\kappa\neq 1/2$. 
For the decoupling solution the divergence
originates in the logarithm $\ln p^2$.
Moreover, from \labelcref{eq:key_relation} it follows that it 
is precisely the sign of the $p^2$-derivative 
in \labelcref{eq:GAlimit} which determines the
sign of the spectral function for low frequencies,
\begin{align}\label{eq:signrhoAGA}
\sign \left[\lim_{\omega\to 0} \rho_A(\omega)\right] = - \sign \left[\lim_{p^2\to 0}\partial_{p^2} G_A(p^2)\right]
\, ,
\end{align}
where we used that the sign of the spectral function and its derivative are identical at low frequencies. 
This follows from the expansion of the spectral function around zero
\begin{align}
\label{eq:rho_expansion}
\rho_A(\omega) = \omega\, \partial_\omega \rho_A(\omega)
\end{align}
for positive frequencies. In \labelcref{eq:rho_expansion} the vanishing of the zeroth order, 
i.e. $\rho_A(0) = 0$, is one of our basic assumptions, c.f. the discussion in \Cref{app:der-analytic}.
\Cref{eq:signrhoAGA} entails that the backbending of the propagator 
leads to a negative spectral function at low frequencies.
Note that a backbending implies the
existence of a gluon propagator maximum at a finite momentum
which indicates positivity violation, see e.g.\ \cite{Alkofer:2000wg}. 

Apart from the low frequency behavior we are also interested in the
analytic structure of the gluon propagator. The latter is relevant for
an accurate determination of the quasi-particle peak we expect at
frequencies related to the physics scale $\Lambda_\tinytext{QCD}$: The
analytic form of \labelcref{eq:prop_gen} is exact for the scaling
solution, see e.g.
\cite{vonSmekal:1997ohs,Zwanziger:2001kw,Lerche:2002ep,Fischer:2002eq,
  Pawlowski:2003hq,Alkofer:2004it,Fischer:2006vf,Alkofer:2008jy,Fischer:2009tn},
and the discussion in \Cref{sec:Analytical:IR:Scaling}. For the
decoupling solution \labelcref{eq:prop_gen} has to be seen as an ansatz. In
particular, it is one that is motivated from an Euclidean perspective
and it may introduce ambiguities regarding the analytic structure of
the propagator in the complex plane. Different proposals for the
analytic structure of the gluon propagator have already been made in
\cite{Alkofer:2003jj}, one of which is compatible with the scenario
discussed here. We postpone the thorough discussion of the
parametrization to \Cref{sec:Analytical:IR:Decoupling} and
\Cref{sec:DecouplingScenarios}.

\begin{figure}
	\hspace*{0.15cm}
	\includegraphics[width=0.15\textwidth]{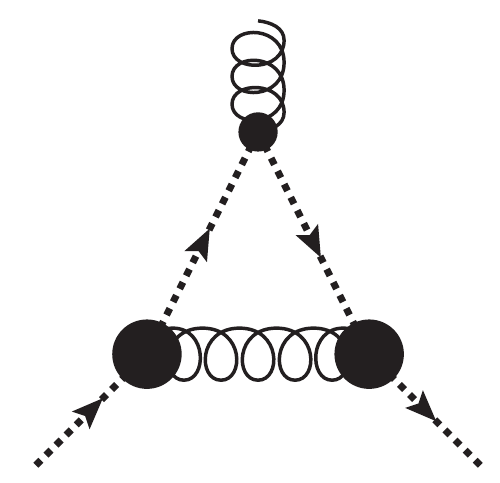}
	\hspace*{1.25cm}
	\includegraphics[width=0.15\textwidth]{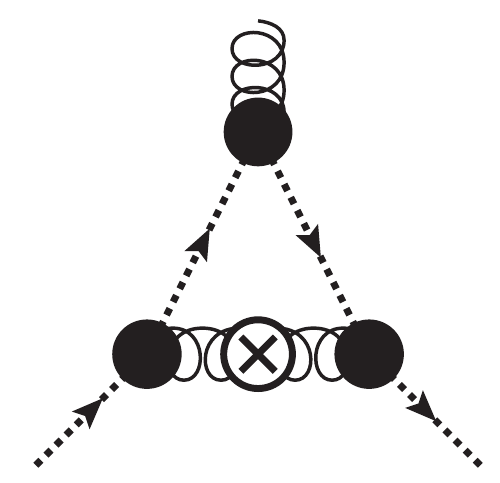}
	\caption{ Typical diagrams that contribute to the ghost-gluon
          vertex DSE (left), \cite{Huber:2012kd} and
          FRG (right), \cite{Cyrol:2016tym}.  The presence of
          the gapped gluon propagator ensures that the ghost-gluon
          vertex is constant in the infrared.  It can be shown that
          this behavior is present at every finite truncation level.
        }
	\label{fig:IrTrivialGhostGluonVertex}
\end{figure}

\subsection{Scaling solution}
\label{sec:Analytical:IR:Scaling}
The scaling solution is obtained by setting $\gamma_G = 0$ in \labelcref{eq:prop_gen}. 
The asymptotic behavior of the gluon propagator then reads
\begin{align} \label{eq:prop_scaling}
\hat G_{\tinytext{A}}^{\tinytext{(sca)}}(p) = Z_\tinytext{IR} \, (\hps)^{-1+2\kappa}
\, .
\end{align}
The scaling coefficient $\kappa$ is constrained by $1/2<\kappa<1$ and recent
numerical calculations suggest $\kappa\approx 0.58$ \cite{Cyrol:2016tym} in 
Yang-Mills theory.

Combining \labelcref{eq:key_relation} and \labelcref{eq:prop_scaling} we obtain the following
low frequency asymptotics of the gluon spectral function for the scaling solution
\begin{align} \label{eq:spectral_scaling}
  \hat \rho_{\tinytext{A}}^{\tinytext{(sca)}}(\omega)=-2\, Z_\tinytext{IR} \sign(\hat{\omega})\,
  (\hat{\omega}^2)^{-1+2\kappa}
\, .
\end{align}
Most notable is the negative sign, i.e. the spectral function is negative for small positive frequencies.
The functional similarity between \labelcref{eq:prop_scaling} and \labelcref{eq:spectral_scaling}
is not very surprising since the scaling solution has a rather simple complex structure, 
a single branch cut at $\mathrm{Re}\, p_0 = 0$.

\begin{figure}
	\hspace*{0.15cm}
	\includegraphics[width=0.15\textwidth]{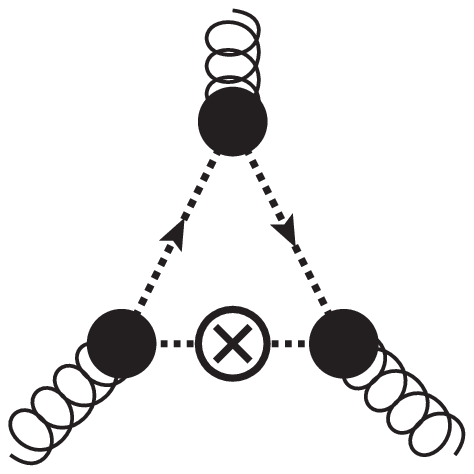}
	\hspace*{1.25cm}
	\includegraphics[width=0.15\textwidth]{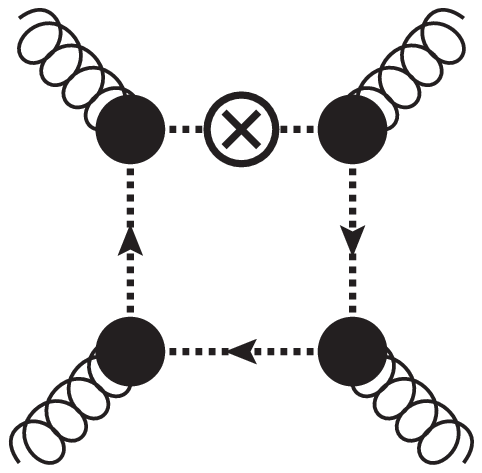}
	\caption{Ghost triangle (left) and ghost box (right) diagrams
	as they appear in the three- and four-gluon vertex flow equations \cite{Cyrol:2016tym}.
	As is well known, these ghost loops generate logarithmic divergences in the vertices.
	Similar diagrams contribute to the respective vertex DSEs, see e.g. \cite{Aguilar:2013vaa,Pelaez:2013cpa,Blum:2014gna,Eichmann:2014xya,Cyrol:2014kca}.}
	\label{fig:GluonicVertices}
\end{figure}

\subsection{Decoupling Solution}
\label{sec:Analytical:IR:Decoupling}
In this section we discuss in detail the infrared behavior of the decoupling
solution. In contradistinction to the scaling solution where the
analytic structure follows directly from scaling in the Euclidean regime, in the decoupling
case this necessitates to monitor the infrared leading logarithms. While 
the leading logarithms are fully accessible, a complete analysis requires 
to take into account the back-coupling of the quantum corrections in 
the functional equations. 

To begin with, the
leading behavior for the decoupling solution strictly speaking reads
\begin{align} \label{eq:Gdecouple}
\hat G_{\tinytext{A}}^{\tinytext{(dec)}}(p_0) \sim \frac{1}{x}
\, ,
\end{align} 
with $x$ given by \labelcref{eq:x} and
$\hat G_{\tinytext{A}}=\Lambda^2_\tinytext{QCD}
G_{\tinytext{A}}$. \labelcref{eq:Gdecouple} is the leading term of
\labelcref{eq:prop_gen} in an expansion around $p_0 = 0$.  The log-term in
\labelcref{eq:x} arises naturally from the momentum integration of the ghost
loop in the IR, see \Cref{fig:GluonPropLogs}. There we have
depicted the ghost loop in both the functional renormalization group
(FRG) equation for the gluon propagator, see e.g.~\cite{Cyrol:2016tym}
for more diagrammatic details, and the Dyson-Schwinger equations (DSE), see
e.g.~\cite{Alkofer:2000wg} for more diagrammatic
details. Both depend on the ghost propagator and the ghost-gluon
vertex. In the decoupling case the ghost propagator has a trivial
infrared behavior proportional to $1/p^2$.

As a side remark
we mention that the ghost propagator is not augmented with a leading order 
logarithmic IR running, even though this would not change the present
analysis. The absence of a leading order logarithmic IR running in the
ghost propagator can be shown along a similar line of arguments as
done here for the gluon propagator. 

\subsubsection{Sources for infrared logarithms}  
We now proceed with the discussion of the IR behavior of the gluon
propagator.  The ghost-gluon vertex dressing tends towards a constant
value with a small angular dependence for small momenta, while the
ghost propagator dressing also tends towards a constant. 
The low momentum triviality of
the ghost-gluon vertex is related to the non-renormalization theorem
for the ghost-gluon vertex in the Landau gauge. It also can be seen
from \Cref{fig:IrTrivialGhostGluonVertex} which features the gapped
gluon propagator and hence is infrared suppressed. This property holds
for all mixed ghost-gluon correlations. For a very detailed and
extensive discussion in the context of a perturbative one-loop
analysis of the Curci-Ferrari setup we refer to
\cite{Reinosa:2017qtf}. In summary the ghost loop depicted in
\Cref{fig:GluonPropLogs} gives rise to a $p^2\ln p^2$ contribution
with a negative prefactor in the IR. This is reflected by $z_G>0$ in
\labelcref{eq:x}.  The other diagrams contain the gapped gluon propagator or
non-classical vertices. Consequently these diagrams cannot contributed 
to the logarithmic running at one-loop. 

Beyond one-loop further contributions to the IR logarithm could
originate from the logarithmic running of the vertices. This scenario
was behind the discussion of the Higgs phase for large explicit gluon
masses in \cite{Cyrol:2016tym}. These contributions would have the
potential of switching the sign of $z_G$. Again such a running can
only be triggered by ghost loops due to the gapping of the
gluon. Hence, from an iterative point of view, they first can only
occur for purely gluonic vertices triggered by the massless ghost loops
contributing to these vertices. If created, they can propagate to all
correlation functions via diagrams with at least one purely gluonic
vertex. For the propagator the three- and four-gluon vertices are
relevant, for the respective diagrams see \Cref{fig:GluonicVertices}.
Indeed these vertices feature logarithmic terms at one loop, see
e.g.~\cite{Cyrol:2016tym,Aguilar:2013vaa,Pelaez:2013cpa,Blum:2014gna,Eichmann:2014xya,Cyrol:2014kca}.

The mere occurrence of logarithmic terms in the vertices is not
sufficient for triggering an additional logarithmic running of the
gluon propagator.  Consider for example a logarithm of the form
$\ln(p^2 +q^2)$, where $q$ is the loop momentum of a given diagram for
the gluon propagator $G_A(p)$. Then the loop integration effectively
removes this logarithm as the gluon in the diagram is
gapped. Consequently only logarithmic terms of the form
$(p_i)_\mu f_i(p_1,p_2) \ln p^2$ for the three gluon vertex, and
$f(p_1,p_2,p_3)\ln p^2$ for the four-gluon vertex would trigger
$p^2 \ln p^2$-terms in the propagator. Here $p$ is one of the 
momenta $p_1,...,p_n$ with $p_n=-(p_1+\cdots +p_{n-1})$ in an $n$-gluon vertex. 

Even though the presence of such terms would be of great interest for
the effective detection of a possible Higgs phase
\cite{Cyrol:2016tym}, a complete analysis is beyond the scope of the
present work. Here we simply remark that the terms of the required form 
are singled out by the infrared limit of one momentum $p\in (p_1,...,p_n)$ with 
\begin{align}\label{eq:logVertex} 
\lim_{\hat p^2\to 0}|\partial_{p^2} \Gamma^{(n)}_\tinytext{gluonic}(p_1,...,p_n) |\to\infty\,, 
\end{align}
for the three- and four gluon vertices, $n=3,4$, at fixed other
momenta. For this limit one can concentrate on the propagators
attaching the ghost-gluon vertex with the momentum $p$. In the above
limit they only carry the loop momentum, but are multiplied by
$q_{\mu_i}$ from the respective ghost-gluon vertices. Hence they
diverge as $1/q^2$. The derivative w.r.t.\ $p^2$ triggers another
$1/q^2$: Applied to th ghost propagator $G_c$ that carries the
external momentum $p$, we are led to
$\partial_{p^2} G_c(q+p)^2 \propto (1/q^2)^2$ in the limit $p\to 0$.
In summary this leaves us with a logarithmic singularity due to
$d^4 q \,1/q^4$.  The other propagators in the diagrams still carry
other external momenta and do not add to the singularity. 

In summary, the kinematic analysis above hints at the existence of the
logarithmic terms in the gluonic vertices that act as additional
sources for the logarithmic running of the propagator. Note however,
that a decisive answer requires an analysis that also takes into
account the underlying gauge symmetry: first of all the Slavnov-Taylor
identities (STIs) connect the different diagrams in the functional
equations for the gluon propagator. Second, the STIs also restrict the
vertex structures themselves and the prefactors of the logarithmic
vertex corrections may even vanish for fully dressed vertices. It goes
without saying that even for being indicative such an analysis
requires at least a full two-loop analysis of the gluon propagator in
the presence of a mass gap. In this context we mention a very careful
complete and illuminating perturbative analysis at one-, two and three
loops in \cite{Gracey:2018fkg,Gracey:2017yfi,Gracey:2014ola} in QCD
and \cite{Bell:2015dbr,Gracey:2014dna} in Curci-Ferrari-type models,
and also references therein. 

Accordingly, the logarithms produced always depend on sums of
combinations of external momenta squared. This kinematic argument
entails that vertex logarithms always depend on loop momenta and hence
do not contribute to $z_G$. Note that this argument, upon iteration,
holds for fully non-perturbative resummations as done within
functional methods. We emphasize that evidently this proof necessitates
both the logarithmic corrections of vertices as well as the
logarithmic corrections that originates from the massless propagators
in the loop. Hence, conclusive arguments should at best make
systematic use of the full iterative structure of resummation schemes
as done here, or exploit perturbation theory at two loop and
beyond. The latter ensures in most cases that the perturbative
structure mimics of the iterative structure of non-perturbative resummations.

\subsubsection{Potential Higgs branch}
For its relevance we come back to the Higgs-phase argument in
\cite{Cyrol:2016tym}, even though it is a bit outside the line of
arguments here. The existence and properties of such a Higgs phase are
not only important for the Standard model but also for finite
temperature QCD, where the temporal gauge field plays the
r$\hat{\rm o}$le of a Higgs field. In~\cite{Cyrol:2016tym} the
dynamics of such a Higgs field was trivially mimicked by an explicit
mass term of the gluon despite of its dynamical structure. The present
analysis makes it apparent why such an argument falls short. In the
presence of a Higgs mechanism one resorts to $R_\xi$-gauges
that leads to massive ghosts in the Higgs phase with the ghost mass
proportional to the expectation value of the Higgs. Within the present
setup this has been discussed in \cite{Fister:2013bh}.  There it has
been also shown that this mechanism has an equivalent in the standard
Landau gauge.  In Landau gauge the Higgs-Kibble dinner is not
apparent. Still, the effect of the massless ghosts is more than
balanced by that of the Goldstone modes. In \cite{Fister:2013bh} it
has been shown that this leads to a deconfining Polyakov loop
potential in the Higgs phase. In the present context it entails that
the Goldstone contributions to the gluon in  Landau gauge are an
additional source of the $p^2\ln p^2$ running of the gluon propagator,
that can turn the sign of $z_G$: this
simply follows from the similarity of the Higgs-gluon vertex to that
of the ghost-gluon vertex and a respective perturbative analysis. A
more detailed analysis is far beyond the scope of the present work and
deferred to future work.

\subsection{Scenarios for analytic structures of the decoupling solution}  \label{sec:DecouplingScenarios}

Now we proceed with our main line of arguments. Even though
sufficiently smooth, the non-trivial angular dependence and the
sub-leading momentum-dependence will still almost certainly modify the
complex structure. Nonetheless \labelcref{eq:x} still provides a very good
parametrization in the infrared. Accordingly, in contradistinction to the
scaling solution it is not possible for the decoupling solution to
determine its analytic structure from the IR asymptotics. The
difference between parameterizations cannot be resolved in currently
available Euclidean data as the effects are sub-leading in the
Euclidean IR domain. Nevertheless, the basic form and generation of
terms is well motivated and an investigation of the IR behavior is
still sensible for the case of the decoupling solutions. It allows us
to classify two likely scenarios for the analytic structure of the
decoupling type gluon propagator:

\subsubsection{Scenario I}\label{sec:scen1}
We start with the parametrization given in \labelcref{eq:prop_gen}
since it is the simplest one capturing the Euclidean behavior.
Keeping a finite $\gamma_G$ in \labelcref{eq:prop_gen}
this parametrization of the decoupling propagator can be reduced to 
\begin{align} \label{eq:prop_decoupling}
  \hat G_{\tinytext{A}}^{\tinytext{(dec)}}(p) = \tilde{Z}_\tinytext{IR}
  \left( \tilde{m}^2_{\text{\tiny{gap}}} + \hps \ln \hps \right)^{-1}
  \, ,
\end{align}
after absorbing $\gamma_g$ and $z_G$ by appropriate redefinitions of
$Z_\tinytext{IR} \to \tilde{Z}_\tinytext{IR}$ and
$m^2_{\text{\tiny{gap}}} \to \tilde{m}^2_{\text{\tiny{gap}}}$.  The
parametrization \labelcref{eq:prop_decoupling} admits complex conjugated
poles, which lead to a modification of the simple spectral
representation \labelcref{eq:specrep}.  Allowing for additional poles, we
make use of the extended spectral representation \labelcref{eq:specrep-gluon}.
This enables us to
separate cut and pole contributions of \labelcref{eq:prop_decoupling}, a
detailed description of the analytic structure can be found in
\Cref{App:dec_poles}. Specializing \labelcref{eq:prop_decoupling} to the
contribution of the cut, i.e. the one contributing to
$\rho_{\tinytext{A}}^{(\tinytext{dec})}$ we obtain with
\labelcref{eq:key_relation}
\begin{align}	\label{eq:spectral_decoupling}
	\hat \rho_A^{\tinytext{(dec)}}(\omega)=
	-\hat{Z}_\tinytext{IR}\sign(\hat{\omega})\frac{2 \pi}{\hat{m}^4_\tinytext{gap}} \hat{\omega}^2
	+\mathcal{O}(\hat{\omega}^4 \ln \hat{\omega})
	\, .
\end{align}
Again, most notable is the negative sign in front of \labelcref{eq:spectral_decoupling},
leading to a negative spectral function at low frequencies for the parametrization
\labelcref{eq:prop_decoupling} of the decoupling solution.

\subsubsection{Scenario II}
\label{sec:scen2}

As already mentioned above, the form \labelcref{eq:prop_decoupling} is not
unique, and cannot be fixed by presently available data. Indeed,
another admissible parametrization removes the additional poles in
\labelcref{eq:prop_decoupling}. Then the propagator exhibits a single cut.
We keep the same leading order expansion in $p_0=0\,$, which renders
all differences sub-leading in the Euclidean data in the IR. A
possible parametrization with these properties is given by 
\begin{align} \label{eq:diff_prop_decoupling}
	\hat G^{\tinytext{(dec)}}(p) =
 	\tilde{Z}_\tinytext{IR} \tilde{m}^{-2}_{\text{\tiny{gap}}} \left(
	1 + \tilde{m}^{-2}_{\text{\tiny{gap}}} \, \logG(\hps)  
	\right)
	\, .
\end{align}
Here, $\logG$ is the logarithmic $\Gamma$ function (not the logarithm
of the $\Gamma$ function).  The logarithmic $\Gamma$ function has a
branch cut for $\mathrm{Re}\, z = 0$ and is analytic everywhere else.
It is defined by
\begin{align} \label{eq:logGamma} \logG(z) =
  \sum_{k=1}^{\infty}\left(\frac{z}{k} -
    \ln\left(1+\frac{z}{k}\right)\right) - \gamma_{\text{\tiny{E}}}\, z - \ln(z) \, , 
\end{align}
with the Euler-Mascheroni constant $\gamma_{\text{\tiny{E}}}$. Both parameterizations
lead to the same leading order term in the propagator,
\begin{align} \label{eq:eucl_IR_expansion} 
\hat G^{\tinytext{(dec)}}(p) =
  \tilde{Z}_\tinytext{IR} \tilde{m}^{-2}_{\text{\tiny{gap}}} \left(1 -
    \tilde{m}^{-2}_{\text{\tiny{gap}}} \, \hps \ln(\hps) \right) +
  \mathcal{O}(\hat{p}^4) \,.
\end{align}
This implies the same low frequency behavior of the
spectral function as in \labelcref{eq:spectral_decoupling} and demonstrates
explicitly the remaining freedom in parameterizing the decoupling
solution while possibly modifying the corresponding spectral
function. It is important to note however that the leading term in
$\partial_{p^2} G$ is the one containing the logarithm, whose sign
cannot be flipped and from which the non-positivity of the 
small frequency spectrum arises. More details regarding the
propagator at zero can be found in \Cref{app:der-analytic}.
We close this section with a word of caution: While a large
class of parameterizations may yield the same spectral function, as it
is the case here, this is by no means guaranteed.


\subsection{Realizations of decoupling solutions} 
In the following subsections several approaches or models that feature
decoupling type solutions are discussed. In most cases the gluon
propagators results in the approaches are worked out in specific
expansion schemes that allow us assigning one the above scenarios described in 
\Cref{sec:scen1}, \Cref{sec:scen2} to
it. Note that this does not necessarily entail the correct analytic
structure of the gluon propagator in the given approach but certainly
that of the given expansion order. Note also, that the systematics of
generic expansion schemes in the analytic functional approaches
suggests the persistence of the analytic structure if resummed
vertices are taken into account. However, a detailed analysis
is beyond the scope of the present work.

\subsubsection{Lattice}\label{sec:lattice}

Our discussion is based on the plethora of lattice results for 
decoupling gluon propagators at vanishing and finite temperature
\cite{Sternbeck:2006cg,Cucchieri:2007ta,Cucchieri:2007rg,Cucchieri:2008fc,Maas:2009ph,Maas:2009ph,Maas:2009se,Aouane:2011fv,Maas:2011ez,Cucchieri:2012gb,Sternbeck:2012mf,Silva:2013maa,Maas:2014xma,Maas:2015nva,Duarte:2016iko}, 
for recent analytic fits to high precision data see,
e.g., \cite{Gao:2017uox} and \cite{Dudal:2018cli}. However, in our
opinion the distinction between the different scenarios
\Cref{sec:scen1}, \Cref{sec:scen2} still requires a far higher
precision. Accordingly, without additional information it is not
possible to differentiate between any of the possible
parameterizations of the decoupling scenario. Therefore statements
about the analytic structure of the gluon propagator based on lattice
data is currently not possible.

Several reconstructions based on lattice data have been performed in
the past years. In \cite{Dudal:2013yva} a reconstruction was
presented using simulations results from low temperature quenched
lattice QCD. The authors deployed the Tikhonov regularization to
extract the spectral function and observed a negative
contribution at low frequencies.

Finite temperature studies have also been carried out. A reconstruction including
finite temperature gluon propagators in quenched QCD based on a modified
Maximum Entropy Method (MEM) was presented in \cite{Haas:2013hpa}. 
The results were mostly positive for small
frequencies  by construction, due to the modified MEM approach for non-positive spectra. 
This method has also been applied to decoupling FRG data, see
\Cref{sec:FRG+DSE}, and the two reconstructions give similar
results. 

\begin{figure}
\includegraphics[trim=0cm 2cm 5cm 0cm, clip=true]{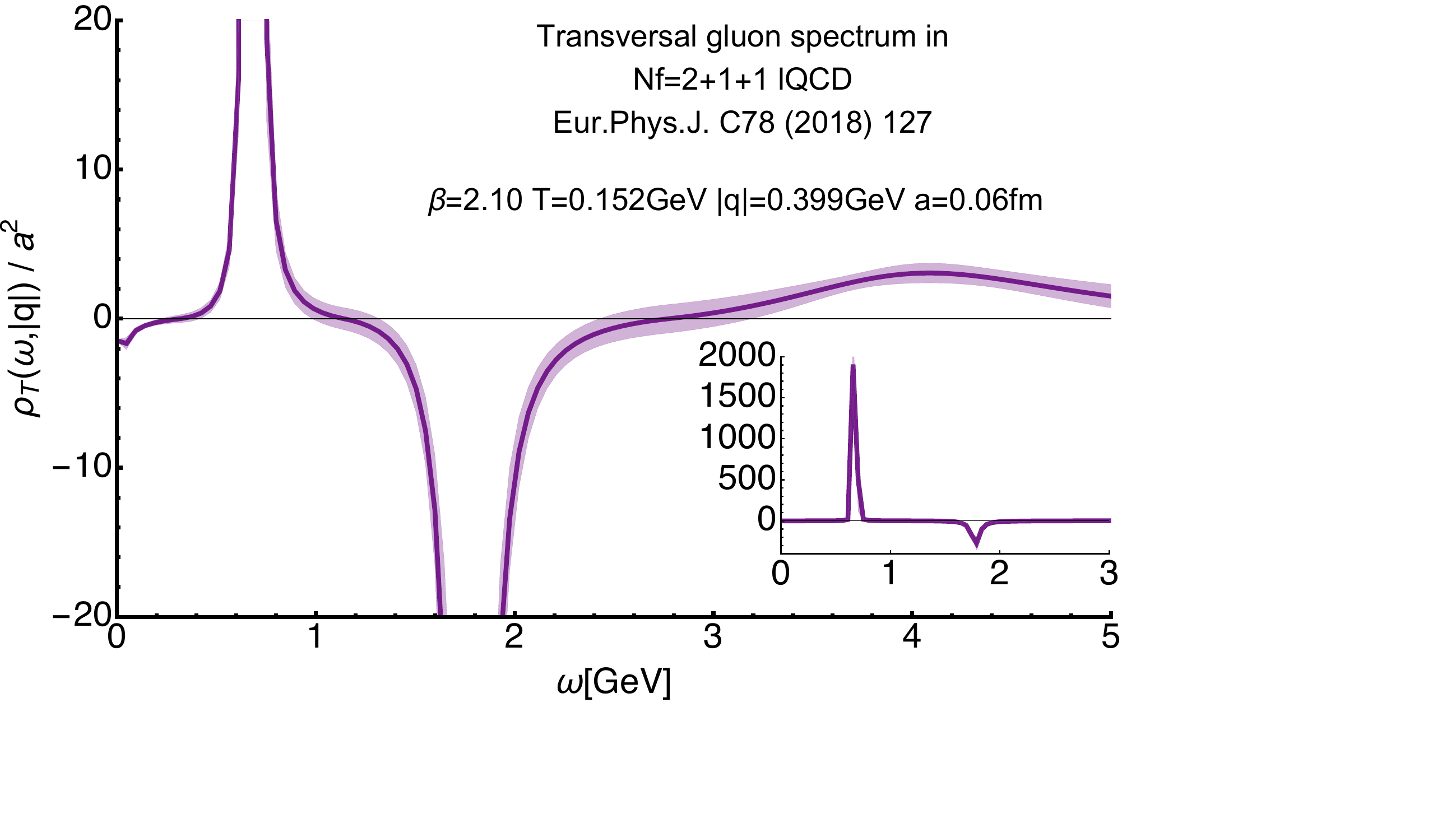}
\caption{Example of a gluon spectral function in the confined phase
  $T=\SI{152}{\MeV}<T_C$ extracted from lattice QCD simulations by the tmft
  collaboration including $N_f=2+1+1$ flavors of quarks. Note that
  while the higher lying negative feature at around $\SI{1.75}{\GeV}$ is
  strongly pronounced, we also find indications for a residual
  negative contribution at small frequencies. The overshoot into
  positive values at higher frequencies originally thought of as a
  Bayesian artifact also emerges in our reconstruction
  presented below. Error bars include both statistical and systematic errors, for
  details see \cite{Ilgenfritz:2017kkp}.}\label{Fig:Nf211Spec}
\end{figure}

Another reconstruction based on a Bayesian approach has been performed
in \cite{Ilgenfritz:2017kkp} using finite temperature lattice QCD
data, featuring $N_f=2+1+1$ quark flavors. The generalized Bayesian
Reconstruction (gBR) approach \cite{Rothkopf:2016luz} deployed in that
study revealed that in the confined phase the gluon spectrum exhibits
a small residual negative contribution at small frequencies, see
\Cref{Fig:Nf211Spec}. Any sign of this negative structure
disappeared at higher temperatures, however the systematic
uncertainties in the study precluded a definite statement, whether
that was a genuine finite temperature effect.
We see the finding of a negative low frequency part as a strong
indication that the Bayesian reconstruction method (gBR) in
\cite{Ilgenfritz:2017kkp} recognizes the low frequency relation
derived in this work. 

In summary, the discussion of the low frequency limit of the gluon spectral
function and of the analytic structure suggests to revisit the
spectral reconstruction of the gluon spectral function based on
improved analytic models that incorporate the logarithmic corrections 
of the gluon propagator. As the logarithmic terms might be difficult to 
extract directly even from the high precision lattice data, it calls for 
a combined lattice-functional methods approach: the logarithmic terms  
could be constrained by using combined propagator and vertices lattice data and 
lattice consistent results from functional methods. In the latter the logarithmic 
infrared terms can easily be extracted.

\subsubsection{DSE and FRG}\label{sec:FRG+DSE}
Decoupling-type propagators have been computed in both DSE and in FRG
calculations in good agreement with the corresponding lattice results,
see e.g. \cite{Aguilar:2008xm,Boucaud:2008ky,Fischer:2008uz} and
\cite{Fischer:2008uz, Cyrol:2016tym}, respectively. Within the DSEs a
direct solution has been computed in the complex plane in
\cite{Strauss:2012dg}, where a single branch cut along the
$\mathrm{Re}\, p_0 = 0$ axis was found. The spectral function found
there stays positive for very small frequencies. Hence the analytic
structure has to violate implicitly our smoothness condition, which 
is very interesting and requires a more detailed investigation.  

As mentioned already in the previous \Cref{sec:lattice}, decoupling
FRG data as well as lattice data for the finite temperature gluon
propagator have been used for a reconstruction of the gluon spectral
function for temperatures $T\geq \SI{100}{\MeV}$ in \cite{Haas:2013hpa}. 
Both, the reconstruction of the lattice data and that of the FRG data
have been in very good agreement with each other. 

Moreover, the results are in qualitative agreement with that of the
direct DSE computation of \cite{Strauss:2012dg}: The finite
temperature data show a thermal broadening. The MEM-type method used
in \cite{Haas:2013hpa} run into accuracy problems for smaller
temperatures. This is a typical sign of a sharp peak in the spectral
function. A low temperature extrapolation of the thermal spectral
functions gives rise to a sharper peak at $T=0$, but no quantitative
statement was possible due to the missing small temperature accuracy.
Note also, that the reconstruction method used in \cite{Haas:2013hpa}
leads to a positive low frequency tail almost by construction. Apart
from this disagreement the results there are also in qualitative agreement 
with the reconstruction of the scaling spectral function presented in 
\Cref{sec:SpecFunc:Gluon} shown in \Cref{fig:MainResult}. 

In summary, as already mentioned at the end of \Cref{sec:lattice},
the situation calls for a combined lattice-functional methods
approach in order to minimize the systematic error of the
reconstruction.

\subsubsection{Gribov-Zwanziger approach}\label{sec:GZ}
The complex structure arising in the Gribov-Zwanziger approach has
been discussed in \cite{Baulieu:2009ha} at the example of a toy model
with complex conjugated poles. The current state of the art comparison
with lattice data \cite{Dudal:2018cli} resorts to a tree-level
propagator with a perturbative RG improvement that captures the
ultraviolet running. It reads
\begin{align}\label{eq:GZ-fit} 
G_A(p^2) = \frac{p^2+M_1^2}{p^4+M_2^2 p^2+M_3^4}\left[
\ln \frac{p^2+m_g(p^2)}{\Lambda^2_{\text{\tiny{QCD}}}}\right]^\gamma
\end{align}
with the one-loop anomalous dimension $\gamma=13/22$ introduced in
\labelcref{eq:gluonandim}. The regularization mass $m_g(p^2)$ is finite in the IR
for $p^2 \to 0$ and either decays or also stays finite in the UV for $p^2\to\infty$.
\labelcref{eq:GZ-fit} is sufficient to capture the high frequency behavior
as well as the non-perturbative gapping of the gluon. Its complex
structure features the perturbative cut as well as complex conjugated
poles. The spectral function that follows from the propagator
\labelcref{eq:GZ-fit} is subject to the infrared relation \labelcref{eq:key_relation}. 
Evidently, the sign of the spectral function depends on the
combination of parameters chosen in \labelcref{eq:GZ-fit}, for the best fits provided in
\cite{Dudal:2018cli} it is negative. 

A one-loop analysis of the GZ approach reveals a logarithmic IR
momentum scaling that originates in the gauge-fixing contributions
similar to the ghost contribution in the Landau gauge. \labelcref{eq:GZ-fit}
lacks this logarithmic IR running that leads to the negative sign of
the spectral function for low frequencies. As it is not built in
naturally in \labelcref{eq:GZ-fit} it suggests to simply restrict the range
of allowed parameters by
\begin{align}
\lim_{p^2\to 0}\partial_{p^2} G(p^2) >0\,, 
\end{align}
which mimics the divergent limit \labelcref{eq:GAlimit} insofar that it
reproduces \labelcref{eq:signrhoAGA}, and hence the correct sign of the
spectral function at low frequencies.  Alternatively the propagator
model \labelcref{eq:GZ-fit} can be amended by an cut.  In either case this
enhances the predictive power of the reconstruction.

\subsubsection{Curci-Ferrari model}\label{sec:CF}
The Curci-Ferrari model \cite{Curci:1976kh} is a massive version of
Yang-Mills theory.  As such it features an additional relevant
coupling, the gluon mass, and reduces to Yang-Mills theory in the
-appropriate- massless limit. In recent years, the model has seen
revived interest in the context of modeling the non-perturbative mass
gap of QCD with a respective choice of the Curci-Ferrari mass
parameter. Then, a perturbative treatment of fluctuations may be
possible.  This reasoning has been introduced in \cite{Tissier:2010ts,
  Tissier:2011ey,Serreau:2012cg} where QCD correlation functions have been modeled using
perturbation theory, for a recent work see \cite{Reinosa:2017qtf}. In
the present context this is particularly interesting, as it also allows
analytically accessing the kinematic arguments given in \Cref{sec:Analytical:IR:Decoupling}. 

The one-loop contribution to the gluon propagator has been calculated
and discussed in \cite{Tissier:2010ts,Reinosa:2017qtf}. It features an
asymptotic IR behavior of the form \labelcref{eq:prop_decoupling}, its
infrared properties and the relation to positivity violation have been
discussed extensively in \cite{Reinosa:2017qtf}. We are led to a
negative low frequency spectral function of the form
\labelcref{eq:spectral_decoupling}. Higher loop considerations may change
the global cut form as discussed in
\Cref{sec:Analytical:IR:Decoupling}, but are not relevant for the
question of the low frequency behavior.

A very detailed analysis of the complex structure of the Curci-Ferrari
model is also found in
\cite{Siringo:2015aka,Siringo:2015wtx,Siringo:2017svp}. In particular
in \cite{Siringo:2017svp} the gluon propagator in the CF-model is
worked out at one loop, leading to a low frequency spectral function
with \labelcref{eq:spectral_decoupling}.

In summary, the detailed one loop analysis in
\cite{Reinosa:2017qtf,Siringo:2017svp} shows the low frequency
properties of the spectral function derived here,
\labelcref{eq:key_relation}. Concerning the global complex structure a
two-loop analysis in the CF-framework with respect to its complex
structure would be very interesting as it features both one-loop
dressed propagators and vertices. In this context we refer the reader
to \cite{Bell:2015dbr,Gracey:2014dna} where Curci-Ferrari-type models
have been studied up to three loop. A respective analysis should 
also provide valuable additional information for
the reconstruction of Landau gauge spectral function in general.

\section{Extracting the spectral function from the Euclidean propagator}
\label{sec:SpecFunc}

The aim of this section is to reconstruct the gluon spectral function from
numerical data of the gluon propagator obtained in the scaling scenario \cite{Cyrol:2016tym}.
The final result for the spectral functions is shown in the left panel of
\Cref{fig:MainResult}.  We would like to emphasize that it exhibits all the analytic
properties discussed above up to numerical uncertainties.

If used to compute the propagator via
\labelcref{eq:specrep} it reproduces the original input with a precision
of $\sim2\%$, as shown in the right panel of \Cref{fig:MainResult}.

To arrive at the spectral function we use a novel approach based on 
an explicitly constructed set of basis functions that is carefully derived from
the analytic properties of two-point correlation functions, see
\Cref{fig:approach} for an illustration.
First, we discuss the underlying analytic structure.
Next, we introduce the explicit form of the basis and finally 
describe how it is applied to extract the gluon spectral function. 

\subsection{Analytic structure of the retarded propagator}
\label{sec:SpecFunc:Approach}

The picture we have used in the preceding sections assumes a specific 
analytic structure for the gluon propagator, i.e. that it contains a single 
branch cut at $\mathrm{Re}\, p_0 = 0$ (for more details see
\Cref{app:der-analytic}).
Therefore, we have two analytic patches in the complex plane, the retarded propagator for $\mathrm{Re}\, p_0 > 0$
and the advanced one for $\mathrm{Re}\, p_0 < 0$. They are related by
the well-known relation
\begin{align} \label{eq:prop_relation_complex}
	G^{(\tinytext{ret})}(-\imag(\omega+\imag\varepsilon)) = \left[G^{(\tinytext{adv})}(-\imag(\omega-\imag\varepsilon))\right]^*
	\, .
\end{align}
In the following we focus on the retarded propagator.
However, all statements hold equivalently 
for the advanced propagator due to \labelcref{eq:prop_relation_complex}.

The finite imaginary part in the retarded propagator at $\mathrm{Re}\, p_0 = 0$ 
signals a branch cut and therefore a finite value of the spectral function, 
which is defined as the discontinuity of the propagator, c.f. \labelcref{eq:spec_from_prop}.  
Being a holomorphic function for $\mathrm{Re}\, p_0 > 0$, the retarded 
correlation function can be analytically continued to the entire complex plane, where it is
a meromorphic function since the propagator must vanish sufficiently fast for $p \to \infty$.

Our reconstruction approach is based on an ansatz for the complex
structure of the analytically continued retarded propagator. This
has the advantage that \labelcref{eq:specrep} holds trivially and it is
possible to enforce \labelcref{eq:OehmeZimmermann}
analytically. Furthermore, the branch cuts describing the IR and UV
asymptotics can be implemented explicitly and in a straightforward manner.

\begin{figure}
	\includegraphics[width=0.4\textwidth]{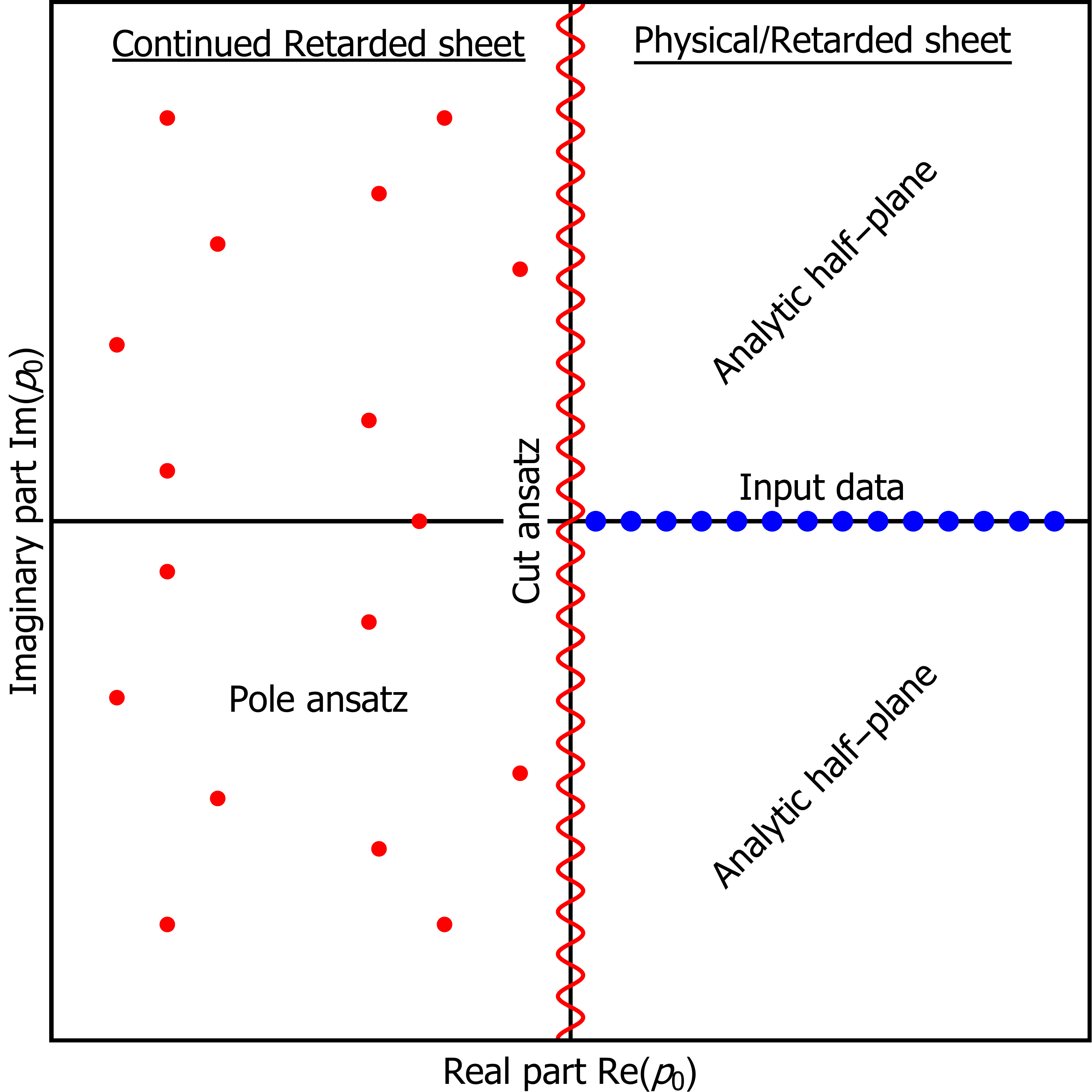}
	\caption{Schematic analytic structure of a retarded propagator. All non-analytic structures are in the $\mathrm{Re}\, p_0 < 0$ half-plane, reflecting the analyticity constraints from the existence of a spectral representation.}
	\label{fig:approach}
\end{figure}

The ansatz is build up from poles and polynomials.
This is possible since the most important, i.e. physically relevant branch 
cuts, e.g. logarithms and square roots, can be constructed from a series of poles.
Of course, branch cuts can also be taken into account directly.
If one is only interested in the reconstruction of a spectral function itself, there is
an additional freedom to choose the branch cut of e.g. logarithms, as
long as they are in the meromorphic half-plane since it does not alter
the result. Therefore a rather generic ansatz is the one depicted in
\Cref{fig:approach}, where all cuts are chosen to be on the
$\mathrm{Re}\, p_0 = 0$ axis.

\subsection{Reconstruction method}
\label{sec:SpecFunc:Reconsturction}

\begin{figure*}
	\includegraphics{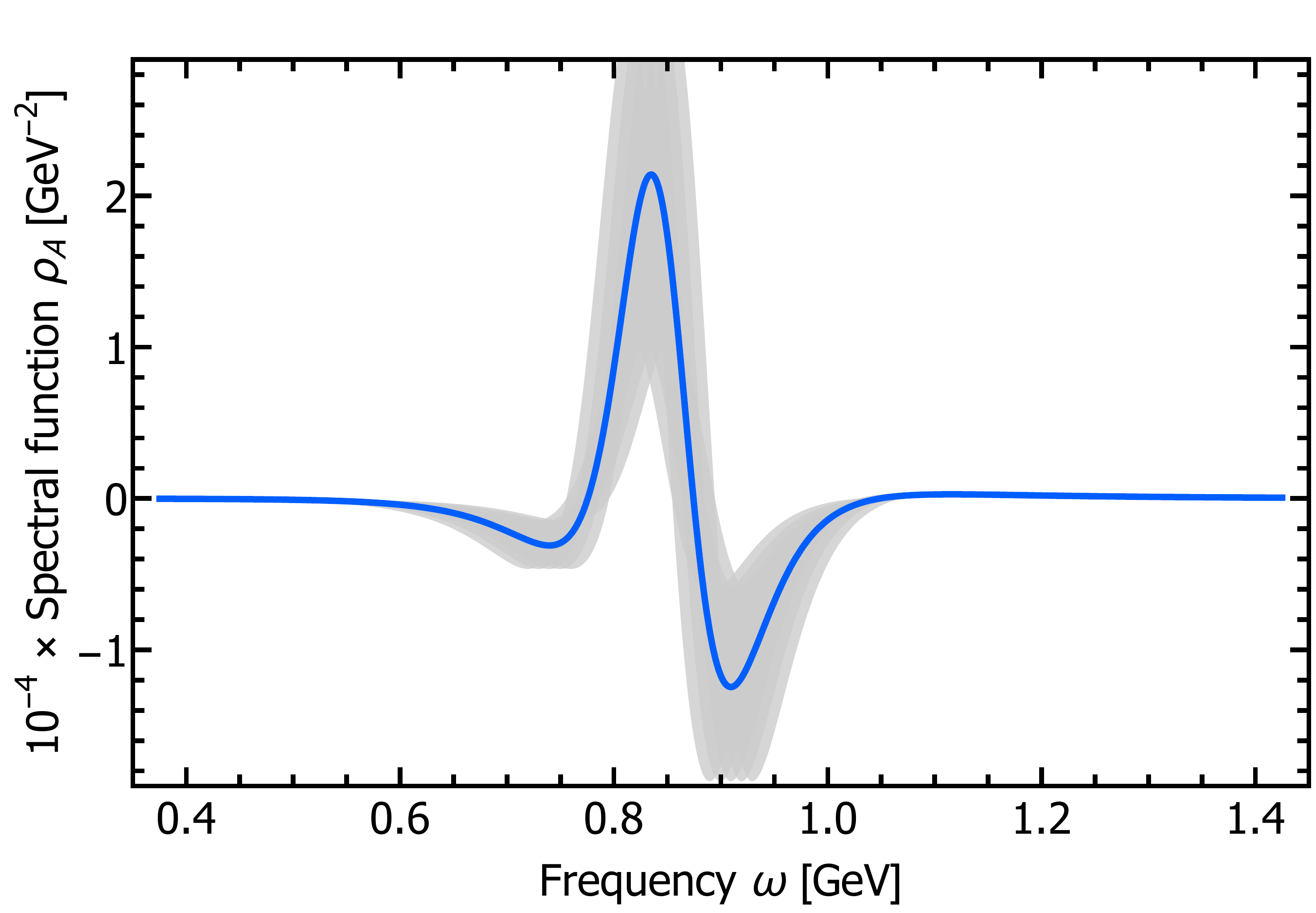}\hfill
	\includegraphics{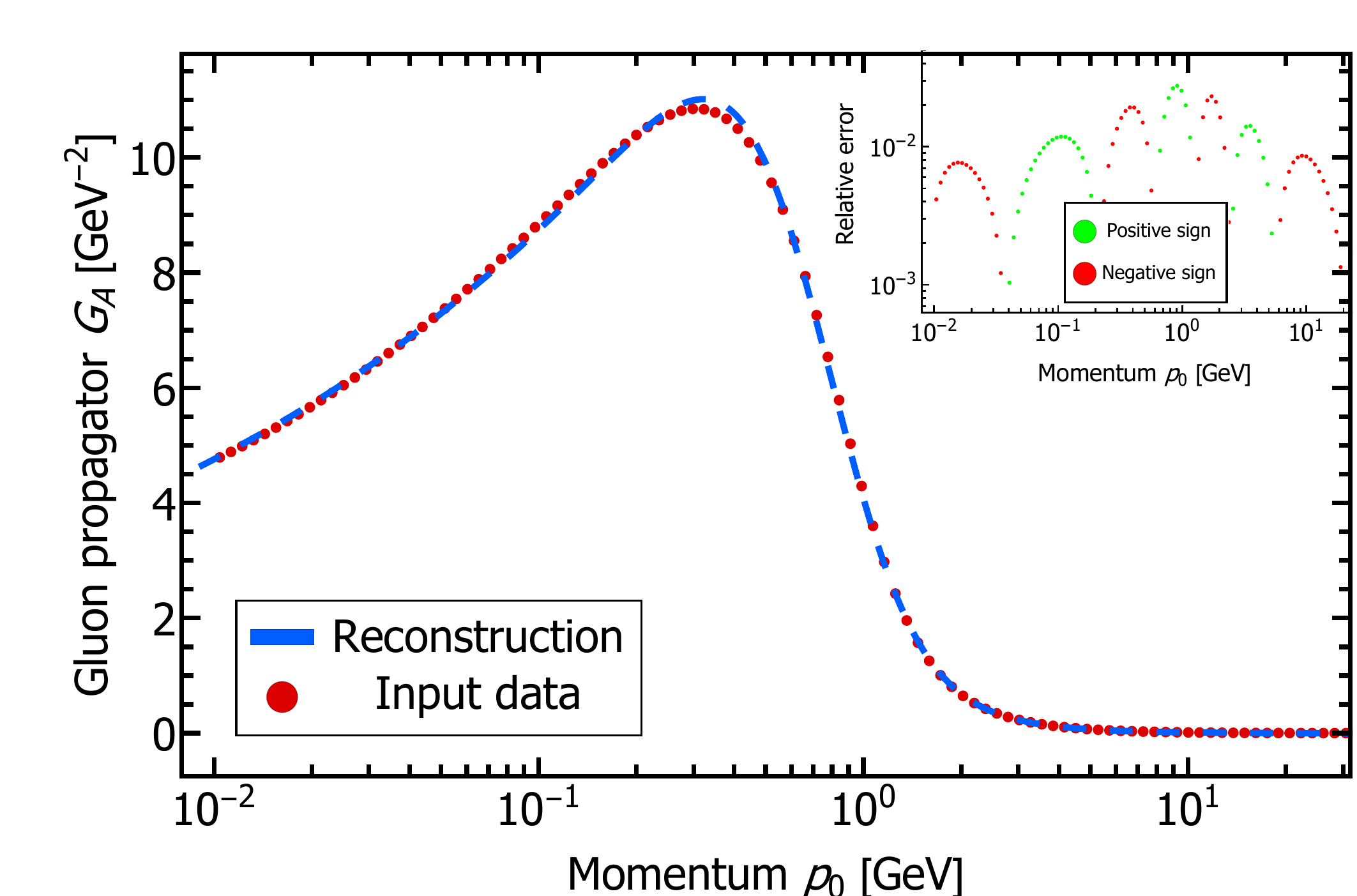}
	\caption{
		Left: 
		Gluon spectral function. The solid blue line shows our best result. The gray band around it indicates our estimate for the systematic error.
		Right: 
		Gluon propagator reconstructed from the spectral function shown in \Cref{fig:spec_limits} in comparison to the original propagator.
	}
	\label{fig:MainResult}
\end{figure*}

Our approach is based on the ability to explicitly select an
appropriate basis.  As a direct consequence, prior knowledge about the
spectral function, e.g. its asymptotics and its functional form in
general, can and should be included into the basis.  In turn, analytic
calculations can serve as a guiding principle for choosing a suitable
basis. Importantly, this does not fix the method used to determine the
coefficients of the basis. Note that the functional bases deployed in most reconstruction
procedures are chosen implicitly, such as e.g. in Bryan's MEM.

One might naively expect that by selecting a basis a priori the
ill-conditioned problem of reconstructing the spectral function from
Euclidean data becomes artificially regularized. In general this is
not the case, as the number of different structures can be chosen
arbitrarily large, as is also the case in most other reconstruction
procedures.  Our specific choice of basis only ensures that the
asymptotic and analytic properties discussed above are met.

If the number of structures permitted by the basis function is larger
than those actually encoded in the Euclidean correlator data, the
problem remains ill-conditioned and Bayesian inference needs to be carried
out, assigning a prior probability to the individual basis
parameters. The state-of-the-art implementation of Bayesian inference,
which provides insight into the full posterior probability
distribution and not simply a maximum a posteriori estimate, rests on
Hamiltonian Monte Carlo (HMC) techniques (for the industry standard
see MC-STAN \cite{JSSv076i01}).

On the other hand one may systematically reduce the number
of allowed structures in the basis ansatz until an ordinary $\chi^2$ fit
becomes stable. If at the same time such a restricted basis still
allows the Euclidean data to be reasonably well reproduced
the corresponding basis parameters constitute a valid solution.
This issue is discussed in a simple mock example in \Cref{app:other}.

\subsection{Construction of a gluon propagator basis}
\label{sec:BasisRec}

We now introduce the explicit functional form of the basis used
in the subsequent reconstruction of the spectral function. 
It consists of the several modular, dimensionless building blocks.
We start with a set of generalized Breit-Wigner structures,
\begin{align}\label{eq:BWpoles}
	\hat{G}^{\tinytext{pole}}_\tinytext{Ans}(p_0) =
	\sum_{k=1}^{N_\tinytext{ps}}
	\prod_{j=1}^{N_\tinytext{pp}^{(k)}}
	\left(\frac{\hat{\mathcal{N}}_k}{(\hat{p}_0+\hat{\Gamma}_{k,j})^2 + \hat{M}_{k,j}^2}\right)^{\delta_{k,j}}
	\, .
\end{align}
In addition, we introduce a polynomial-like structures
\begin{align}\label{eq:polyStructures}
	\hat{G}^{\tinytext{poly}}_\tinytext{Ans}(p_0) =
	\sum_{j=1}^{N_\tinytext{poly}}
	\hat{a}_j \left(\hat{p}_0^2\right)^{\frac{j}{2}}
	\, .
\end{align}
To capture the asymptotic IR and UV behaviors, we introduce the
following factor,
\begin{align}\label{eq:asymAnsatz}
	\hat{G}^{\tinytext{asy}}_\tinytext{Ans}(p_0) = 
	(\hat{p}_0^2)^{-1-2\alpha} \left[\log\left(1+\frac{\hat{p}_0^2}{\hat{\lambda}^2}\right)\right]^{-1-\beta}
	\, .
\end{align} 

The final ansatz is then given by the product of the three individual
contributions \labelcref{eq:BWpoles}, \labelcref{eq:polyStructures} and
\labelcref{eq:asymAnsatz}:
\begin{align} \label{eq:AnsatzFull}
	G_\tinytext{Ans}(p_0) =
	\mathcal{K}\, 
	\hat{G}^{\tinytext{pole}}_\tinytext{Ans}(p_0)\,
	\hat{G}^{\tinytext{poly}}_\tinytext{Ans}(p_0)\,
	\hat{G}^{\tinytext{asy}}_\tinytext{Ans}(p_0)
	\, ,
\end{align}
where $\mathcal{K}$ only carries the appropriate dimension.
The coefficients are constraint such that \labelcref{eq:specrep} holds
analytically. The
superconvergence \labelcref{eq:OehmeZimmermann} is not included
analytically, however it is realized with high accuracy, we get back
to this in \Cref{sec:SpecFunc:Gluon}.

\subsection{Gluon spectral function reconstruction and benchmarking}
\label{sec:SpecFunc:Gluon}

With the explicit form of the basis laid out above, we can continue to 
extract the gluon spectral function from gluon propagator data 
obtained in the scaling scenario~\cite{Cyrol:2016tym}.

As a full HMC analysis of the gluon propagator data is beyond the
scope of this paper. Instead, we choose the simpler strategy of
systematically reducing the number of possible structures allowed by
the ansatz. We arrive at a functional form, which permits us to
reproduce the Euclidean data of the scaling scenario within $\sim2\%$
relative deviation, as shown in the right panel of
\Cref{fig:MainResult}. At the same time this restricted basis is
simple enough that its parameters can be fixed by a standard $\chi^2$
fit.  Our best fit uses $N_\tinytext{ps} = 1$,
$N_\tinytext{pp}^{(1)}=6$ and $N_\tinytext{poly}=5$ and leads to our
final result, the gluon spectral function shown in the left panel of
\Cref{fig:MainResult}. The shape of the result is stable against a
small variation of $N_\tinytext{pp}^{(1)}$ and
$N_\tinytext{poly}$. Nevertheless, we find degenerate solutions
varying in their peak height, this is indicated by the grey band in
the left panel of \Cref{fig:MainResult}.

\begin{figure*}
	\includegraphics{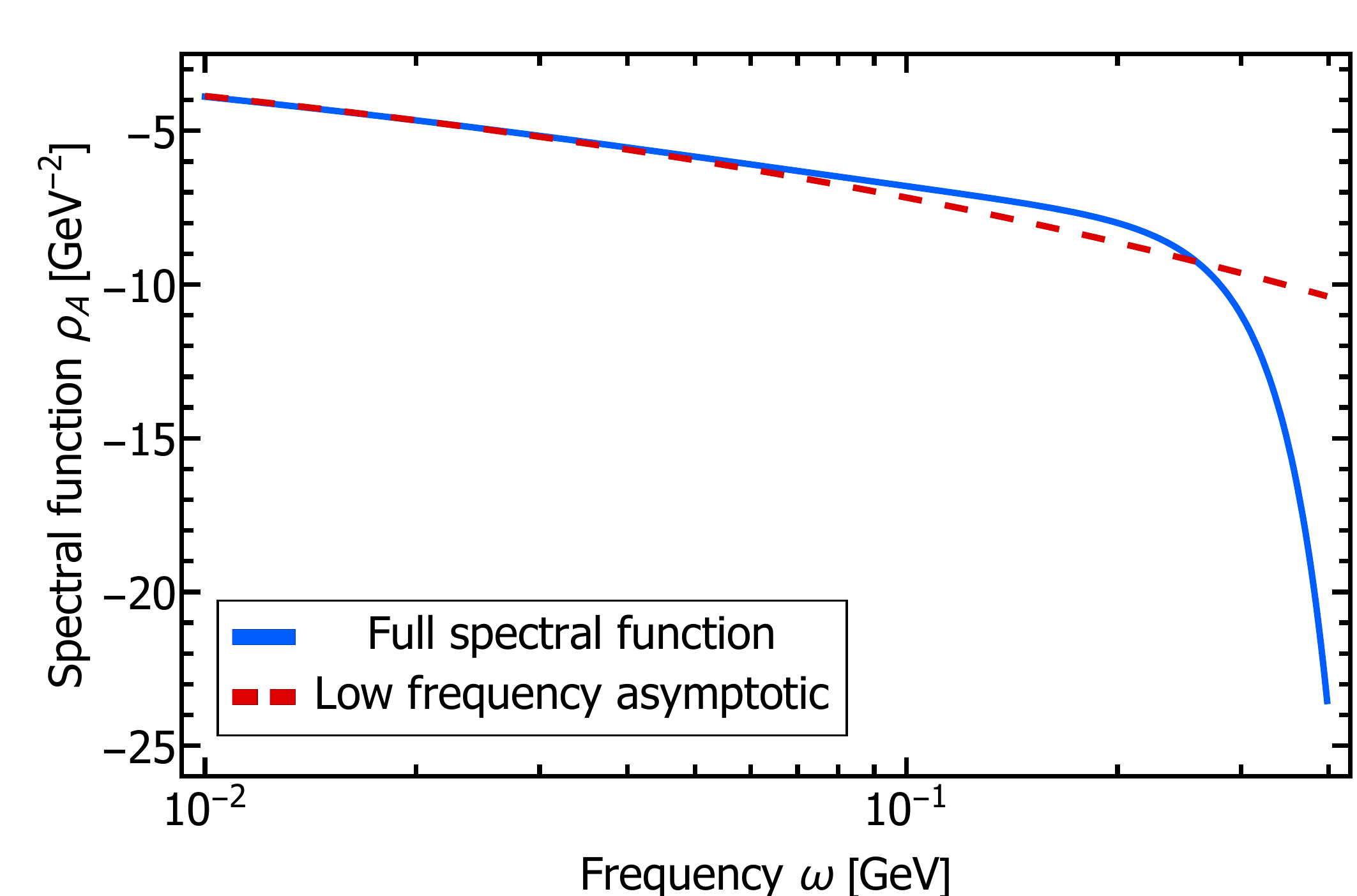}\hfill
	\includegraphics{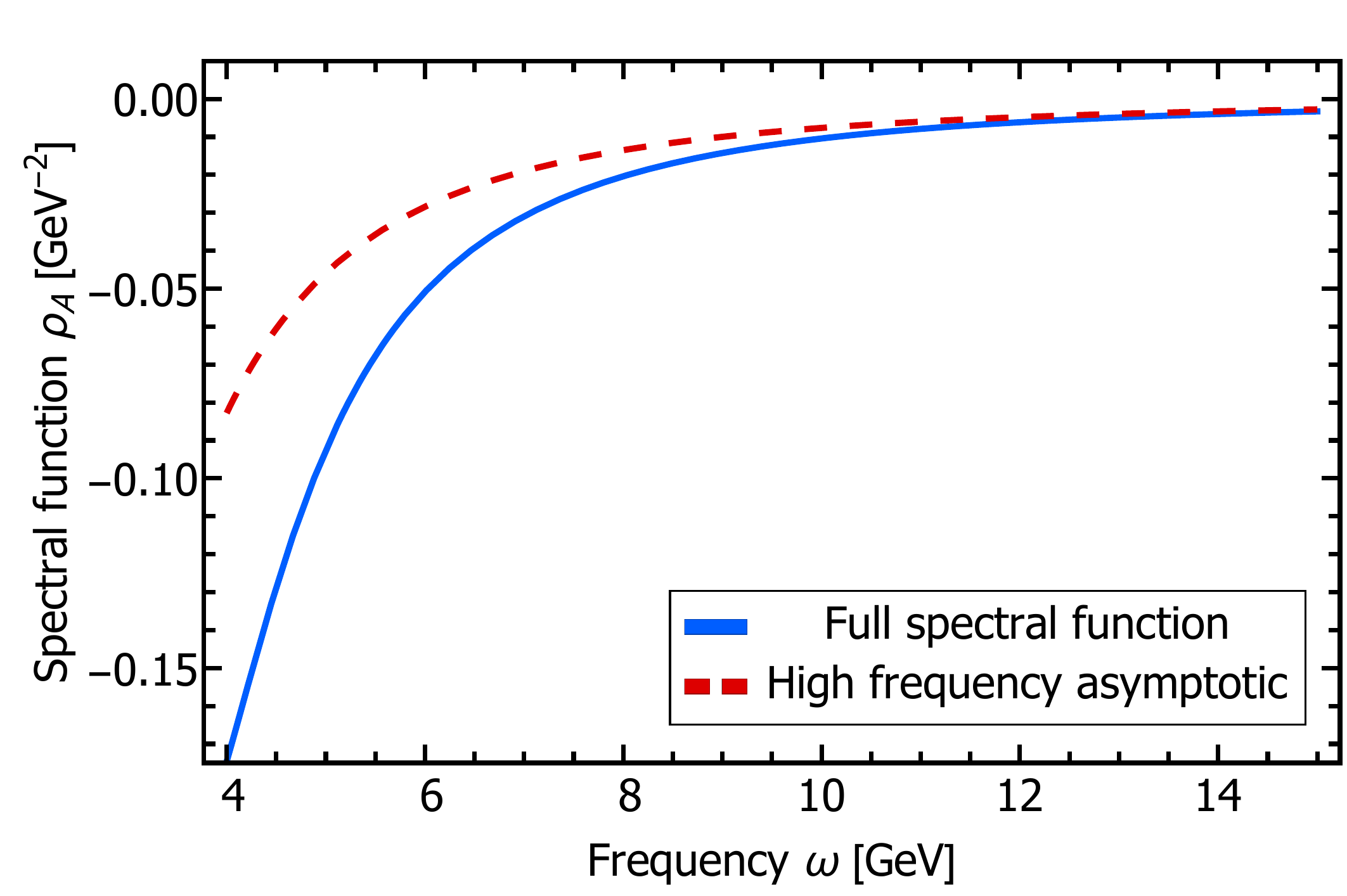}
	\caption{Low (left) and high (right) frequency behavior of our
          result for the spectral function shown in
          \Cref{fig:MainResult}.  The dashed red lines show the
          asymptotic limits given by \labelcref{eq:spectral_scaling} and
          \labelcref{eq:spectral_UV}.  }
	\label{fig:spec_limits}
\end{figure*}

The red dots in the right panel of \Cref{fig:MainResult} denote 
the numerically evaluated input data from \cite{Cyrol:2016tym}, while
the dashed line represents the the Euclidean correlator corresponding
to our reconstructed spectral function.  The inset shows the relative
error on a logarithmic scale where deviations with positive sign are
colored green, those with negative sign are colored red. While the
coefficients $\alpha$ and $\beta$ in the asymptotic part of the basis
functions are related to $\kappa$ and $\gamma$ of the IR and UV
asymptotics, respectively, we note that they do not need to match
exactly, since the former may be partially absorbed by some of the
$\delta_{k,j}$'s. We list their values in \Cref{tab:fit_params} for
completeness.  The fit is heavily constraint, as we must enforce the
complex structure as well as the correct asymptotics. As a consequence
a useful and reliable error estimation is not possible.

Let us inspect the behavior of the reconstructed spectrum in more
detail. From \Cref{fig:MainResult} we infer that the fitted
propagator, by construction, is able to reproduce the asymptotics of
the UV and the IR very well. This directly translates into the correct
asymptotic behavior of the spectral function in the IR and the UV, as
shown in the left and right panel of \Cref{fig:spec_limits},
respectively. The asymptotics are closely reproduced either below
$\omega \approx \SI{20}{\MeV}$ and above
$\omega\approx \SI{12}{\GeV}$.

\begin{table}[b]
\begin{tabular}{|c|c|c|c|c|c|}
	\hline
	$\hat{\mathcal{N}}_1$ & $\alpha$ & $\beta$ & $\hat{\lambda}$ & & \\
	1.33678 & -0.428714 & -0.777213 & 1.75049 & & \\ \hline
	$\hat{a}_1$ & $\hat{a}_2$ & $\hat{a}_3$ & $\hat{a}_4$ & $\hat{a}_5$ & \\
	0.454024 & 0.241017 & 3.10257 & -1.30804 & 0.63701 & \\ \hline
	$\hat{\Gamma}_{1,1}$ & $\hat{\Gamma}_{1,2}$ & $\hat{\Gamma}_{1,3}$ & $\hat{\Gamma}_{1,4}$ & $\hat{\Gamma}_{1,5}$ & $\hat{\Gamma}_{1,6}$ \\
	0.100169 & 0.100141 & 2.36445 & 1.5564 & 1.22013 & 1.15102 \\ \hline
	$\hat{M}_{1,1}$ & $\hat{M}_{1,2}$ & $\hat{M}_{1,3}$ & $\hat{M}_{1,4}$ & $\hat{M}_{1,5}$ & $\hat{M}_{1,6}$ \\
	0.849883 & 0.849902 & 2.52171 & 2.44035 & 3.6016 & 2.36723 \\ \hline
	$\delta_{1,1}$ & $\delta_{1,2}$ & $\delta_{1,3}$ & $\delta_{1,4}$ & $\delta_{1,5}$ & $\delta_{1,6}$ \\
	1.61116 & 1.94095 & -2.54586 & 1.89765 & 0.168592 & 0.296215 \\ \hline
\end{tabular}
\caption{Parameters obtained in our best fit for the ansatz~\labelcref{eq:AnsatzFull}.}
\label{tab:fit_params}
\end{table}

Note that the well pronounced negative trough above the main positive
peak in \Cref{fig:MainResult} does not connect directly to the
negative asymptotics but instead the spectrum returns into the
positive once more before eventually becoming negative for good,
i.e. approaching the frequency axis from below asymptotically.  This
behavior is reminiscent to what has been found in a previous lattice
QCD study \cite{Ilgenfritz:2017kkp}. While the data there was not
precise enough to capture the asymptotic behavior reliably,
indications for a similar positive bump structure above a deep
negative trough were found (see \Cref{Fig:Nf211Spec}).

Superconvergence \labelcref{eq:OehmeZimmermann} is not enforced
analytically as it would unnecessarily complicate our ansatz while
being realized already very well on a numerical level as it is only
violated by the branch point of the perturbative cut
\begin{align}
	\left( \int_{0}^{\infty} \mathrm{d}\eta\ \left| \eta \rho_{\tinytext{A}}(\eta) \right| \right)^{-1}
	 \left( \int_{0}^{\infty} \mathrm{d}\eta\ \eta \rho_{\tinytext{A}}(\eta) \right) \approx 10^{-4}
	 \, .
\end{align}

\begin{figure*}
	\includegraphics{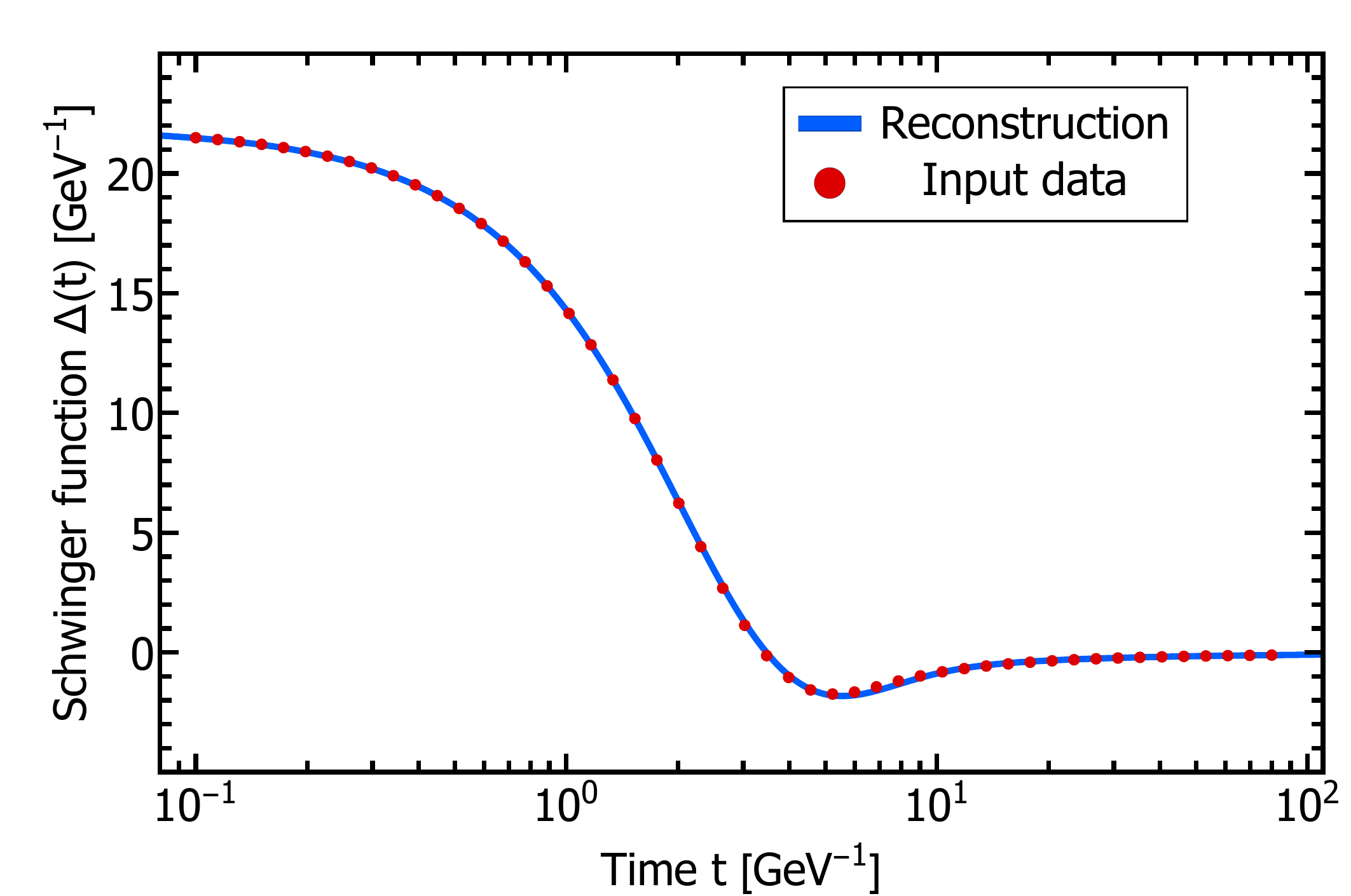}\hfill
	\includegraphics{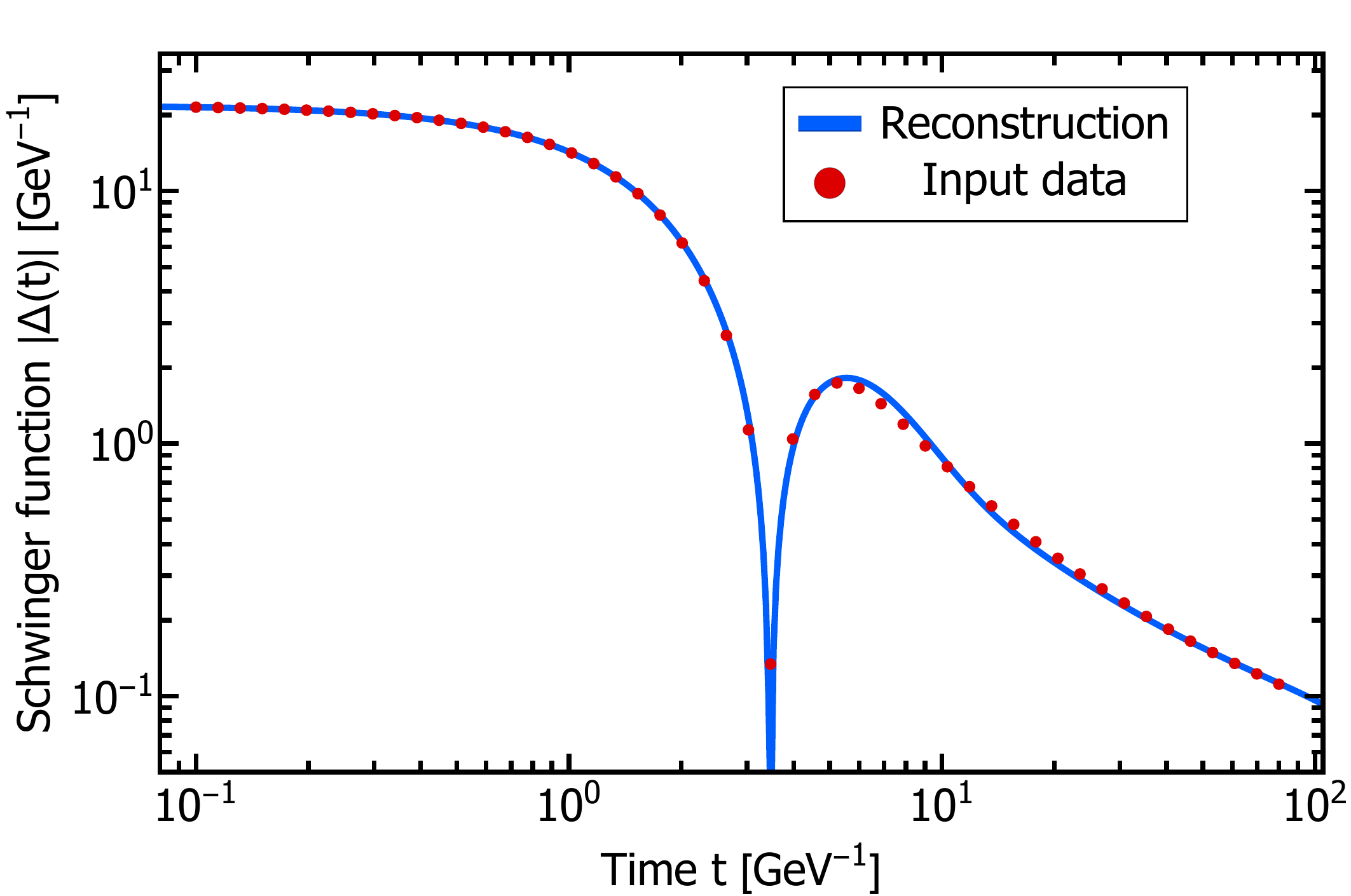}
	\caption{Schwinger function obtained from the reconstruction, blue line, and from the Euclidean input, red dots, in a semi-log (left) and log-log (right) depiction. See \Cref{fig:MainResult} for the corresponding Euclidean propagators in momentum space.}
	\label{fig:schwinger_function}
\end{figure*}

The height of the main positive and negative structure still show
rather sizable uncertainties, which is related to our fit having been
applied only to a precision of $\sim2\%$. 

The Schwinger function
\begin{align}
	\Delta (t) = 2 \int_0^{\infty} \mathrm{d}p_0\ e^{\imag p_0 t} G(p_0)
\end{align}
is potentially more sensible to differences in the peak height of the spectral function as it corresponds to a Laplace transform of the spectral function, see e.g. \cite{Pawlowski:2017gxj}.
The Schwinger functions from the reconstruction and the input data are shown in \Cref{fig:schwinger_function}. 
The point of the zero crossing between both result matches very well and the overall agreement is of the same level as for the Euclidean propagator.
We interpret this as further evidence for our successful reconstruction
of the gluon spectral function.

Performing a full Bayesian
analysis, which allows for a robust reconstruction including more
analytic structures in the basis, we expect the uncertainties of the reconstruction
to reduce further. This is however postponed to future work.

\section{Conclusion}
\label{sec:conclusion}

We have discussed the reconstruction of the gluon spectral function in
Landau gauge QCD from numerical Euclidean data, as well as its analytic properties.
In particular, we have put forward a novel reconstruction approach,
which possesses these analytic properties. It satisfies the
Oehme-Zimmermann superconvergence relation, has the correct low and
high frequency asymptotics, and reproduces the Euclidean gluon
propagator data with $\sim\SI{2}{\percent}$ accuracy.
The key to this successful reconstruction lies in two novel
ingredients:

The first one is the use of the analytic low frequency asymptotics
of the gluon spectral function in the reconstruction. The
latter is related to the IR asymptotics of the Euclidean propagator 
through the novel general relation \labelcref{eq:key_relation} that has been
derived in \Cref{sec:anspec}. The
analytic knowledge of the spectral function for $\omega\to 0$
eliminates the typically large systematic uncertainty in
reconstruction methods at low frequencies, and hence may significantly
improve the spectral reconstruction, independently of the used method.

The second novel ingredient in our approach originates in the careful
analysis of the analytic structure of two-point correlation
functions, and is described in \Cref{sec:SpecFunc}. 
This analysis leads to an ansatz for the
propagator in the complex plane that takes into account the generic
pole and cut structure. The parameters of our quite general ansatz 
can then be determined from Euclidean data.

In our opinion these two novel ingredients will improve 
the accuracy of spectral reconstructions in general,
and should be incorporated into existing Bayesian and non-Bayesian
frameworks. This is briefly discussed in \Cref{app:other}. 

We currently extent the present analysis to the finite temperature 
Euclidean data from \cite{Cyrol:2017qkl}, and the QCD 
correlation functions from \cite{Cyrol:2017ewj}.
This allows for an improved determination of 
transport coefficients
following up on~\cite{Haas:2013hpa,Christiansen:2014ypa}
as well as an access to hadronic observables.

\acknowledgements

We thank M.~Betancourt, D.~Binosi, E.~Grossi, A.~Maas, D.~Rosenblüh,
A.~Tripolt and F.~Ziegler for discussions. This work is supported by
the ExtreMe Matter Institute (EMMI) and the grant BMBF 05P12VHCTG.  It
is part of and supported by the DFG Collaborative Research Centre "SFB
1225 (ISOQUANT)".

\appendix

\section{Details of the derivation of \labelcref{eq:key_relation}}
\label{app:der-analytic} 

Our analysis that lead to the derivation of \labelcref{eq:key_relation} in
\Cref{sec:anspec} is based on the existence of a spectral
representation \labelcref{eq:specrep}. The inverse relation
\labelcref{eq:spec_from_prop} together with the existence of a spectral
representation \labelcref{eq:specrep}, that holds for the entire complex
plane, has strong implications on the analytic structure of the
propagator on the left hand side.  As a consequence the holomorphicity
of the associated retarded and advanced propagators in their respective
domain of definition, c.f.~\Cref{sec:SpecFunc:Approach}, follows
directly from Cauchy's theorem.  As we define our spectral function as
the cut in the propagator along the $\mathrm{Re}\, p_0$ axis,
i.e.~\labelcref{eq:spec_from_prop}, the spectral representation is
sufficient as condition.  Having established analyticity in one
half-plane, \labelcref{eq:key_relation} holds already by means of the
Cauchy-Riemann equations, where the additional factor two comes from
\labelcref{eq:spec_from_prop}.

In general neither the Minkowski propagator nor the spectral function
is a function in the classical sense, but a tempered distribution.
Nevertheless, the arguments about the analytic properties hold, see
e.g.~\cite{glimm1981quantum, Cuniberti:2001hm}.  Another prerequisite mentioned in the
main text is the smoothness of the spectral function at zero. In the
absence of truly distributional contribution to the integral for small
$p_0$ in \labelcref{eq:specrep}, the spectral function must go to zero
sufficiently fast, otherwise the propagator obtains at least a
log-divergence in $p_0$. On the other hand, if such contributions
would be present this argument might not hold.  A detailed discussion
about the issue of these contributions regarding the gluon and their
relation to functional methods can be found in \cite{Lowdon:2017uqe,
  Lowdon:2018mbn}. Mathematically rigorous statements in the context
of axiomatic QFT's about this issues and the relation to our
derivation are beyond the scope of this work.

\section{Poles of decoupling scenario I} \label{App:dec_poles}

More details about the poles of \labelcref{eq:prop_decoupling} are collected here
We drop the normalization $Z_\tinytext{IR}$ and additional notation to keep things simple
\begin{align} \label{eq:prop_decoupling_appendix}
\hat G^{\tinytext{(dec)}}(p) =  \left( \hat{m}^2_\tinytext{gap} + \hps \ln \hps \right)^{-1}
\end{align}
Using the Lambert W-Function, with the usual index notation for the
different branches, the roots can be expressed as
\begin{align}
z^{(r)}_{0,\pm} &= \pm \sqrt{\psi_{0}} \\[2ex]
z^{(r)}_{-1,\pm} &= \pm \sqrt{\psi_{-1}}
\, ,
\label{eq:decoupling_roots}\end{align}
with
\begin{align}
	\psi_k = e^{W_{k}(-\hat{m}^2_\tinytext{gap})}
	\, .
\end{align}
There are now three different cases for the additional poles

\begin{description}
	\item[$\hat{m}^2_\tinytext{gap} < 1/e$]
	\begin{itemize}
		\item[]
		\item[] First order poles, located pairwise on the Euclidean axis ($\Im(z^{(r)}_{k,\pm})=0$).
	\end{itemize}
	\item[$\hat{m}^2_\tinytext{gap} = 1/e$]
	\begin{itemize}
		\item[]
		\item[] Two second order poles at $p=\pm1/\sqrt{e}$
	\end{itemize}
	\item[$\hat{m}^2_\tinytext{gap} > 1/e$]
	\begin{itemize}
		\item[]  
		\item[] Complex first order poles which are linked by complex conjugation (as required by Lorentz invariance)
		\begin{align}
			z^{(r)}_{0,\pm} = \left(z^{(r)}_{-1,\pm}\right)^*
			\, .
		\end{align}
	\end{itemize}
\end{description} 

Introducing the additional function
\begin{align}
	\chi_k = 1+\ln \psi_k\,,
\end{align}
the residues can be computed in a straight forward manner
\begin{align} \label{decoupling_residues}
	\mathrm{Res}[\hat G^{\tinytext{(dec)}}, z^{(r)}_{k,\pm}] = (2 \chi_k z^{(r)}_{k,\pm})^{-1}
\end{align}
for $k=-1,0$.
The additional term in \labelcref{eq:specrep-gluon} can then conveniently be written as
\begin{align} \label{decoupling_add_contr}
	\text{pole term} = \sum_{k\in=\{-1,0\}} \frac{1}{\chi_{k}\left(\hps-e^{W_{k}(-\hat{m}^2_\tinytext{gap})}\right)}
	\, .
\end{align}

\section{Other reconstruction approaches}
\label{app:other}

In this appendix we discuss two reconstruction strategies, the
Bayesian Reconstruction using a quadratic prior (Tikhonov) and the
Schlessinger point method, which we had explored previous to
developing the method presented in this paper. The reason that a new
approach became necessary for the reconstruction of the gluon spectral
function lies in the individual shortcomings of the above mentioned
methods.

Our goal is to reconstruct a spectral function, which, as discussed
in the main part of this paper, should exhibit the following properties:
\begin{enumerate}
\item Normalization, as required by Oehme-Zimmermann superconvergence
  \labelcref{eq:OehmeZimmermann}
	\item Correct low frequency asymptotics,
          c.f. \Cref{sec:lowfreq}
	\item Correct high frequency asymptotics, c.f. \Cref{sec:Analytical}
	\item Respect the K{\"a}ll{\'e}n–Lehmann spectral representation \labelcref{eq:specrep}
	\item Antisymmetric around $\omega=0$, c.f. \labelcref{eq:antisym}
	\item A smooth function without drastic oscillations
\end{enumerate}
A successful reconstruction based on Euclidean data should fulfill
these requirement. We observed that neither the gBR method nor the
Schlessinger point method was able to meet all the requirements in a
satisfactory manner.

\begin{figure}[t]
\includegraphics{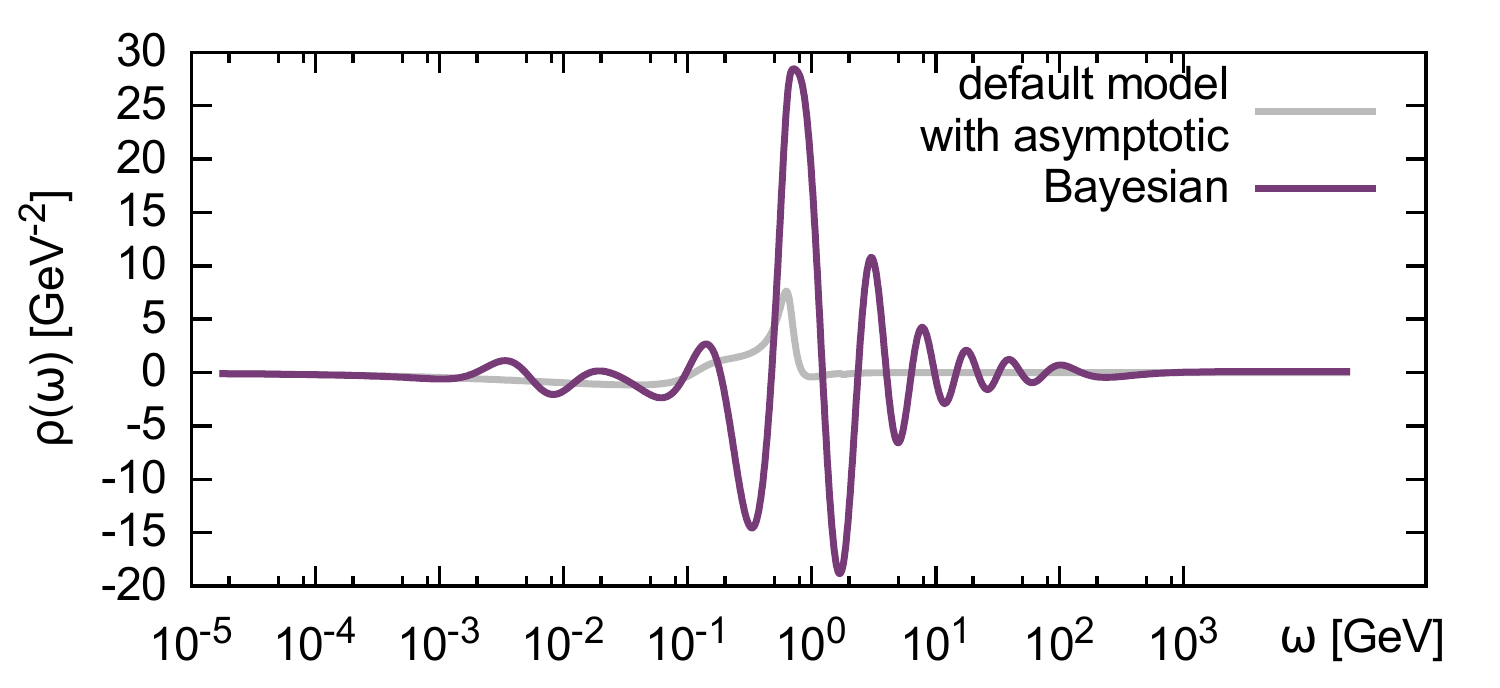}
\includegraphics{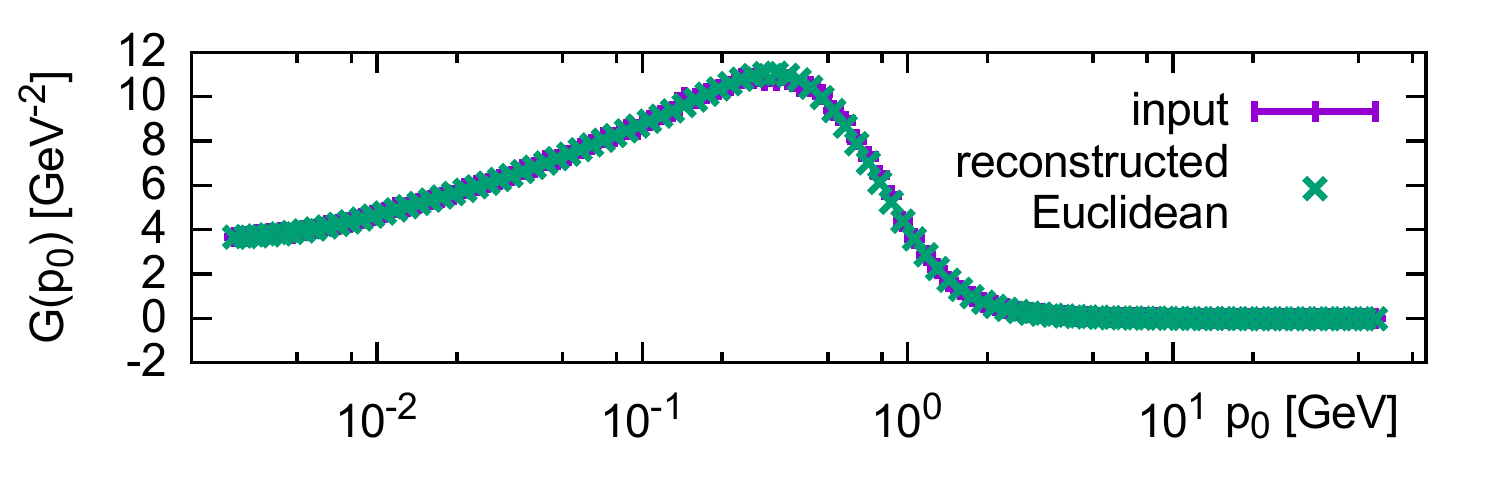}
\caption{(top) Representative example of a Bayesian reconstruction of
  the gluon spectral function based on FRG Euclidean data in the
  scaling scenario. One hundred data points with a relative error of
  $10^{-2}$ were supplied, as well as a default model (gray solid),
  which produces the correct asymptotics of the correlator.  We find
  that while the asymptotics are correctly recovered in the final
  result (solid violet) eventually, the agreement emerges far later
  than in our new method. In addition we find that the
  non-perturbative region at around $\omega\sim\SI{1}{\GeV}$ is
  contaminated by strong ringing artifacts (bottom) Comparison
  between the reconstructed Euclidean correlator (green) and the input
  data (violet).
}
\label{Fig:BayesGluon}
\end{figure}

The underlying reason for these difficulties is related to the fact
that both methods indirectly or directly choose a set of basis
functions not adapted to the problem at hand.

\subsubsection{Bayesian reconstruction}
\label{sec:SpecFunc:Nonworking:Bayes}

The Bayesian reconstruction usually selects a set of basis
functions simply by introducing a numerical integration scheme to
represent the discretized spectral function. This choice of basis is
naturally unaware of the analytic structure of the correlator and thus
of the functional form admissible for the spectral function. In
particular this basis does not prevent highly oscillatory and thus
unphysical structures to manifest themselves in the end result. In the
spirit of the Bayesian approach the prior probability then needs to be
constructed such that these oscillatory solutions are suppressed. The
quadratic prior, we have found is not efficient in doing so and thus
unphysical ringing may persists in the end result. Similar ringing
also manifests itself in case of the generalized BR method.

We have performed reconstructions based on the Euclidean data in the
scaling scenario, using different default models, which were endowed
with the correct asymptotic form. One example is shown in
\Cref{Fig:BayesGluon}, where the green solid line denotes the
default model and the solid violet line corresponds to the Bayesian
end result. One hundred data points with a relative error of $10^{-2}$
were used. Due to the asymptotics supplied in the default model, as
well as the fact that the Bayesian method uses the
K{\"a}ll{\'e}n–Lehmann representation to translate the spectrum into a
Euclidean correlator, items (2-5) from our list are fulfilled
here. Item (1) is not fulfilled since the ratio between the area under
the curve and the area under the absolute value of the spectrum is
around $0.8$. The most striking drawback of this reconstruction
however is the strong oscillatory behavior found, which renders an
interpretation of the non-perturbative region at intermediate
frequencies at best difficult. Note that with currently available
lattice QCD data such a strong oscillatory
behavior was not observed when investigating the gluon spectrum.

It will be interesting future work to include the prior information on
the analytic structure of the gluon propagator into the prior
probability of these Bayesian approaches.

\subsubsection{Backus-Gilbert reconstruction}
\label{sec:SpecFunc:Nonworking:BG}

\begin{figure}
\includegraphics{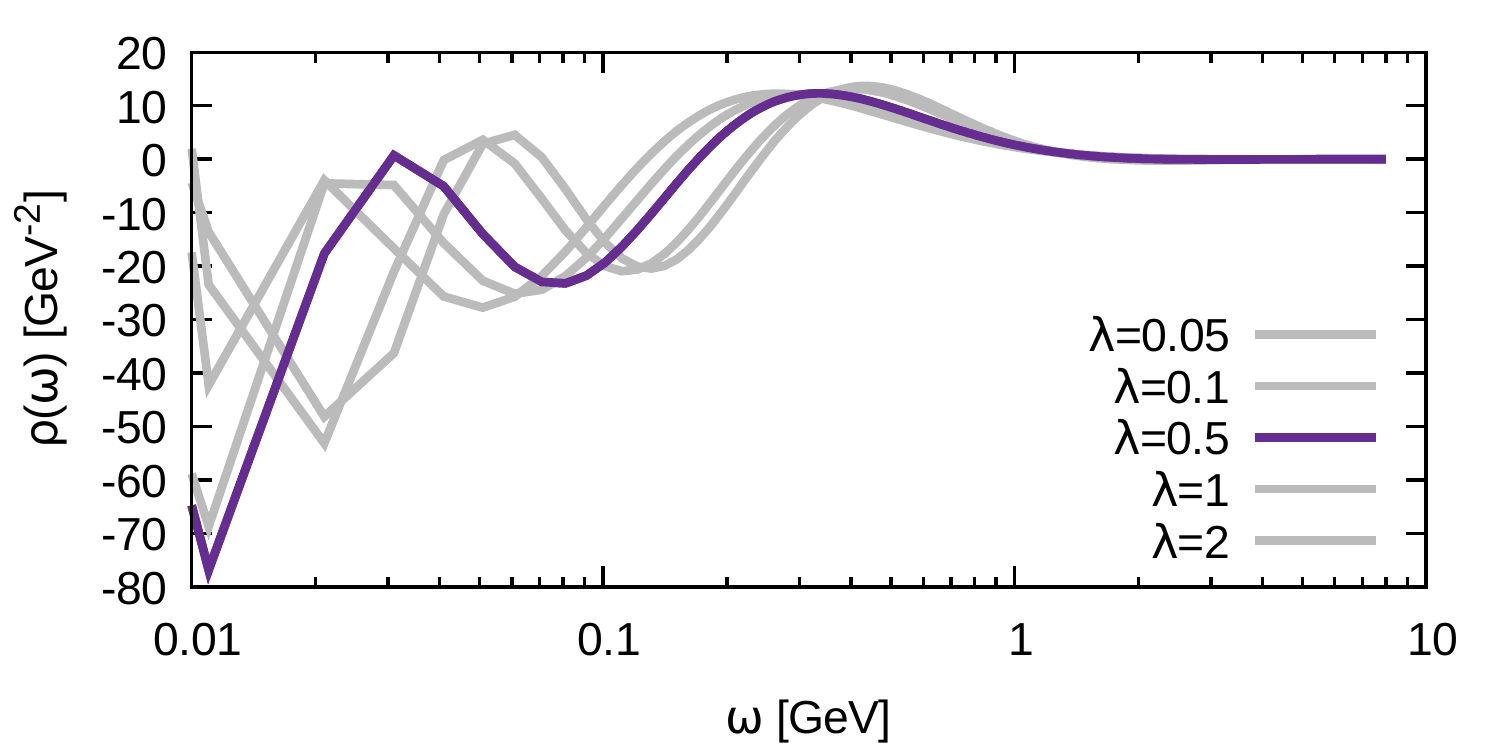}
\includegraphics{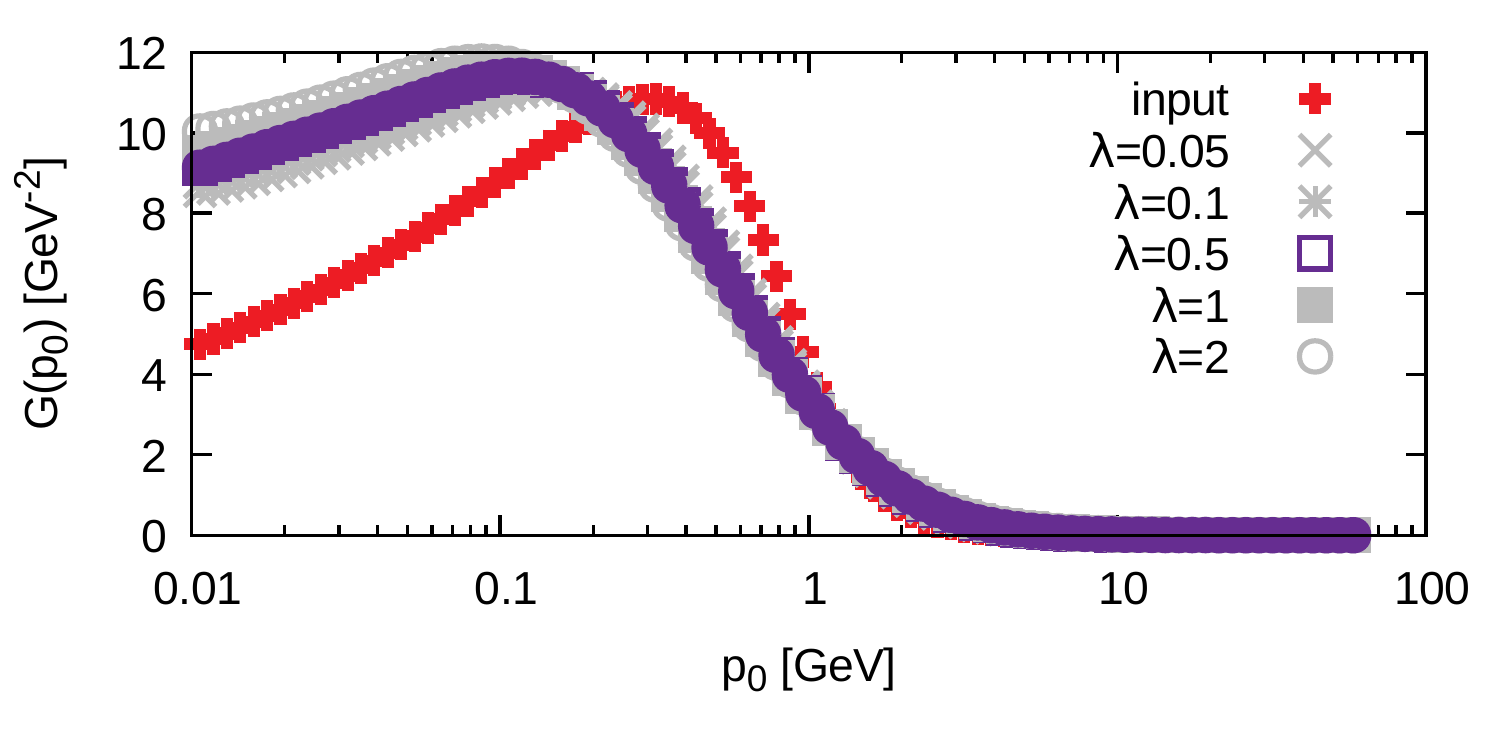}
\includegraphics{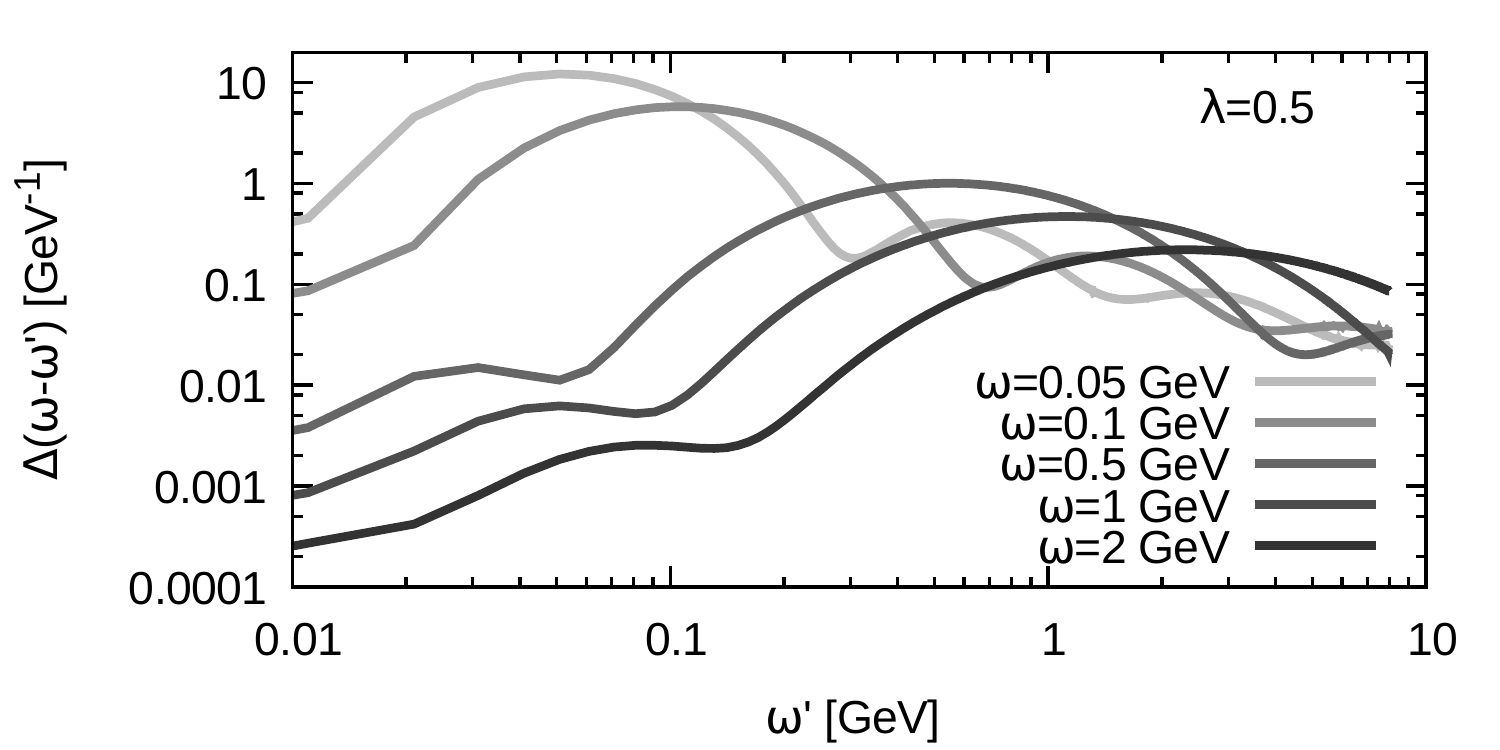}
\caption{(top) Series of Backus-Gilbert reconstructed spectral functions
using the Tickhonov regularization prescription for different regulator parameter 
$\lambda=[0.05..2]$. Note that the reconstruction identifies the presence of both
the negative trough close to the origin and the one above the main peak at around
$\omega>\SI{1}{\GeV}$. Ringing at small frequencies is not cured by simply increasing the value of $\lambda$. The best choice $\lambda=0.5$ is given in dark violet, while the other
values are denoted by light gray curves. (center) Euclidean data of the reconstructed
spectra compared to the original input (red). The result corresponding to $\lambda=0.5$ (dark violet)
works relatively well above $p_0>\SI{1}{\GeV}$ but misses the position of the main peak structure and
exhibits too weak of a backbending. (bottom) Resolution function $\Delta(\omega-\omega')$ 
for the best choice $\lambda=0.5$ plotted for completeness.
}
\label{Fig:BGGluon}
\end{figure}

\begin{figure*}[t]
	\includegraphics{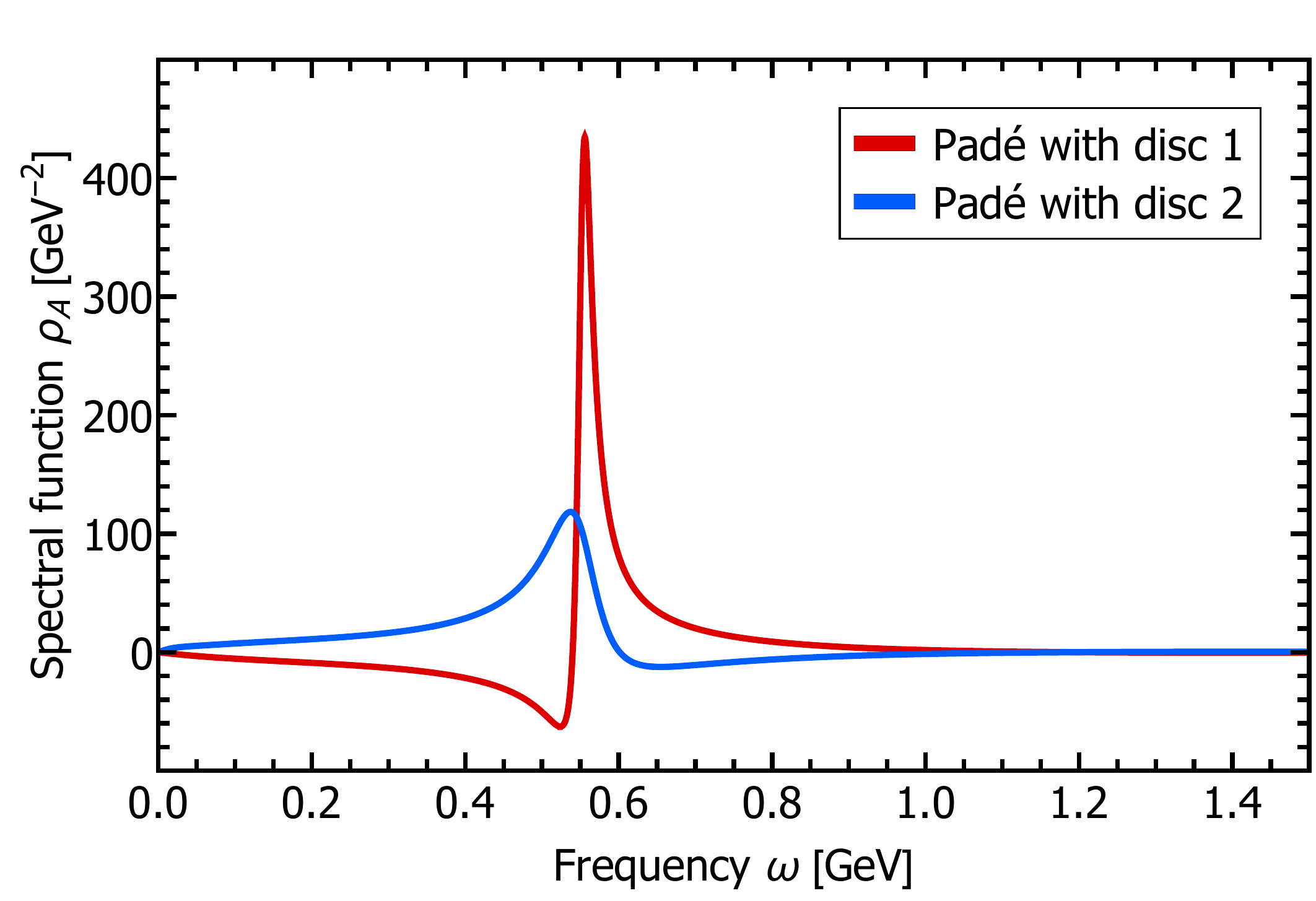}\hfill
	\includegraphics{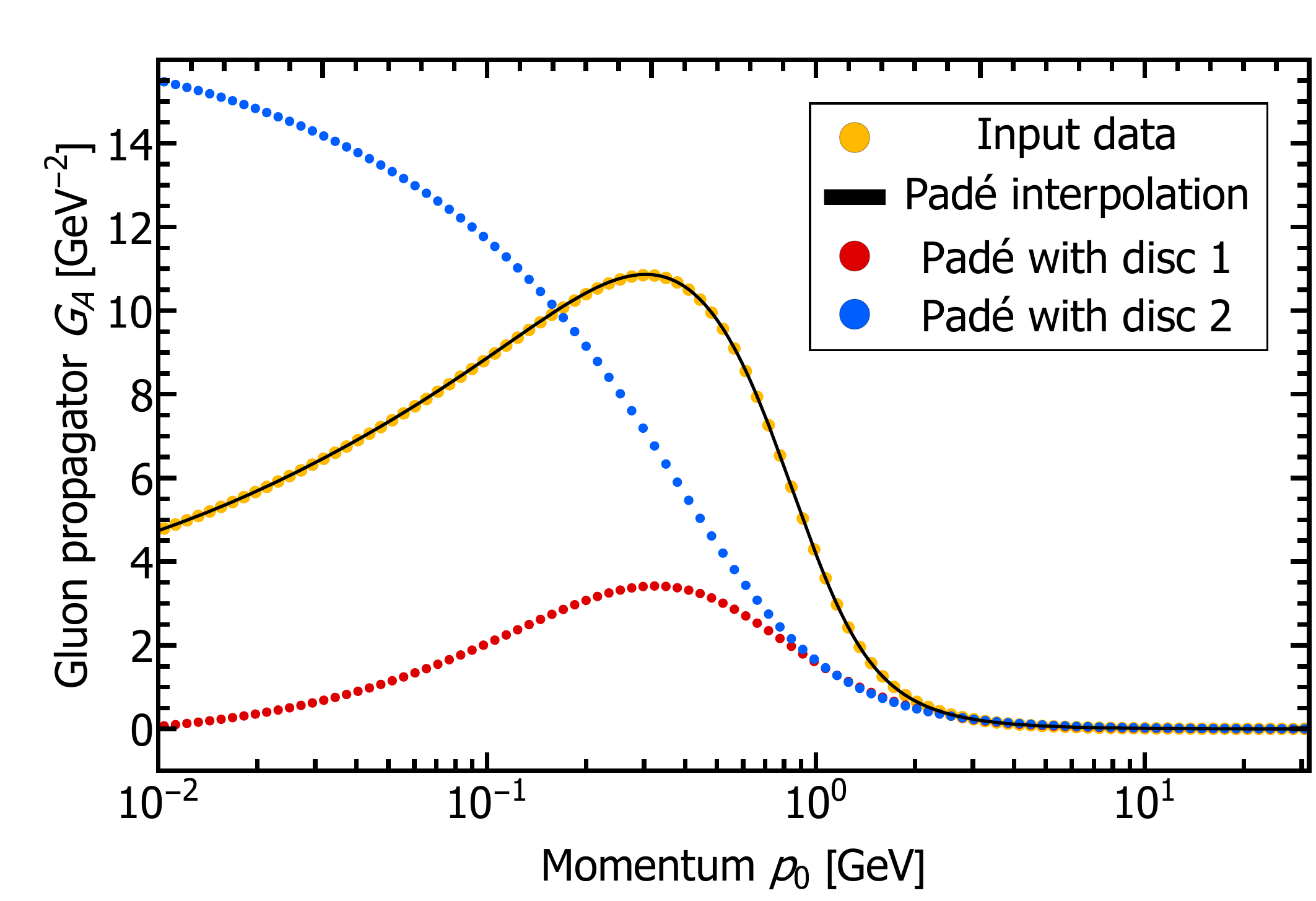}
	\caption{Reconstruction of the gluon propagator via Pad{\'e}, the difference between the two spectral functions is described in the main text. The left panel shows the spectral functions while the right panel shows the input data, the Pad{\'e} interpolant and the reconstructions obtained from the spectral functions.}
	\label{fig:pade}
\end{figure*}

The Backus-Gilbert approach to spectral function reconstruction operates \cite{BG1,BG2} with an
implicit set of basis functions, which are characterized by the resolution function
$\Delta(\omega-\omega')$. Also in this case the basis is not aware of the 
analytic structure admissible for the correlator under investigation. The naive Backus-Gilbert
method in addition requires a regularization prescription, for which we here choose
the Tickhonov approach with regulator parameter $\lambda$ \cite{Press:1997nr}. 

No explicit default model enters the BG approach, i.e. the prior information needed
to give meaning to the ill-conditioned inverse problem enters through the definition of the
optimization functions of which the BG spectrum is an extrema. This functional 
is designed such that it selects a solution for which the resolution function $\Delta(\omega-\omega')$
is most sharply peaked.

Note that in order to carry out the BG reconstruction the so called response kernels 
need to be computed \cite{Press:1997nr}. Here in case of the gluon spectrum these correspond to 
integrals over the K{\"a}ll{\'e}n–Lehmann without the spectrum multiplied. As these
functions are not well defined, we instead choose to reconstruct the function $\rho(\omega)\omega$
with the corresponding response Kernel 
\begin{align}
R(p_0)=\int_0^\infty  \frac{ d\omega }{\pi} \frac{1}{\omega^2 + p_0^2}
\end{align}
which is finite. The spectrum is the obtained from dividing out $\omega$ from the raw reconstruction.

A series of results for the BG spectral reconstruction based on one-hundred ideal Euclidean input data points is shown in the top panel of \Cref{Fig:BGGluon} for several different values of $\lambda$. We find that a main peak close to $\omega=\SI{1}{\GeV}$ is consistently found and in addition the reconstructions show a negative trough close to the origin and above the main peak. However it is also clear from the top panel that close to the origin ringing artifacts again impede the physics interpretation of the result.

The BG solution by construction is not required to reproduce the input data, which then also leads to significant deviations as shown in the center panel of \Cref{Fig:BGGluon}. 
Thus, we find that the BG approach in case of the gluon spectral function is challenged to meet, in particular, criteria (1-4) that we require for a successful reconstruction.

\subsubsection{Pad{\'e}-type approaches}
\label{sec:SpecFunc:Nonworking:Pade}

Pad{\'e} approaches, e.g. Schlessinger's point method, obtain the spectral function from analytically continuing an interpolating or fitted rational function. The nomenclature \textit{Pad{\'e}} is used here in a loose sense, referring to all approaches based on matching a rational function of arbitrary degree to Euclidean data. Usually, \labelcref{eq:spec_from_prop} is used in order to obtain the spectral function. It has the advantage of being easy to apply and gives reasonable results for the position of the lowest lying resonances, see e.g. \cite{Tripolt:2018xeo}.
Pad{\'e} approaches describe the analytic structure of the analytically continued retarded propagator in terms of poles and is therefore naturally contained in our approach outlined in \Cref{sec:SpecFunc}.

However, it does not respect the holomorphicity of the retarded propagator in the given half-plane, c.f. \Cref{fig:approach} and the corresponding discussion, by design. While Pad{\'e} approaches will still converge for infinite precision and infinitely many data points, violations of the holomorphicity at finite precision might be severe and make an unambiguous reconstruction impossible. 
As the spectral function obtained from \labelcref{eq:spec_from_prop} does not respect \labelcref{eq:specrep} anymore, if the holomorphicity of the right half-plane is violated, the obtained spectral function will fail to reproduce the Euclidean propagator. The violation might be acceptable if the violation is small, suitably defined via the reconstruction. But is most certainly not, if a pole dominantly contributing in \labelcref{eq:specrep} is missing.
A more realistic spectral function can then still be obtained by suitably modifying \labelcref{eq:spec_from_prop}, but
there is no consistent, unambiguous way of doing so.
\newline\indent
This discussion is especially relevant when turning to the reconstruction of the gluon, since here the dominant pole violates the holomorphicity of the retarded propagator. We employed two choices for the reconstruction, "Pad{\'e} with disc 2", where we keep \labelcref{eq:spec_from_prop} with a flipped sign to account for the dominant pole being in the wrong half-plane, and "Pad{\'e} with disc 1", but evaluated it at an argument with a finite real part slightly larger than the position of the pole. The corresponding spectral functions are shown in the left panel of \Cref{fig:pade}. Both methods get the roughly a similar shape which is also loosely compatibly with our main result \Cref{fig:MainResult}. However, "Pad{\'e} with disc 2" fails to reproduce the general shape in the reconstruction and "Pad{\'e} with disc 1" gets a significantly too small propagator, but describes the shape correctly. Additionally several other requirements listed above are not fulfilled.
While some requirements might be fixed by manipulating poles in the Pad{\'e} interpolant, any systematic way of doing so will lead to an approach very similar to the one described in the main text in \Cref{sec:SpecFunc}.

\subsection{Mock reconstruction benchmark}
\label{sec:SpecFunc:Bench}

In this section we demonstrate with the simple example of a two
Breit-Wigner spectral function how incorporating our novel approach
into a Bayesian framework allows to straight forwardly improve the
spectral reconstruction. As mock spectral function we take the
parametrization of \labelcref{eq:spec_BW} with a direct sum of two
Breit-Wigner peaks
\begin{align}
\rho_\tinytext{mock}(\omega) = \rho^\tinytext{(BW)}_\tinytext{1}(\omega) + \rho^\tinytext{(BW)}_\tinytext{2}(\omega)
\end{align}
and the following values for peak positions and widths
\begin{table}[h]
\begin{tabular}{|l|c|c|c|}
\hline
Parameter & $A$ & $M$ & $\Gamma$ \\ \hline
Peak 1 & 0.35 & 1.0 & 0.25 \\
Peak 2 & 0.65 & 3.0 & 0.25 \\ \hline
\end{tabular}
\label{tab:params_BW}
\end{table}

From this spectrum we evaluate 60 equidistantly spaced Euclidean
correlator data points in the imaginary frequency interval between
$\unit[0-45]{GeV}$, which are subsequently salted with Gaussian noise
with a strength leading to $10^{-3}$ relative errors.

\begin{figure}[t]
	\includegraphics{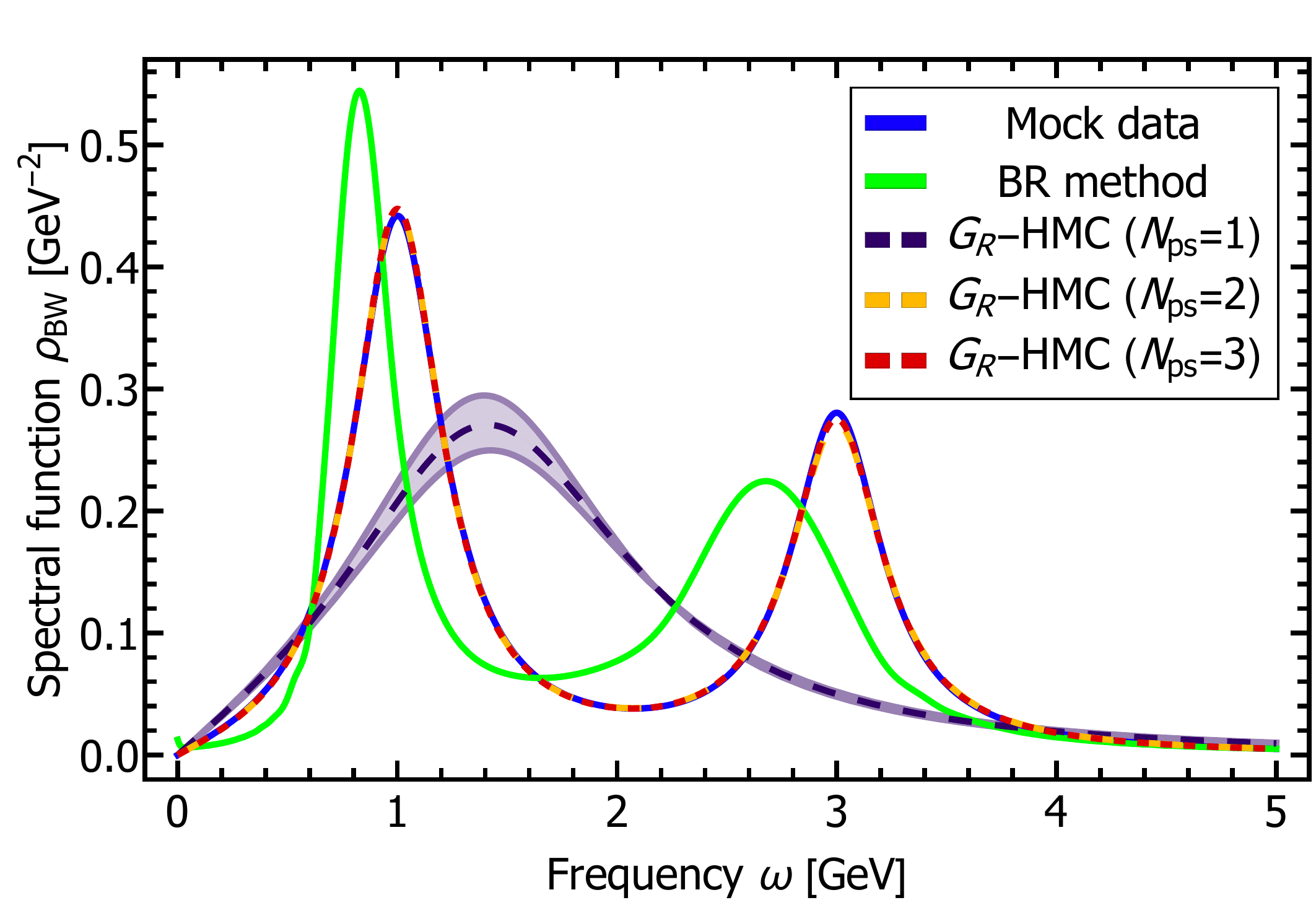}
	\caption{
		Reconstruction benchmark with a double Breit-Wigner peak.
	}
	\label{fig:mock_spectra}
\end{figure}

In the absence of any prior knowledge about the analytic structure
(i.e. the two poles) contained in our example data we may choose to
deploy a standard Bayesian method, such as the BR method \cite{Burnier:2013nla}, which only
enforces positive definiteness and smoothness.  As shown in
\Cref{fig:mock_spectra} as green solid line, with the provided
quality of the Euclidean data, this method manages to correctly
identify the number of peak present but only achieves an accuracy of
the peak positions of around $75-80$\%.

Now we can proceed to deploy our novel ansatz for the functional basis
in \labelcref{eq:AnsatzFull}. Three different cases are
possible. Depending on the number $N_{\rm ps}$ of pole structures
chosen in the basis, we may either have less structure, exactly the
same amount of pole structures, or irrelevant additional structure
present compared to what is actually encoded in the Euclidean data.

In the first case, where only one pole is contained in the basis,
fitting its parameters is a well posed problem and can be carried out
using both a naive $\chi^2$ fit or a full Bayesian analysis, where
each parameter is endowed with an additional prior probability. In
\Cref{fig:mock_spectra} we have carried out a full Bayesian
estimation of the posterior for the parameters in the single pole
basis (purple dashed line) using the Hamiltonian Monte Carlo framework
implemented by the MC-STAN \cite{JSSv076i01, rstan} library. The used
prior only enforces the finiteness of the peak position. As expected
this too strongly restricted basis yields a reconstruction, which
cannot correctly account for the pole structure and instead positions
one peak in between the actual two peaks present.

Increasing the number of available poles in the basis to two, we have
the same number of structures in our basis, as we have in our
data. This case is still well-posed and again admits a solution both
via $\chi^2$ fit and a full Bayesian analysis. Since we are using a
parametrization that can exactly match our input data it is not
surprising that now the end result from the HMC analysis lies
spot on the mock spectrum (orange dashed curve). Note that for the
$\chi^2$ fit the result is slightly worse and only reproduces the mock
spectrum with around $5\%$ deviation.

Please note that as soon as two or more poles are present in the basis
an ordering prescription must be enforced, otherwise the posterior
distribution degenerates by means of a simple reparameterization.

The more interesting case, relevant in practice concerns allowing more
structure in the basis than encoded in the Euclidean data, which we
here tests with a three pole basis ansatz. In view of the $\chi^2$ fit
this problem is now ill-conditioned and indeed carrying out a naive fit
shows that the obtained parameters become unstable. Here the HMC
approach plays out its strength, maintaining stability even if the
problem becomes ill-conditioned. Indeed the position and width of the lowest
two pole structures is again recovered excellently (red dashed curve),
while the posterior of the position of the third peak shows that it is
highly unconstrained and thus irrelevant. The pole ordering naturally
achieves that any excess pole structure beyond what is encoded in the
data is simply pushed to infinity, not contributing to the end result.

We find that incorporating our novel basis may significantly improve
the reconstruction result compared to those methods, which do not make
any assumptions on the analytic structure of the underlying data. The
example used here of course is simplistic but captures a main aspect,
a multi pole structure, encountered also in more realistic cases.

\bibliography{bib_lat-spec}

\begin{thebibliography}{92}%
\makeatletter
\providecommand \@ifxundefined [1]{%
 \@ifx{#1\undefined}
}%
\providecommand \@ifnum [1]{%
 \ifnum #1\expandafter \@firstoftwo
 \else \expandafter \@secondoftwo
 \fi
}%
\providecommand \@ifx [1]{%
 \ifx #1\expandafter \@firstoftwo
 \else \expandafter \@secondoftwo
 \fi
}%
\providecommand \natexlab [1]{#1}%
\providecommand \enquote  [1]{``#1''}%
\providecommand \bibnamefont  [1]{#1}%
\providecommand \bibfnamefont [1]{#1}%
\providecommand \citenamefont [1]{#1}%
\providecommand \href@noop [0]{\@secondoftwo}%
\providecommand \href [0]{\begingroup \@sanitize@url \@href}%
\providecommand \@href[1]{\@@startlink{#1}\@@href}%
\providecommand \@@href[1]{\endgroup#1\@@endlink}%
\providecommand \@sanitize@url [0]{\catcode `\\12\catcode `\$12\catcode
  `\&12\catcode `\#12\catcode `\^12\catcode `\_12\catcode `\%12\relax}%
\providecommand \@@startlink[1]{}%
\providecommand \@@endlink[0]{}%
\providecommand \url  [0]{\begingroup\@sanitize@url \@url }%
\providecommand \@url [1]{\endgroup\@href {#1}{\urlprefix }}%
\providecommand \urlprefix  [0]{URL }%
\providecommand \Eprint [0]{\href }%
\providecommand \doibase [0]{http://dx.doi.org/}%
\providecommand \selectlanguage [0]{\@gobble}%
\providecommand \bibinfo  [0]{\@secondoftwo}%
\providecommand \bibfield  [0]{\@secondoftwo}%
\providecommand \translation [1]{[#1]}%
\providecommand \BibitemOpen [0]{}%
\providecommand \bibitemStop [0]{}%
\providecommand \bibitemNoStop [0]{.\EOS\space}%
\providecommand \EOS [0]{\spacefactor3000\relax}%
\providecommand \BibitemShut  [1]{\csname bibitem#1\endcsname}%
\let\auto@bib@innerbib\@empty
\bibitem [{\citenamefont {Haas}\ \emph {et~al.}(2014)\citenamefont {Haas},
  \citenamefont {Fister},\ and\ \citenamefont {Pawlowski}}]{Haas:2013hpa}%
  \BibitemOpen
  \bibfield  {author} {\bibinfo {author} {\bibfnamefont {M.}~\bibnamefont
  {Haas}}, \bibinfo {author} {\bibfnamefont {L.}~\bibnamefont {Fister}}, \ and\
  \bibinfo {author} {\bibfnamefont {J.~M.}\ \bibnamefont {Pawlowski}},\ }\href
  {\doibase 10.1103/PhysRevD.90.091501} {\bibfield  {journal} {\bibinfo
  {journal} {Phys.Rev.}\ }\textbf {\bibinfo {volume} {D90}},\ \bibinfo {pages}
  {091501} (\bibinfo {year} {2014})},\ \Eprint {http://arxiv.org/abs/1308.4960}
  {arXiv:1308.4960 [hep-ph]} \BibitemShut {NoStop}%
\bibitem [{\citenamefont {Qin}\ and\ \citenamefont
  {Rischke}(2013)}]{Qin:2013ufa}%
  \BibitemOpen
  \bibfield  {author} {\bibinfo {author} {\bibfnamefont {S.-x.}\ \bibnamefont
  {Qin}}\ and\ \bibinfo {author} {\bibfnamefont {D.~H.}\ \bibnamefont
  {Rischke}},\ }\href {\doibase 10.1103/PhysRevD.88.056007} {\bibfield
  {journal} {\bibinfo  {journal} {Phys. Rev.}\ }\textbf {\bibinfo {volume}
  {D88}},\ \bibinfo {pages} {056007} (\bibinfo {year} {2013})},\ \Eprint
  {http://arxiv.org/abs/1304.6547} {arXiv:1304.6547 [nucl-th]} \BibitemShut
  {NoStop}%
\bibitem [{\citenamefont {Dudal}\ \emph {et~al.}(2014)\citenamefont {Dudal},
  \citenamefont {Oliveira},\ and\ \citenamefont {Silva}}]{Dudal:2013yva}%
  \BibitemOpen
  \bibfield  {author} {\bibinfo {author} {\bibfnamefont {D.}~\bibnamefont
  {Dudal}}, \bibinfo {author} {\bibfnamefont {O.}~\bibnamefont {Oliveira}}, \
  and\ \bibinfo {author} {\bibfnamefont {P.~J.}\ \bibnamefont {Silva}},\ }\href
  {\doibase 10.1103/PhysRevD.89.014010} {\bibfield  {journal} {\bibinfo
  {journal} {Phys. Rev.}\ }\textbf {\bibinfo {volume} {D89}},\ \bibinfo {pages}
  {014010} (\bibinfo {year} {2014})},\ \Eprint {http://arxiv.org/abs/1310.4069}
  {arXiv:1310.4069 [hep-lat]} \BibitemShut {NoStop}%
\bibitem [{\citenamefont {Christiansen}\ \emph {et~al.}(2015)\citenamefont
  {Christiansen}, \citenamefont {Haas}, \citenamefont {Pawlowski},\ and\
  \citenamefont {Strodthoff}}]{Christiansen:2014ypa}%
  \BibitemOpen
  \bibfield  {author} {\bibinfo {author} {\bibfnamefont {N.}~\bibnamefont
  {Christiansen}}, \bibinfo {author} {\bibfnamefont {M.}~\bibnamefont {Haas}},
  \bibinfo {author} {\bibfnamefont {J.~M.}\ \bibnamefont {Pawlowski}}, \ and\
  \bibinfo {author} {\bibfnamefont {N.}~\bibnamefont {Strodthoff}},\ }\href
  {\doibase 10.1103/PhysRevLett.115.112002} {\bibfield  {journal} {\bibinfo
  {journal} {Phys. Rev. Lett.}\ }\textbf {\bibinfo {volume} {115}},\ \bibinfo
  {pages} {112002} (\bibinfo {year} {2015})},\ \Eprint
  {http://arxiv.org/abs/1411.7986} {arXiv:1411.7986 [hep-ph]} \BibitemShut
  {NoStop}%
\bibitem [{\citenamefont {Rothkopf}(2017)}]{Rothkopf:2016luz}%
  \BibitemOpen
  \bibfield  {author} {\bibinfo {author} {\bibfnamefont {A.}~\bibnamefont
  {Rothkopf}},\ }\href {\doibase 10.1103/PhysRevD.95.056016} {\bibfield
  {journal} {\bibinfo  {journal} {Phys. Rev.}\ }\textbf {\bibinfo {volume}
  {D95}},\ \bibinfo {pages} {056016} (\bibinfo {year} {2017})},\ \Eprint
  {http://arxiv.org/abs/1611.00482} {arXiv:1611.00482 [hep-ph]} \BibitemShut
  {NoStop}%
\bibitem [{\citenamefont {Ilgenfritz}\ \emph {et~al.}(2018)\citenamefont
  {Ilgenfritz}, \citenamefont {Pawlowski}, \citenamefont {Rothkopf},\ and\
  \citenamefont {Trunin}}]{Ilgenfritz:2017kkp}%
  \BibitemOpen
  \bibfield  {author} {\bibinfo {author} {\bibfnamefont {E.-M.}\ \bibnamefont
  {Ilgenfritz}}, \bibinfo {author} {\bibfnamefont {J.~M.}\ \bibnamefont
  {Pawlowski}}, \bibinfo {author} {\bibfnamefont {A.}~\bibnamefont {Rothkopf}},
  \ and\ \bibinfo {author} {\bibfnamefont {A.}~\bibnamefont {Trunin}},\ }\href
  {\doibase 10.1140/epjc/s10052-018-5593-7} {\bibfield  {journal} {\bibinfo
  {journal} {Eur. Phys. J.}\ }\textbf {\bibinfo {volume} {C78}},\ \bibinfo
  {pages} {127} (\bibinfo {year} {2018})},\ \Eprint
  {http://arxiv.org/abs/1701.08610} {arXiv:1701.08610 [hep-lat]} \BibitemShut
  {NoStop}%
\bibitem [{\citenamefont {Fischer}\ \emph {et~al.}(2018)\citenamefont
  {Fischer}, \citenamefont {Pawlowski}, \citenamefont {Rothkopf},\ and\
  \citenamefont {Welzbacher}}]{Fischer:2017kbq}%
  \BibitemOpen
  \bibfield  {author} {\bibinfo {author} {\bibfnamefont {C.~S.}\ \bibnamefont
  {Fischer}}, \bibinfo {author} {\bibfnamefont {J.~M.}\ \bibnamefont
  {Pawlowski}}, \bibinfo {author} {\bibfnamefont {A.}~\bibnamefont {Rothkopf}},
  \ and\ \bibinfo {author} {\bibfnamefont {C.~A.}\ \bibnamefont {Welzbacher}},\
  }\href {\doibase 10.1103/PhysRevD.98.014009} {\bibfield  {journal} {\bibinfo
  {journal} {Phys. Rev.}\ }\textbf {\bibinfo {volume} {D98}},\ \bibinfo {pages}
  {014009} (\bibinfo {year} {2018})},\ \Eprint
  {http://arxiv.org/abs/1705.03207} {arXiv:1705.03207 [hep-ph]} \BibitemShut
  {NoStop}%
\bibitem [{\citenamefont {Silva}\ \emph {et~al.}(2018)\citenamefont {Silva},
  \citenamefont {Oliveira}, \citenamefont {Dudal},\ and\ \citenamefont
  {Roelfs}}]{Silva:2017mds}%
  \BibitemOpen
  \bibfield  {author} {\bibinfo {author} {\bibfnamefont {P.~J.}\ \bibnamefont
  {Silva}}, \bibinfo {author} {\bibfnamefont {O.}~\bibnamefont {Oliveira}},
  \bibinfo {author} {\bibfnamefont {D.}~\bibnamefont {Dudal}}, \ and\ \bibinfo
  {author} {\bibfnamefont {M.}~\bibnamefont {Roelfs}},\ }\bibfield  {booktitle}
  {\emph {\bibinfo {booktitle} {{35th International Symposium on Lattice Field
  Theory (Lattice 2017) Granada, Spain, June 18-24, 2017}}},\ }\href {\doibase
  10.1051/epjconf/201817507038} {\bibfield  {journal} {\bibinfo  {journal} {EPJ
  Web Conf.}\ }\textbf {\bibinfo {volume} {175}},\ \bibinfo {pages} {07038}
  (\bibinfo {year} {2018})},\ \Eprint {http://arxiv.org/abs/1711.02584}
  {arXiv:1711.02584 [hep-lat]} \BibitemShut {NoStop}%
\bibitem [{\citenamefont {Astrakhantsev}\ \emph {et~al.}(2017)\citenamefont
  {Astrakhantsev}, \citenamefont {Braguta},\ and\ \citenamefont
  {Kotov}}]{Astrakhantsev:2017nrs}%
  \BibitemOpen
  \bibfield  {author} {\bibinfo {author} {\bibfnamefont {N.}~\bibnamefont
  {Astrakhantsev}}, \bibinfo {author} {\bibfnamefont {V.}~\bibnamefont
  {Braguta}}, \ and\ \bibinfo {author} {\bibfnamefont {A.}~\bibnamefont
  {Kotov}},\ }\href {\doibase 10.1007/JHEP04(2017)101} {\bibfield  {journal}
  {\bibinfo  {journal} {JHEP}\ }\textbf {\bibinfo {volume} {04}},\ \bibinfo
  {pages} {101} (\bibinfo {year} {2017})},\ \Eprint
  {http://arxiv.org/abs/1701.02266} {arXiv:1701.02266 [hep-lat]} \BibitemShut
  {NoStop}%
\bibitem [{\citenamefont {Borsányi}\ \emph {et~al.}(2018)\citenamefont
  {Borsányi}, \citenamefont {Fodor}, \citenamefont {Giordano}, \citenamefont
  {Katz}, \citenamefont {Pasztor}, \citenamefont {Ratti}, \citenamefont
  {Schäfer}, \citenamefont {Szabo},\ and\ \citenamefont
  {Tóth}}]{Pasztor:2018yae}%
  \BibitemOpen
  \bibfield  {author} {\bibinfo {author} {\bibfnamefont {S.}~\bibnamefont
  {Borsányi}}, \bibinfo {author} {\bibfnamefont {Z.}~\bibnamefont {Fodor}},
  \bibinfo {author} {\bibfnamefont {M.}~\bibnamefont {Giordano}}, \bibinfo
  {author} {\bibfnamefont {S.~D.}\ \bibnamefont {Katz}}, \bibinfo {author}
  {\bibfnamefont {A.}~\bibnamefont {Pasztor}}, \bibinfo {author} {\bibfnamefont
  {C.}~\bibnamefont {Ratti}}, \bibinfo {author} {\bibfnamefont
  {A.}~\bibnamefont {Schäfer}}, \bibinfo {author} {\bibfnamefont {K.~K.}\
  \bibnamefont {Szabo}}, \ and\ \bibinfo {author} {\bibfnamefont {B.~C.}\
  \bibnamefont {Tóth}},\ }\href {\doibase 10.1103/PhysRevD.98.014512}
  {\bibfield  {journal} {\bibinfo  {journal} {Phys. Rev.}\ }\textbf {\bibinfo
  {volume} {D98}},\ \bibinfo {pages} {014512} (\bibinfo {year} {2018})},\
  \Eprint {http://arxiv.org/abs/1802.07718} {arXiv:1802.07718 [hep-lat]}
  \BibitemShut {NoStop}%
\bibitem [{\citenamefont {Mages}\ \emph {et~al.}(2015)\citenamefont {Mages},
  \citenamefont {Borsányi}, \citenamefont {Fodor}, \citenamefont {Schaefer},\
  and\ \citenamefont {Szabó}}]{Mages:2015rea}%
  \BibitemOpen
  \bibfield  {author} {\bibinfo {author} {\bibfnamefont {S.~W.}\ \bibnamefont
  {Mages}}, \bibinfo {author} {\bibfnamefont {S.}~\bibnamefont {Borsányi}},
  \bibinfo {author} {\bibfnamefont {Z.}~\bibnamefont {Fodor}}, \bibinfo
  {author} {\bibfnamefont {A.}~\bibnamefont {Schaefer}}, \ and\ \bibinfo
  {author} {\bibfnamefont {K.}~\bibnamefont {Szabó}},\ }\bibfield  {booktitle}
  {\emph {\bibinfo {booktitle} {{Proceedings, 32nd International Symposium on
  Lattice Field Theory (Lattice 2014): Brookhaven, NY, USA, June 23-28,
  2014}}},\ }\href@noop {} {\bibfield  {journal} {\bibinfo  {journal} {PoS}\
  }\textbf {\bibinfo {volume} {LATTICE2014}},\ \bibinfo {pages} {232} (\bibinfo
  {year} {2015})}\BibitemShut {NoStop}%
\bibitem [{\citenamefont {Astrakhantsev}\ \emph {et~al.}(2015)\citenamefont
  {Astrakhantsev}, \citenamefont {Braguta},\ and\ \citenamefont
  {Kotov}}]{Astrakhantsev:2015jta}%
  \BibitemOpen
  \bibfield  {author} {\bibinfo {author} {\bibfnamefont {N.~{\relax Yu}.}\
  \bibnamefont {Astrakhantsev}}, \bibinfo {author} {\bibfnamefont {V.~V.}\
  \bibnamefont {Braguta}}, \ and\ \bibinfo {author} {\bibfnamefont {A.~{\relax
  Yu}.}\ \bibnamefont {Kotov}},\ }\href {\doibase 10.1007/JHEP09(2015)082}
  {\bibfield  {journal} {\bibinfo  {journal} {JHEP}\ }\textbf {\bibinfo
  {volume} {09}},\ \bibinfo {pages} {082} (\bibinfo {year} {2015})},\ \Eprint
  {http://arxiv.org/abs/1507.06225} {arXiv:1507.06225 [hep-lat]} \BibitemShut
  {NoStop}%
\bibitem [{\citenamefont {Meyer}(2008)}]{Meyer:2007dy}%
  \BibitemOpen
  \bibfield  {author} {\bibinfo {author} {\bibfnamefont {H.~B.}\ \bibnamefont
  {Meyer}},\ }\href {\doibase 10.1103/PhysRevLett.100.162001} {\bibfield
  {journal} {\bibinfo  {journal} {Phys. Rev. Lett.}\ }\textbf {\bibinfo
  {volume} {100}},\ \bibinfo {pages} {162001} (\bibinfo {year} {2008})},\
  \Eprint {http://arxiv.org/abs/0710.3717} {arXiv:0710.3717 [hep-lat]}
  \BibitemShut {NoStop}%
\bibitem [{\citenamefont {Meyer}(2007)}]{Meyer:2007ic}%
  \BibitemOpen
  \bibfield  {author} {\bibinfo {author} {\bibfnamefont {H.~B.}\ \bibnamefont
  {Meyer}},\ }\href {\doibase 10.1103/PhysRevD.76.101701} {\bibfield  {journal}
  {\bibinfo  {journal} {Phys.Rev.}\ }\textbf {\bibinfo {volume} {D76}},\
  \bibinfo {pages} {101701} (\bibinfo {year} {2007})},\ \Eprint
  {http://arxiv.org/abs/0704.1801} {arXiv:0704.1801 [hep-lat]} \BibitemShut
  {NoStop}%
\bibitem [{\citenamefont {Hobson}\ and\ \citenamefont
  {Lasenby}(1998)}]{Hobson:1998bz}%
  \BibitemOpen
  \bibfield  {author} {\bibinfo {author} {\bibfnamefont {M.}~\bibnamefont
  {Hobson}}\ and\ \bibinfo {author} {\bibfnamefont {A.}~\bibnamefont
  {Lasenby}},\ }\href {\doibase 10.1046/j.1365-8711.1998.01707.x} {\bibfield
  {journal} {\bibinfo  {journal} {Mon. Not. Roy. Astron. Soc.}\ }\textbf
  {\bibinfo {volume} {298}},\ \bibinfo {pages} {905} (\bibinfo {year}
  {1998})},\ \Eprint {http://arxiv.org/abs/astro-ph/9810240}
  {arXiv:astro-ph/9810240 [astro-ph]} \BibitemShut {NoStop}%
\bibitem [{\citenamefont {Jarrell}\ and\ \citenamefont
  {Gubernatis}(1996)}]{Jarrell:1996rrw}%
  \BibitemOpen
  \bibfield  {author} {\bibinfo {author} {\bibfnamefont {M.}~\bibnamefont
  {Jarrell}}\ and\ \bibinfo {author} {\bibfnamefont {J.~E.}\ \bibnamefont
  {Gubernatis}},\ }\href {\doibase 10.1016/0370-1573(95)00074-7} {\bibfield
  {journal} {\bibinfo  {journal} {Phys. Rept.}\ }\textbf {\bibinfo {volume}
  {269}},\ \bibinfo {pages} {133} (\bibinfo {year} {1996})}\BibitemShut
  {NoStop}%
\bibitem [{\citenamefont {Asakawa}\ \emph {et~al.}(2001)\citenamefont
  {Asakawa}, \citenamefont {Hatsuda},\ and\ \citenamefont
  {Nakahara}}]{Asakawa:2000tr}%
  \BibitemOpen
  \bibfield  {author} {\bibinfo {author} {\bibfnamefont {M.}~\bibnamefont
  {Asakawa}}, \bibinfo {author} {\bibfnamefont {T.}~\bibnamefont {Hatsuda}}, \
  and\ \bibinfo {author} {\bibfnamefont {Y.}~\bibnamefont {Nakahara}},\ }\href
  {\doibase 10.1016/S0146-6410(01)00150-8} {\bibfield  {journal} {\bibinfo
  {journal} {Prog.Part.Nucl.Phys.}\ }\textbf {\bibinfo {volume} {46}},\
  \bibinfo {pages} {459} (\bibinfo {year} {2001})},\ \Eprint
  {http://arxiv.org/abs/hep-lat/0011040} {arXiv:hep-lat/0011040 [hep-lat]}
  \BibitemShut {NoStop}%
\bibitem [{\citenamefont {Cyrol}\ \emph {et~al.}(2016)\citenamefont {Cyrol},
  \citenamefont {Fister}, \citenamefont {Mitter}, \citenamefont {Pawlowski},\
  and\ \citenamefont {Strodthoff}}]{Cyrol:2016tym}%
  \BibitemOpen
  \bibfield  {author} {\bibinfo {author} {\bibfnamefont {A.~K.}\ \bibnamefont
  {Cyrol}}, \bibinfo {author} {\bibfnamefont {L.}~\bibnamefont {Fister}},
  \bibinfo {author} {\bibfnamefont {M.}~\bibnamefont {Mitter}}, \bibinfo
  {author} {\bibfnamefont {J.~M.}\ \bibnamefont {Pawlowski}}, \ and\ \bibinfo
  {author} {\bibfnamefont {N.}~\bibnamefont {Strodthoff}},\ }\href {\doibase
  10.1103/PhysRevD.94.054005} {\bibfield  {journal} {\bibinfo  {journal} {Phys.
  Rev.}\ }\textbf {\bibinfo {volume} {D94}},\ \bibinfo {pages} {054005}
  (\bibinfo {year} {2016})},\ \Eprint {http://arxiv.org/abs/1605.01856}
  {arXiv:1605.01856 [hep-ph]} \BibitemShut {NoStop}%
\bibitem [{\citenamefont {Pawlowski}\ \emph {et~al.}(2017)\citenamefont
  {Pawlowski}, \citenamefont {Strodthoff},\ and\ \citenamefont
  {Wink}}]{Pawlowski:2017gxj}%
  \BibitemOpen
  \bibfield  {author} {\bibinfo {author} {\bibfnamefont {J.~M.}\ \bibnamefont
  {Pawlowski}}, \bibinfo {author} {\bibfnamefont {N.}~\bibnamefont
  {Strodthoff}}, \ and\ \bibinfo {author} {\bibfnamefont {N.}~\bibnamefont
  {Wink}},\ }\href@noop {} {\  (\bibinfo {year} {2017})},\ \Eprint
  {http://arxiv.org/abs/1711.07444} {arXiv:1711.07444 [hep-th]} \BibitemShut
  {NoStop}%
\bibitem [{\citenamefont {Aguilar}\ and\ \citenamefont
  {Papavassiliou}(2006)}]{Aguilar:2006gr}%
  \BibitemOpen
  \bibfield  {author} {\bibinfo {author} {\bibfnamefont {A.~C.}\ \bibnamefont
  {Aguilar}}\ and\ \bibinfo {author} {\bibfnamefont {J.}~\bibnamefont
  {Papavassiliou}},\ }\href {\doibase 10.1088/1126-6708/2006/12/012} {\bibfield
   {journal} {\bibinfo  {journal} {JHEP}\ }\textbf {\bibinfo {volume} {12}},\
  \bibinfo {pages} {012} (\bibinfo {year} {2006})},\ \Eprint
  {http://arxiv.org/abs/hep-ph/0610040} {arXiv:hep-ph/0610040 [hep-ph]}
  \BibitemShut {NoStop}%
\bibitem [{\citenamefont {Qin}\ \emph {et~al.}(2011)\citenamefont {Qin},
  \citenamefont {Chang}, \citenamefont {Liu},\ and\ \citenamefont
  {Roberts}}]{Qin:2010pc}%
  \BibitemOpen
  \bibfield  {author} {\bibinfo {author} {\bibfnamefont {S.-x.}\ \bibnamefont
  {Qin}}, \bibinfo {author} {\bibfnamefont {L.}~\bibnamefont {Chang}}, \bibinfo
  {author} {\bibfnamefont {Y.-x.}\ \bibnamefont {Liu}}, \ and\ \bibinfo
  {author} {\bibfnamefont {C.~D.}\ \bibnamefont {Roberts}},\ }\href {\doibase
  10.1103/PhysRevD.84.014017} {\bibfield  {journal} {\bibinfo  {journal} {Phys.
  Rev.}\ }\textbf {\bibinfo {volume} {D84}},\ \bibinfo {pages} {014017}
  (\bibinfo {year} {2011})},\ \Eprint {http://arxiv.org/abs/1010.4231}
  {arXiv:1010.4231 [nucl-th]} \BibitemShut {NoStop}%
\bibitem [{\citenamefont {Su}\ and\ \citenamefont
  {Tywoniuk}(2015)}]{Su:2014rma}%
  \BibitemOpen
  \bibfield  {author} {\bibinfo {author} {\bibfnamefont {N.}~\bibnamefont
  {Su}}\ and\ \bibinfo {author} {\bibfnamefont {K.}~\bibnamefont {Tywoniuk}},\
  }\href {\doibase 10.1103/PhysRevLett.114.161601} {\bibfield  {journal}
  {\bibinfo  {journal} {Phys. Rev. Lett.}\ }\textbf {\bibinfo {volume} {114}},\
  \bibinfo {pages} {161601} (\bibinfo {year} {2015})},\ \Eprint
  {http://arxiv.org/abs/1409.3203} {arXiv:1409.3203 [hep-ph]} \BibitemShut
  {NoStop}%
\bibitem [{\citenamefont {Oehme}\ and\ \citenamefont
  {Zimmermann}(1980)}]{oehme1980gauge}%
  \BibitemOpen
  \bibfield  {author} {\bibinfo {author} {\bibfnamefont {R.}~\bibnamefont
  {Oehme}}\ and\ \bibinfo {author} {\bibfnamefont {W.}~\bibnamefont
  {Zimmermann}},\ }\href@noop {} {\bibfield  {journal} {\bibinfo  {journal}
  {Physical Review D}\ }\textbf {\bibinfo {volume} {21}},\ \bibinfo {pages}
  {1661} (\bibinfo {year} {1980})}\BibitemShut {NoStop}%
\bibitem [{\citenamefont {Oehme}(1990)}]{Oehme:1990kd}%
  \BibitemOpen
  \bibfield  {author} {\bibinfo {author} {\bibfnamefont {R.}~\bibnamefont
  {Oehme}},\ }\href {\doibase 10.1016/0370-2693(90)90499-V} {\bibfield
  {journal} {\bibinfo  {journal} {Phys. Lett.}\ }\textbf {\bibinfo {volume}
  {B252}},\ \bibinfo {pages} {641} (\bibinfo {year} {1990})}\BibitemShut
  {NoStop}%
\bibitem [{\citenamefont {Oehme}\ and\ \citenamefont
  {Xu}(1994)}]{Oehme:1994hf}%
  \BibitemOpen
  \bibfield  {author} {\bibinfo {author} {\bibfnamefont {R.}~\bibnamefont
  {Oehme}}\ and\ \bibinfo {author} {\bibfnamefont {W.-T.}\ \bibnamefont {Xu}},\
  }\href {\doibase 10.1016/0370-2693(94)91025-1} {\bibfield  {journal}
  {\bibinfo  {journal} {Phys. Lett.}\ }\textbf {\bibinfo {volume} {B333}},\
  \bibinfo {pages} {172} (\bibinfo {year} {1994})},\ \Eprint
  {http://arxiv.org/abs/hep-th/9406081} {arXiv:hep-th/9406081 [hep-th]}
  \BibitemShut {NoStop}%
\bibitem [{\citenamefont {Alkofer}\ and\ \citenamefont {von
  Smekal}(2001)}]{Alkofer:2000wg}%
  \BibitemOpen
  \bibfield  {author} {\bibinfo {author} {\bibfnamefont {R.}~\bibnamefont
  {Alkofer}}\ and\ \bibinfo {author} {\bibfnamefont {L.}~\bibnamefont {von
  Smekal}},\ }\href {\doibase 10.1016/S0370-1573(01)00010-2} {\bibfield
  {journal} {\bibinfo  {journal} {Phys. Rept.}\ }\textbf {\bibinfo {volume}
  {353}},\ \bibinfo {pages} {281} (\bibinfo {year} {2001})},\ \Eprint
  {http://arxiv.org/abs/hep-ph/0007355} {arXiv:hep-ph/0007355} \BibitemShut
  {NoStop}%
\bibitem [{\citenamefont {Cornwall}(2013)}]{Cornwall:2013zra}%
  \BibitemOpen
  \bibfield  {author} {\bibinfo {author} {\bibfnamefont {J.~M.}\ \bibnamefont
  {Cornwall}},\ }\href {\doibase 10.1142/S0217732313300358} {\bibfield
  {journal} {\bibinfo  {journal} {Mod. Phys. Lett.}\ }\textbf {\bibinfo
  {volume} {A28}},\ \bibinfo {pages} {1330035} (\bibinfo {year} {2013})},\
  \Eprint {http://arxiv.org/abs/1310.7897} {arXiv:1310.7897 [hep-ph]}
  \BibitemShut {NoStop}%
\bibitem [{\citenamefont {Fischer}\ \emph {et~al.}(2009)\citenamefont
  {Fischer}, \citenamefont {Maas},\ and\ \citenamefont
  {Pawlowski}}]{Fischer:2008uz}%
  \BibitemOpen
  \bibfield  {author} {\bibinfo {author} {\bibfnamefont {C.~S.}\ \bibnamefont
  {Fischer}}, \bibinfo {author} {\bibfnamefont {A.}~\bibnamefont {Maas}}, \
  and\ \bibinfo {author} {\bibfnamefont {J.~M.}\ \bibnamefont {Pawlowski}},\
  }\href {\doibase 10.1016/j.aop.2009.07.009} {\bibfield  {journal} {\bibinfo
  {journal} {Annals Phys.}\ }\textbf {\bibinfo {volume} {324}},\ \bibinfo
  {pages} {2408} (\bibinfo {year} {2009})},\ \Eprint
  {http://arxiv.org/abs/0810.1987} {arXiv:0810.1987 [hep-ph]} \BibitemShut
  {NoStop}%
\bibitem [{\citenamefont {Kugo}\ and\ \citenamefont
  {Ojima}(1979)}]{Kugo:1979gm}%
  \BibitemOpen
  \bibfield  {author} {\bibinfo {author} {\bibfnamefont {T.}~\bibnamefont
  {Kugo}}\ and\ \bibinfo {author} {\bibfnamefont {I.}~\bibnamefont {Ojima}},\
  }\href@noop {} {\bibfield  {journal} {\bibinfo  {journal} {Prog. Theor. Phys.
  Suppl.}\ }\textbf {\bibinfo {volume} {66}},\ \bibinfo {pages} {1} (\bibinfo
  {year} {1979})}\BibitemShut {NoStop}%
\bibitem [{\citenamefont {von Smekal}\ \emph {et~al.}(1997)\citenamefont {von
  Smekal}, \citenamefont {Alkofer},\ and\ \citenamefont
  {Hauck}}]{vonSmekal:1997ohs}%
  \BibitemOpen
  \bibfield  {author} {\bibinfo {author} {\bibfnamefont {L.}~\bibnamefont {von
  Smekal}}, \bibinfo {author} {\bibfnamefont {R.}~\bibnamefont {Alkofer}}, \
  and\ \bibinfo {author} {\bibfnamefont {A.}~\bibnamefont {Hauck}},\ }\href
  {\doibase 10.1103/PhysRevLett.79.3591} {\bibfield  {journal} {\bibinfo
  {journal} {Phys. Rev. Lett.}\ }\textbf {\bibinfo {volume} {79}},\ \bibinfo
  {pages} {3591} (\bibinfo {year} {1997})},\ \Eprint
  {http://arxiv.org/abs/hep-ph/9705242} {arXiv:hep-ph/9705242 [hep-ph]}
  \BibitemShut {NoStop}%
\bibitem [{\citenamefont {Zwanziger}(2002)}]{Zwanziger:2001kw}%
  \BibitemOpen
  \bibfield  {author} {\bibinfo {author} {\bibfnamefont {D.}~\bibnamefont
  {Zwanziger}},\ }\href {\doibase 10.1103/PhysRevD.65.094039} {\bibfield
  {journal} {\bibinfo  {journal} {Phys. Rev.}\ }\textbf {\bibinfo {volume}
  {D65}},\ \bibinfo {pages} {094039} (\bibinfo {year} {2002})},\ \Eprint
  {http://arxiv.org/abs/hep-th/0109224} {arXiv:hep-th/0109224} \BibitemShut
  {NoStop}%
\bibitem [{\citenamefont {Lerche}\ and\ \citenamefont {von
  Smekal}(2002)}]{Lerche:2002ep}%
  \BibitemOpen
  \bibfield  {author} {\bibinfo {author} {\bibfnamefont {C.}~\bibnamefont
  {Lerche}}\ and\ \bibinfo {author} {\bibfnamefont {L.}~\bibnamefont {von
  Smekal}},\ }\href {\doibase 10.1103/PhysRevD.65.125006} {\bibfield  {journal}
  {\bibinfo  {journal} {Phys. Rev.}\ }\textbf {\bibinfo {volume} {D65}},\
  \bibinfo {pages} {125006} (\bibinfo {year} {2002})},\ \Eprint
  {http://arxiv.org/abs/hep-ph/0202194} {arXiv:hep-ph/0202194} \BibitemShut
  {NoStop}%
\bibitem [{\citenamefont {Fischer}\ \emph {et~al.}(2002)\citenamefont
  {Fischer}, \citenamefont {Alkofer},\ and\ \citenamefont
  {Reinhardt}}]{Fischer:2002eq}%
  \BibitemOpen
  \bibfield  {author} {\bibinfo {author} {\bibfnamefont {C.~S.}\ \bibnamefont
  {Fischer}}, \bibinfo {author} {\bibfnamefont {R.}~\bibnamefont {Alkofer}}, \
  and\ \bibinfo {author} {\bibfnamefont {H.}~\bibnamefont {Reinhardt}},\ }\href
  {\doibase 10.1103/PhysRevD.65.094008} {\bibfield  {journal} {\bibinfo
  {journal} {Phys. Rev.}\ }\textbf {\bibinfo {volume} {D65}},\ \bibinfo {pages}
  {094008} (\bibinfo {year} {2002})},\ \Eprint
  {http://arxiv.org/abs/hep-ph/0202195} {arXiv:hep-ph/0202195} \BibitemShut
  {NoStop}%
\bibitem [{\citenamefont {Pawlowski}\ \emph {et~al.}(2004)\citenamefont
  {Pawlowski}, \citenamefont {Litim}, \citenamefont {Nedelko},\ and\
  \citenamefont {von Smekal}}]{Pawlowski:2003hq}%
  \BibitemOpen
  \bibfield  {author} {\bibinfo {author} {\bibfnamefont {J.~M.}\ \bibnamefont
  {Pawlowski}}, \bibinfo {author} {\bibfnamefont {D.~F.}\ \bibnamefont
  {Litim}}, \bibinfo {author} {\bibfnamefont {S.}~\bibnamefont {Nedelko}}, \
  and\ \bibinfo {author} {\bibfnamefont {L.}~\bibnamefont {von Smekal}},\
  }\href {\doibase 10.1103/PhysRevLett.93.152002} {\bibfield  {journal}
  {\bibinfo  {journal} {Phys.Rev.Lett.}\ }\textbf {\bibinfo {volume} {93}},\
  \bibinfo {pages} {152002} (\bibinfo {year} {2004})},\ \Eprint
  {http://arxiv.org/abs/hep-th/0312324} {arXiv:hep-th/0312324 [hep-th]}
  \BibitemShut {NoStop}%
\bibitem [{\citenamefont {Alkofer}\ \emph {et~al.}(2005)\citenamefont
  {Alkofer}, \citenamefont {Fischer},\ and\ \citenamefont
  {Llanes-Estrada}}]{Alkofer:2004it}%
  \BibitemOpen
  \bibfield  {author} {\bibinfo {author} {\bibfnamefont {R.}~\bibnamefont
  {Alkofer}}, \bibinfo {author} {\bibfnamefont {C.~S.}\ \bibnamefont
  {Fischer}}, \ and\ \bibinfo {author} {\bibfnamefont {F.~J.}\ \bibnamefont
  {Llanes-Estrada}},\ }\href {\doibase 10.1016/j.physletb.2005.02.043}
  {\bibfield  {journal} {\bibinfo  {journal} {Phys. Lett.}\ }\textbf {\bibinfo
  {volume} {B611}},\ \bibinfo {pages} {279} (\bibinfo {year} {2005})},\ \Eprint
  {http://arxiv.org/abs/hep-th/0412330} {arXiv:hep-th/0412330} \BibitemShut
  {NoStop}%
\bibitem [{\citenamefont {Fischer}\ and\ \citenamefont
  {Pawlowski}(2007)}]{Fischer:2006vf}%
  \BibitemOpen
  \bibfield  {author} {\bibinfo {author} {\bibfnamefont {C.~S.}\ \bibnamefont
  {Fischer}}\ and\ \bibinfo {author} {\bibfnamefont {J.~M.}\ \bibnamefont
  {Pawlowski}},\ }\href {\doibase 10.1103/PhysRevD.75.025012} {\bibfield
  {journal} {\bibinfo  {journal} {Phys. Rev.}\ }\textbf {\bibinfo {volume}
  {D75}},\ \bibinfo {pages} {025012} (\bibinfo {year} {2007})},\ \Eprint
  {http://arxiv.org/abs/hep-th/0609009} {arXiv:hep-th/0609009} \BibitemShut
  {NoStop}%
\bibitem [{\citenamefont {Alkofer}\ \emph {et~al.}(2010)\citenamefont
  {Alkofer}, \citenamefont {Huber},\ and\ \citenamefont
  {Schwenzer}}]{Alkofer:2008jy}%
  \BibitemOpen
  \bibfield  {author} {\bibinfo {author} {\bibfnamefont {R.}~\bibnamefont
  {Alkofer}}, \bibinfo {author} {\bibfnamefont {M.~Q.}\ \bibnamefont {Huber}},
  \ and\ \bibinfo {author} {\bibfnamefont {K.}~\bibnamefont {Schwenzer}},\
  }\href {\doibase 10.1103/PhysRevD.81.105010} {\bibfield  {journal} {\bibinfo
  {journal} {Phys. Rev.}\ }\textbf {\bibinfo {volume} {D81}},\ \bibinfo {pages}
  {105010} (\bibinfo {year} {2010})},\ \Eprint {http://arxiv.org/abs/0801.2762}
  {arXiv:0801.2762 [hep-th]} \BibitemShut {NoStop}%
\bibitem [{\citenamefont {Fischer}\ and\ \citenamefont
  {Pawlowski}(2009)}]{Fischer:2009tn}%
  \BibitemOpen
  \bibfield  {author} {\bibinfo {author} {\bibfnamefont {C.~S.}\ \bibnamefont
  {Fischer}}\ and\ \bibinfo {author} {\bibfnamefont {J.~M.}\ \bibnamefont
  {Pawlowski}},\ }\href {\doibase 10.1103/PhysRevD.80.025023} {\bibfield
  {journal} {\bibinfo  {journal} {Phys. Rev.}\ }\textbf {\bibinfo {volume}
  {D80}},\ \bibinfo {pages} {025023} (\bibinfo {year} {2009})},\ \Eprint
  {http://arxiv.org/abs/0903.2193} {arXiv:0903.2193 [hep-th]} \BibitemShut
  {NoStop}%
\bibitem [{\citenamefont {Alkofer}\ \emph {et~al.}(2004)\citenamefont
  {Alkofer}, \citenamefont {Detmold}, \citenamefont {Fischer},\ and\
  \citenamefont {Maris}}]{Alkofer:2003jj}%
  \BibitemOpen
  \bibfield  {author} {\bibinfo {author} {\bibfnamefont {R.}~\bibnamefont
  {Alkofer}}, \bibinfo {author} {\bibfnamefont {W.}~\bibnamefont {Detmold}},
  \bibinfo {author} {\bibfnamefont {C.~S.}\ \bibnamefont {Fischer}}, \ and\
  \bibinfo {author} {\bibfnamefont {P.}~\bibnamefont {Maris}},\ }\href
  {\doibase 10.1103/PhysRevD.70.014014} {\bibfield  {journal} {\bibinfo
  {journal} {Phys. Rev.}\ }\textbf {\bibinfo {volume} {D70}},\ \bibinfo {pages}
  {014014} (\bibinfo {year} {2004})},\ \Eprint
  {http://arxiv.org/abs/hep-ph/0309077} {arXiv:hep-ph/0309077} \BibitemShut
  {NoStop}%
\bibitem [{\citenamefont {Huber}\ and\ \citenamefont {von
  Smekal}(2013)}]{Huber:2012kd}%
  \BibitemOpen
  \bibfield  {author} {\bibinfo {author} {\bibfnamefont {M.~Q.}\ \bibnamefont
  {Huber}}\ and\ \bibinfo {author} {\bibfnamefont {L.}~\bibnamefont {von
  Smekal}},\ }\href {\doibase 10.1007/JHEP04(2013)149} {\bibfield  {journal}
  {\bibinfo  {journal} {JHEP}\ }\textbf {\bibinfo {volume} {1304}},\ \bibinfo
  {pages} {149} (\bibinfo {year} {2013})},\ \Eprint
  {http://arxiv.org/abs/1211.6092} {arXiv:1211.6092 [hep-th]} \BibitemShut
  {NoStop}%
\bibitem [{\citenamefont {Aguilar}\ \emph {et~al.}(2014)\citenamefont
  {Aguilar}, \citenamefont {Binosi}, \citenamefont {Ib{\'a}{\~n}ez},\ and\
  \citenamefont {Papavassiliou}}]{Aguilar:2013vaa}%
  \BibitemOpen
  \bibfield  {author} {\bibinfo {author} {\bibfnamefont {A.}~\bibnamefont
  {Aguilar}}, \bibinfo {author} {\bibfnamefont {D.}~\bibnamefont {Binosi}},
  \bibinfo {author} {\bibfnamefont {D.}~\bibnamefont {Ib{\'a}{\~n}ez}}, \ and\
  \bibinfo {author} {\bibfnamefont {J.}~\bibnamefont {Papavassiliou}},\ }\href
  {\doibase 10.1103/PhysRevD.89.085008} {\bibfield  {journal} {\bibinfo
  {journal} {Phys.Rev.}\ }\textbf {\bibinfo {volume} {D89}},\ \bibinfo {pages}
  {085008} (\bibinfo {year} {2014})},\ \Eprint {http://arxiv.org/abs/1312.1212}
  {arXiv:1312.1212 [hep-ph]} \BibitemShut {NoStop}%
\bibitem [{\citenamefont {Pelaez}\ \emph {et~al.}(2013)\citenamefont {Pelaez},
  \citenamefont {Tissier},\ and\ \citenamefont {Wschebor}}]{Pelaez:2013cpa}%
  \BibitemOpen
  \bibfield  {author} {\bibinfo {author} {\bibfnamefont {M.}~\bibnamefont
  {Pelaez}}, \bibinfo {author} {\bibfnamefont {M.}~\bibnamefont {Tissier}}, \
  and\ \bibinfo {author} {\bibfnamefont {N.}~\bibnamefont {Wschebor}},\ }\href
  {\doibase 10.1103/PhysRevD.88.125003} {\bibfield  {journal} {\bibinfo
  {journal} {Phys.Rev.}\ }\textbf {\bibinfo {volume} {D88}},\ \bibinfo {pages}
  {125003} (\bibinfo {year} {2013})},\ \Eprint {http://arxiv.org/abs/1310.2594}
  {arXiv:1310.2594 [hep-th]} \BibitemShut {NoStop}%
\bibitem [{\citenamefont {Blum}\ \emph {et~al.}(2014)\citenamefont {Blum},
  \citenamefont {Huber}, \citenamefont {Mitter},\ and\ \citenamefont {von
  Smekal}}]{Blum:2014gna}%
  \BibitemOpen
  \bibfield  {author} {\bibinfo {author} {\bibfnamefont {A.}~\bibnamefont
  {Blum}}, \bibinfo {author} {\bibfnamefont {M.~Q.}\ \bibnamefont {Huber}},
  \bibinfo {author} {\bibfnamefont {M.}~\bibnamefont {Mitter}}, \ and\ \bibinfo
  {author} {\bibfnamefont {L.}~\bibnamefont {von Smekal}},\ }\href {\doibase
  10.1103/PhysRevD.89.061703} {\bibfield  {journal} {\bibinfo  {journal}
  {Phys.Rev.}\ }\textbf {\bibinfo {volume} {D89}},\ \bibinfo {pages} {061703}
  (\bibinfo {year} {2014})},\ \Eprint {http://arxiv.org/abs/1401.0713}
  {arXiv:1401.0713 [hep-ph]} \BibitemShut {NoStop}%
\bibitem [{\citenamefont {Eichmann}\ \emph {et~al.}(2014)\citenamefont
  {Eichmann}, \citenamefont {Williams}, \citenamefont {Alkofer},\ and\
  \citenamefont {Vujinovic}}]{Eichmann:2014xya}%
  \BibitemOpen
  \bibfield  {author} {\bibinfo {author} {\bibfnamefont {G.}~\bibnamefont
  {Eichmann}}, \bibinfo {author} {\bibfnamefont {R.}~\bibnamefont {Williams}},
  \bibinfo {author} {\bibfnamefont {R.}~\bibnamefont {Alkofer}}, \ and\
  \bibinfo {author} {\bibfnamefont {M.}~\bibnamefont {Vujinovic}},\ }\href
  {\doibase 10.1103/PhysRevD.89.105014} {\bibfield  {journal} {\bibinfo
  {journal} {Phys.Rev.}\ }\textbf {\bibinfo {volume} {D89}},\ \bibinfo {pages}
  {105014} (\bibinfo {year} {2014})},\ \Eprint {http://arxiv.org/abs/1402.1365}
  {arXiv:1402.1365 [hep-ph]} \BibitemShut {NoStop}%
\bibitem [{\citenamefont {Cyrol}\ \emph {et~al.}(2015)\citenamefont {Cyrol},
  \citenamefont {Huber},\ and\ \citenamefont {von Smekal}}]{Cyrol:2014kca}%
  \BibitemOpen
  \bibfield  {author} {\bibinfo {author} {\bibfnamefont {A.~K.}\ \bibnamefont
  {Cyrol}}, \bibinfo {author} {\bibfnamefont {M.~Q.}\ \bibnamefont {Huber}}, \
  and\ \bibinfo {author} {\bibfnamefont {L.}~\bibnamefont {von Smekal}},\
  }\href {\doibase 10.1140/epjc/s10052-015-3312-1} {\bibfield  {journal}
  {\bibinfo  {journal} {Eur. Phys. J.}\ }\textbf {\bibinfo {volume} {C75}},\
  \bibinfo {pages} {102} (\bibinfo {year} {2015})},\ \Eprint
  {http://arxiv.org/abs/1408.5409} {arXiv:1408.5409 [hep-ph]} \BibitemShut
  {NoStop}%
\bibitem [{\citenamefont {Reinosa}\ \emph {et~al.}(2017)\citenamefont
  {Reinosa}, \citenamefont {Serreau}, \citenamefont {Tissier},\ and\
  \citenamefont {Wschebor}}]{Reinosa:2017qtf}%
  \BibitemOpen
  \bibfield  {author} {\bibinfo {author} {\bibfnamefont {U.}~\bibnamefont
  {Reinosa}}, \bibinfo {author} {\bibfnamefont {J.}~\bibnamefont {Serreau}},
  \bibinfo {author} {\bibfnamefont {M.}~\bibnamefont {Tissier}}, \ and\
  \bibinfo {author} {\bibfnamefont {N.}~\bibnamefont {Wschebor}},\ }\href
  {\doibase 10.1103/PhysRevD.96.014005} {\bibfield  {journal} {\bibinfo
  {journal} {Phys. Rev.}\ }\textbf {\bibinfo {volume} {D96}},\ \bibinfo {pages}
  {014005} (\bibinfo {year} {2017})},\ \Eprint
  {http://arxiv.org/abs/1703.04041} {arXiv:1703.04041 [hep-th]} \BibitemShut
  {NoStop}%
\bibitem [{\citenamefont {Gracey}\ and\ \citenamefont
  {Simms}(2018)}]{Gracey:2018fkg}%
  \BibitemOpen
  \bibfield  {author} {\bibinfo {author} {\bibfnamefont {J.~A.}\ \bibnamefont
  {Gracey}}\ and\ \bibinfo {author} {\bibfnamefont {R.~M.}\ \bibnamefont
  {Simms}},\ }\href {\doibase 10.1103/PhysRevD.97.085016} {\bibfield  {journal}
  {\bibinfo  {journal} {Phys. Rev.}\ }\textbf {\bibinfo {volume} {D97}},\
  \bibinfo {pages} {085016} (\bibinfo {year} {2018})},\ \Eprint
  {http://arxiv.org/abs/1801.10415} {arXiv:1801.10415 [hep-th]} \BibitemShut
  {NoStop}%
\bibitem [{\citenamefont {Gracey}(2017)}]{Gracey:2017yfi}%
  \BibitemOpen
  \bibfield  {author} {\bibinfo {author} {\bibfnamefont {J.~A.}\ \bibnamefont
  {Gracey}},\ }\href {\doibase 10.1103/PhysRevD.95.065013} {\bibfield
  {journal} {\bibinfo  {journal} {Phys. Rev.}\ }\textbf {\bibinfo {volume}
  {D95}},\ \bibinfo {pages} {065013} (\bibinfo {year} {2017})},\ \Eprint
  {http://arxiv.org/abs/1703.01094} {arXiv:1703.01094 [hep-ph]} \BibitemShut
  {NoStop}%
\bibitem [{\citenamefont {Gracey}(2014{\natexlab{a}})}]{Gracey:2014ola}%
  \BibitemOpen
  \bibfield  {author} {\bibinfo {author} {\bibfnamefont {J.~A.}\ \bibnamefont
  {Gracey}},\ }\href {\doibase 10.1103/PhysRevD.90.025011} {\bibfield
  {journal} {\bibinfo  {journal} {Phys. Rev.}\ }\textbf {\bibinfo {volume}
  {D90}},\ \bibinfo {pages} {025011} (\bibinfo {year} {2014}{\natexlab{a}})},\
  \Eprint {http://arxiv.org/abs/1406.1618} {arXiv:1406.1618 [hep-ph]}
  \BibitemShut {NoStop}%
\bibitem [{\citenamefont {Bell}\ and\ \citenamefont
  {Gracey}(2015)}]{Bell:2015dbr}%
  \BibitemOpen
  \bibfield  {author} {\bibinfo {author} {\bibfnamefont {J.~M.}\ \bibnamefont
  {Bell}}\ and\ \bibinfo {author} {\bibfnamefont {J.~A.}\ \bibnamefont
  {Gracey}},\ }\href {\doibase 10.1103/PhysRevD.92.125001} {\bibfield
  {journal} {\bibinfo  {journal} {Phys. Rev.}\ }\textbf {\bibinfo {volume}
  {D92}},\ \bibinfo {pages} {125001} (\bibinfo {year} {2015})},\ \Eprint
  {http://arxiv.org/abs/1511.00854} {arXiv:1511.00854 [hep-th]} \BibitemShut
  {NoStop}%
\bibitem [{\citenamefont {Gracey}(2014{\natexlab{b}})}]{Gracey:2014dna}%
  \BibitemOpen
  \bibfield  {author} {\bibinfo {author} {\bibfnamefont {J.~A.}\ \bibnamefont
  {Gracey}},\ }\href {\doibase 10.1088/1751-8113/47/44/445401} {\bibfield
  {journal} {\bibinfo  {journal} {J. Phys.}\ }\textbf {\bibinfo {volume}
  {A47}},\ \bibinfo {pages} {445401} (\bibinfo {year} {2014}{\natexlab{b}})},\
  \Eprint {http://arxiv.org/abs/1409.0455} {arXiv:1409.0455 [hep-ph]}
  \BibitemShut {NoStop}%
\bibitem [{\citenamefont {Fister}\ and\ \citenamefont
  {Pawlowski}(2013)}]{Fister:2013bh}%
  \BibitemOpen
  \bibfield  {author} {\bibinfo {author} {\bibfnamefont {L.}~\bibnamefont
  {Fister}}\ and\ \bibinfo {author} {\bibfnamefont {J.~M.}\ \bibnamefont
  {Pawlowski}},\ }\href {\doibase 10.1103/PhysRevD.88.045010} {\bibfield
  {journal} {\bibinfo  {journal} {Phys.Rev.}\ }\textbf {\bibinfo {volume}
  {D88}},\ \bibinfo {pages} {045010} (\bibinfo {year} {2013})},\ \Eprint
  {http://arxiv.org/abs/1301.4163} {arXiv:1301.4163 [hep-ph]} \BibitemShut
  {NoStop}%
\bibitem [{\citenamefont {Sternbeck}\ \emph {et~al.}(2006)\citenamefont
  {Sternbeck}, \citenamefont {Ilgenfritz}, \citenamefont {Muller-Preussker},
  \citenamefont {Schiller},\ and\ \citenamefont
  {Bogolubsky}}]{Sternbeck:2006cg}%
  \BibitemOpen
  \bibfield  {author} {\bibinfo {author} {\bibfnamefont {A.}~\bibnamefont
  {Sternbeck}}, \bibinfo {author} {\bibfnamefont {E.~M.}\ \bibnamefont
  {Ilgenfritz}}, \bibinfo {author} {\bibfnamefont {M.}~\bibnamefont
  {Muller-Preussker}}, \bibinfo {author} {\bibfnamefont {A.}~\bibnamefont
  {Schiller}}, \ and\ \bibinfo {author} {\bibfnamefont {I.~L.}\ \bibnamefont
  {Bogolubsky}},\ }\href@noop {} {\bibfield  {journal} {\bibinfo  {journal}
  {PoS}\ }\textbf {\bibinfo {volume} {LAT2006}},\ \bibinfo {pages} {076}
  (\bibinfo {year} {2006})},\ \Eprint {http://arxiv.org/abs/hep-lat/0610053}
  {arXiv:hep-lat/0610053} \BibitemShut {NoStop}%
\bibitem [{\citenamefont {Cucchieri}\ \emph {et~al.}(2007)\citenamefont
  {Cucchieri}, \citenamefont {Maas},\ and\ \citenamefont
  {Mendes}}]{Cucchieri:2007ta}%
  \BibitemOpen
  \bibfield  {author} {\bibinfo {author} {\bibfnamefont {A.}~\bibnamefont
  {Cucchieri}}, \bibinfo {author} {\bibfnamefont {A.}~\bibnamefont {Maas}}, \
  and\ \bibinfo {author} {\bibfnamefont {T.}~\bibnamefont {Mendes}},\ }\href
  {\doibase 10.1103/PhysRevD.75.076003} {\bibfield  {journal} {\bibinfo
  {journal} {Phys. Rev.}\ }\textbf {\bibinfo {volume} {D75}},\ \bibinfo {pages}
  {076003} (\bibinfo {year} {2007})},\ \Eprint
  {http://arxiv.org/abs/hep-lat/0702022} {arXiv:hep-lat/0702022} \BibitemShut
  {NoStop}%
\bibitem [{\citenamefont {Cucchieri}\ and\ \citenamefont
  {Mendes}(2008{\natexlab{a}})}]{Cucchieri:2007rg}%
  \BibitemOpen
  \bibfield  {author} {\bibinfo {author} {\bibfnamefont {A.}~\bibnamefont
  {Cucchieri}}\ and\ \bibinfo {author} {\bibfnamefont {T.}~\bibnamefont
  {Mendes}},\ }\href {\doibase 10.1103/PhysRevLett.100.241601} {\bibfield
  {journal} {\bibinfo  {journal} {Phys. Rev. Lett.}\ }\textbf {\bibinfo
  {volume} {100}},\ \bibinfo {pages} {241601} (\bibinfo {year}
  {2008}{\natexlab{a}})},\ \Eprint {http://arxiv.org/abs/0712.3517}
  {arXiv:0712.3517 [hep-lat]} \BibitemShut {NoStop}%
\bibitem [{\citenamefont {Cucchieri}\ and\ \citenamefont
  {Mendes}(2008{\natexlab{b}})}]{Cucchieri:2008fc}%
  \BibitemOpen
  \bibfield  {author} {\bibinfo {author} {\bibfnamefont {A.}~\bibnamefont
  {Cucchieri}}\ and\ \bibinfo {author} {\bibfnamefont {T.}~\bibnamefont
  {Mendes}},\ }\href {\doibase 10.1103/PhysRevD.78.094503} {\bibfield
  {journal} {\bibinfo  {journal} {Phys. Rev.}\ }\textbf {\bibinfo {volume}
  {D78}},\ \bibinfo {pages} {094503} (\bibinfo {year} {2008}{\natexlab{b}})},\
  \Eprint {http://arxiv.org/abs/0804.2371} {arXiv:0804.2371 [hep-lat]}
  \BibitemShut {NoStop}%
\bibitem [{\citenamefont {Maas}\ \emph {et~al.}(2010)\citenamefont {Maas},
  \citenamefont {Pawlowski}, \citenamefont {Spielmann}, \citenamefont
  {Sternbeck},\ and\ \citenamefont {von Smekal}}]{Maas:2009ph}%
  \BibitemOpen
  \bibfield  {author} {\bibinfo {author} {\bibfnamefont {A.}~\bibnamefont
  {Maas}}, \bibinfo {author} {\bibfnamefont {J.~M.}\ \bibnamefont {Pawlowski}},
  \bibinfo {author} {\bibfnamefont {D.}~\bibnamefont {Spielmann}}, \bibinfo
  {author} {\bibfnamefont {A.}~\bibnamefont {Sternbeck}}, \ and\ \bibinfo
  {author} {\bibfnamefont {L.}~\bibnamefont {von Smekal}},\ }\href {\doibase
  10.1140/epjc/s10052-010-1306-6} {\bibfield  {journal} {\bibinfo  {journal}
  {Eur.Phys.J.}\ }\textbf {\bibinfo {volume} {C68}},\ \bibinfo {pages} {183}
  (\bibinfo {year} {2010})},\ \Eprint {http://arxiv.org/abs/0912.4203}
  {arXiv:0912.4203 [hep-lat]} \BibitemShut {NoStop}%
\bibitem [{\citenamefont {Maas}(2010)}]{Maas:2009se}%
  \BibitemOpen
  \bibfield  {author} {\bibinfo {author} {\bibfnamefont {A.}~\bibnamefont
  {Maas}},\ }\href {\doibase 10.1016/j.physletb.2010.04.052} {\bibfield
  {journal} {\bibinfo  {journal} {Phys.Lett.}\ }\textbf {\bibinfo {volume}
  {B689}},\ \bibinfo {pages} {107} (\bibinfo {year} {2010})},\ \Eprint
  {http://arxiv.org/abs/0907.5185} {arXiv:0907.5185 [hep-lat]} \BibitemShut
  {NoStop}%
\bibitem [{\citenamefont {Aouane}\ \emph {et~al.}(2012)\citenamefont {Aouane},
  \citenamefont {Bornyakov}, \citenamefont {Ilgenfritz}, \citenamefont
  {Mitrjushkin}, \citenamefont {Muller-Preussker},\ and\ \citenamefont
  {Sternbeck}}]{Aouane:2011fv}%
  \BibitemOpen
  \bibfield  {author} {\bibinfo {author} {\bibfnamefont {R.}~\bibnamefont
  {Aouane}}, \bibinfo {author} {\bibfnamefont {V.~G.}\ \bibnamefont
  {Bornyakov}}, \bibinfo {author} {\bibfnamefont {E.~M.}\ \bibnamefont
  {Ilgenfritz}}, \bibinfo {author} {\bibfnamefont {V.~K.}\ \bibnamefont
  {Mitrjushkin}}, \bibinfo {author} {\bibfnamefont {M.}~\bibnamefont
  {Muller-Preussker}}, \ and\ \bibinfo {author} {\bibfnamefont
  {A.}~\bibnamefont {Sternbeck}},\ }\href {\doibase 10.1103/PhysRevD.85.034501}
  {\bibfield  {journal} {\bibinfo  {journal} {Phys. Rev.}\ }\textbf {\bibinfo
  {volume} {D85}},\ \bibinfo {pages} {034501} (\bibinfo {year} {2012})},\
  \Eprint {http://arxiv.org/abs/1108.1735} {arXiv:1108.1735 [hep-lat]}
  \BibitemShut {NoStop}%
\bibitem [{\citenamefont {Maas}\ \emph {et~al.}(2012)\citenamefont {Maas},
  \citenamefont {Pawlowski}, \citenamefont {von Smekal},\ and\ \citenamefont
  {Spielmann}}]{Maas:2011ez}%
  \BibitemOpen
  \bibfield  {author} {\bibinfo {author} {\bibfnamefont {A.}~\bibnamefont
  {Maas}}, \bibinfo {author} {\bibfnamefont {J.~M.}\ \bibnamefont {Pawlowski}},
  \bibinfo {author} {\bibfnamefont {L.}~\bibnamefont {von Smekal}}, \ and\
  \bibinfo {author} {\bibfnamefont {D.}~\bibnamefont {Spielmann}},\ }\href
  {\doibase 10.1103/PhysRevD.85.034037} {\bibfield  {journal} {\bibinfo
  {journal} {Phys.Rev.}\ }\textbf {\bibinfo {volume} {D85}},\ \bibinfo {pages}
  {034037} (\bibinfo {year} {2012})},\ \Eprint {http://arxiv.org/abs/1110.6340}
  {arXiv:1110.6340 [hep-lat]} \BibitemShut {NoStop}%
\bibitem [{\citenamefont {Cucchieri}\ \emph {et~al.}(2012)\citenamefont
  {Cucchieri}, \citenamefont {Dudal}, \citenamefont {Mendes},\ and\
  \citenamefont {Vandersickel}}]{Cucchieri:2012gb}%
  \BibitemOpen
  \bibfield  {author} {\bibinfo {author} {\bibfnamefont {A.}~\bibnamefont
  {Cucchieri}}, \bibinfo {author} {\bibfnamefont {D.}~\bibnamefont {Dudal}},
  \bibinfo {author} {\bibfnamefont {T.}~\bibnamefont {Mendes}}, \ and\ \bibinfo
  {author} {\bibfnamefont {N.}~\bibnamefont {Vandersickel}},\ }\href@noop {} {\
   (\bibinfo {year} {2012})},\ \Eprint {http://arxiv.org/abs/1202.0639}
  {arXiv:1202.0639 [hep-lat]} \BibitemShut {NoStop}%
\bibitem [{\citenamefont {Sternbeck}\ and\ \citenamefont
  {M\"uller-Preussker}(2013)}]{Sternbeck:2012mf}%
  \BibitemOpen
  \bibfield  {author} {\bibinfo {author} {\bibfnamefont {A.}~\bibnamefont
  {Sternbeck}}\ and\ \bibinfo {author} {\bibfnamefont {M.}~\bibnamefont
  {M\"uller-Preussker}},\ }\href {\doibase 10.1016/j.physletb.2013.08.017}
  {\bibfield  {journal} {\bibinfo  {journal} {Phys. Lett.}\ }\textbf {\bibinfo
  {volume} {B726}},\ \bibinfo {pages} {396} (\bibinfo {year} {2013})},\ \Eprint
  {http://arxiv.org/abs/1211.3057} {arXiv:1211.3057 [hep-lat]} \BibitemShut
  {NoStop}%
\bibitem [{\citenamefont {Silva}\ \emph {et~al.}(2014)\citenamefont {Silva},
  \citenamefont {Oliveira}, \citenamefont {Bicudo},\ and\ \citenamefont
  {Cardoso}}]{Silva:2013maa}%
  \BibitemOpen
  \bibfield  {author} {\bibinfo {author} {\bibfnamefont {P.~J.}\ \bibnamefont
  {Silva}}, \bibinfo {author} {\bibfnamefont {O.}~\bibnamefont {Oliveira}},
  \bibinfo {author} {\bibfnamefont {P.}~\bibnamefont {Bicudo}}, \ and\ \bibinfo
  {author} {\bibfnamefont {N.}~\bibnamefont {Cardoso}},\ }\href {\doibase
  10.1103/PhysRevD.89.074503} {\bibfield  {journal} {\bibinfo  {journal} {Phys.
  Rev.}\ }\textbf {\bibinfo {volume} {D89}},\ \bibinfo {pages} {074503}
  (\bibinfo {year} {2014})},\ \Eprint {http://arxiv.org/abs/1310.5629}
  {arXiv:1310.5629 [hep-lat]} \BibitemShut {NoStop}%
\bibitem [{\citenamefont {Maas}(2015)}]{Maas:2014xma}%
  \BibitemOpen
  \bibfield  {author} {\bibinfo {author} {\bibfnamefont {A.}~\bibnamefont
  {Maas}},\ }\href {\doibase 10.1103/PhysRevD.91.034502} {\bibfield  {journal}
  {\bibinfo  {journal} {Phys. Rev.}\ }\textbf {\bibinfo {volume} {D91}},\
  \bibinfo {pages} {034502} (\bibinfo {year} {2015})},\ \Eprint
  {http://arxiv.org/abs/1402.5050} {arXiv:1402.5050 [hep-lat]} \BibitemShut
  {NoStop}%
\bibitem [{\citenamefont {Maas}(2016)}]{Maas:2015nva}%
  \BibitemOpen
  \bibfield  {author} {\bibinfo {author} {\bibfnamefont {A.}~\bibnamefont
  {Maas}},\ }\href {\doibase 10.1103/PhysRevD.93.054504} {\bibfield  {journal}
  {\bibinfo  {journal} {Phys. Rev.}\ }\textbf {\bibinfo {volume} {D93}},\
  \bibinfo {pages} {054504} (\bibinfo {year} {2016})},\ \Eprint
  {http://arxiv.org/abs/1510.08407} {arXiv:1510.08407 [hep-lat]} \BibitemShut
  {NoStop}%
\bibitem [{\citenamefont {Duarte}\ \emph {et~al.}(2016)\citenamefont {Duarte},
  \citenamefont {Oliveira},\ and\ \citenamefont {Silva}}]{Duarte:2016iko}%
  \BibitemOpen
  \bibfield  {author} {\bibinfo {author} {\bibfnamefont {A.~G.}\ \bibnamefont
  {Duarte}}, \bibinfo {author} {\bibfnamefont {O.}~\bibnamefont {Oliveira}}, \
  and\ \bibinfo {author} {\bibfnamefont {P.~J.}\ \bibnamefont {Silva}},\ }\href
  {\doibase 10.1103/PhysRevD.94.014502} {\bibfield  {journal} {\bibinfo
  {journal} {Phys. Rev.}\ }\textbf {\bibinfo {volume} {D94}},\ \bibinfo {pages}
  {014502} (\bibinfo {year} {2016})},\ \Eprint
  {http://arxiv.org/abs/1605.00594} {arXiv:1605.00594 [hep-lat]} \BibitemShut
  {NoStop}%
\bibitem [{\citenamefont {Gao}\ \emph {et~al.}(2018)\citenamefont {Gao},
  \citenamefont {Qin}, \citenamefont {Roberts},\ and\ \citenamefont
  {Rodriguez-Quintero}}]{Gao:2017uox}%
  \BibitemOpen
  \bibfield  {author} {\bibinfo {author} {\bibfnamefont {F.}~\bibnamefont
  {Gao}}, \bibinfo {author} {\bibfnamefont {S.-X.}\ \bibnamefont {Qin}},
  \bibinfo {author} {\bibfnamefont {C.~D.}\ \bibnamefont {Roberts}}, \ and\
  \bibinfo {author} {\bibfnamefont {J.}~\bibnamefont {Rodriguez-Quintero}},\
  }\href {\doibase 10.1103/PhysRevD.97.034010} {\bibfield  {journal} {\bibinfo
  {journal} {Phys. Rev.}\ }\textbf {\bibinfo {volume} {D97}},\ \bibinfo {pages}
  {034010} (\bibinfo {year} {2018})},\ \Eprint
  {http://arxiv.org/abs/1706.04681} {arXiv:1706.04681 [hep-ph]} \BibitemShut
  {NoStop}%
\bibitem [{\citenamefont {Dudal}\ \emph {et~al.}(2018)\citenamefont {Dudal},
  \citenamefont {Oliveira},\ and\ \citenamefont {Silva}}]{Dudal:2018cli}%
  \BibitemOpen
  \bibfield  {author} {\bibinfo {author} {\bibfnamefont {D.}~\bibnamefont
  {Dudal}}, \bibinfo {author} {\bibfnamefont {O.}~\bibnamefont {Oliveira}}, \
  and\ \bibinfo {author} {\bibfnamefont {P.~J.}\ \bibnamefont {Silva}},\
  }\href@noop {} {\  (\bibinfo {year} {2018})},\ \Eprint
  {http://arxiv.org/abs/1803.02281} {arXiv:1803.02281 [hep-lat]} \BibitemShut
  {NoStop}%
\bibitem [{\citenamefont {Aguilar}\ \emph {et~al.}(2008)\citenamefont
  {Aguilar}, \citenamefont {Binosi},\ and\ \citenamefont
  {Papavassiliou}}]{Aguilar:2008xm}%
  \BibitemOpen
  \bibfield  {author} {\bibinfo {author} {\bibfnamefont {A.~C.}\ \bibnamefont
  {Aguilar}}, \bibinfo {author} {\bibfnamefont {D.}~\bibnamefont {Binosi}}, \
  and\ \bibinfo {author} {\bibfnamefont {J.}~\bibnamefont {Papavassiliou}},\
  }\href {\doibase 10.1103/PhysRevD.78.025010} {\bibfield  {journal} {\bibinfo
  {journal} {Phys. Rev.}\ }\textbf {\bibinfo {volume} {D78}},\ \bibinfo {pages}
  {025010} (\bibinfo {year} {2008})},\ \Eprint {http://arxiv.org/abs/0802.1870}
  {arXiv:0802.1870 [hep-ph]} \BibitemShut {NoStop}%
\bibitem [{\citenamefont {Boucaud}\ \emph {et~al.}(2008)\citenamefont {Boucaud}
  \emph {et~al.}}]{Boucaud:2008ky}%
  \BibitemOpen
  \bibfield  {author} {\bibinfo {author} {\bibfnamefont {P.}~\bibnamefont
  {Boucaud}} \emph {et~al.},\ }\href {\doibase 10.1088/1126-6708/2008/06/099}
  {\bibfield  {journal} {\bibinfo  {journal} {JHEP}\ }\textbf {\bibinfo
  {volume} {06}},\ \bibinfo {pages} {099} (\bibinfo {year} {2008})},\ \Eprint
  {http://arxiv.org/abs/0803.2161} {arXiv:0803.2161 [hep-ph]} \BibitemShut
  {NoStop}%
\bibitem [{\citenamefont {Strauss}\ \emph {et~al.}(2012)\citenamefont
  {Strauss}, \citenamefont {Fischer},\ and\ \citenamefont
  {Kellermann}}]{Strauss:2012dg}%
  \BibitemOpen
  \bibfield  {author} {\bibinfo {author} {\bibfnamefont {S.}~\bibnamefont
  {Strauss}}, \bibinfo {author} {\bibfnamefont {C.~S.}\ \bibnamefont
  {Fischer}}, \ and\ \bibinfo {author} {\bibfnamefont {C.}~\bibnamefont
  {Kellermann}},\ }\href {\doibase 10.1103/PhysRevLett.109.252001} {\bibfield
  {journal} {\bibinfo  {journal} {Phys.Rev.Lett.}\ }\textbf {\bibinfo {volume}
  {109}},\ \bibinfo {pages} {252001} (\bibinfo {year} {2012})},\ \Eprint
  {http://arxiv.org/abs/1208.6239} {arXiv:1208.6239 [hep-ph]} \BibitemShut
  {NoStop}%
\bibitem [{\citenamefont {Baulieu}\ \emph {et~al.}(2010)\citenamefont
  {Baulieu}, \citenamefont {Dudal}, \citenamefont {Guimaraes}, \citenamefont
  {Huber}, \citenamefont {Sorella}, \citenamefont {Vandersickel},\ and\
  \citenamefont {Zwanziger}}]{Baulieu:2009ha}%
  \BibitemOpen
  \bibfield  {author} {\bibinfo {author} {\bibfnamefont {L.}~\bibnamefont
  {Baulieu}}, \bibinfo {author} {\bibfnamefont {D.}~\bibnamefont {Dudal}},
  \bibinfo {author} {\bibfnamefont {M.~S.}\ \bibnamefont {Guimaraes}}, \bibinfo
  {author} {\bibfnamefont {M.~Q.}\ \bibnamefont {Huber}}, \bibinfo {author}
  {\bibfnamefont {S.~P.}\ \bibnamefont {Sorella}}, \bibinfo {author}
  {\bibfnamefont {N.}~\bibnamefont {Vandersickel}}, \ and\ \bibinfo {author}
  {\bibfnamefont {D.}~\bibnamefont {Zwanziger}},\ }\href {\doibase
  10.1103/PhysRevD.82.025021} {\bibfield  {journal} {\bibinfo  {journal} {Phys.
  Rev.}\ }\textbf {\bibinfo {volume} {D82}},\ \bibinfo {pages} {025021}
  (\bibinfo {year} {2010})},\ \Eprint {http://arxiv.org/abs/0912.5153}
  {arXiv:0912.5153 [hep-th]} \BibitemShut {NoStop}%
\bibitem [{\citenamefont {Curci}\ and\ \citenamefont
  {Ferrari}(1976)}]{Curci:1976kh}%
  \BibitemOpen
  \bibfield  {author} {\bibinfo {author} {\bibfnamefont {G.}~\bibnamefont
  {Curci}}\ and\ \bibinfo {author} {\bibfnamefont {R.}~\bibnamefont
  {Ferrari}},\ }\href {\doibase 10.1007/BF02730056} {\bibfield  {journal}
  {\bibinfo  {journal} {Nuovo Cim.}\ }\textbf {\bibinfo {volume} {A35}},\
  \bibinfo {pages} {1} (\bibinfo {year} {1976})},\ \bibinfo {note} {[Erratum:
  Nuovo Cim.A47,555(1978)]}\BibitemShut {NoStop}%
\bibitem [{\citenamefont {Tissier}\ and\ \citenamefont
  {Wschebor}(2010)}]{Tissier:2010ts}%
  \BibitemOpen
  \bibfield  {author} {\bibinfo {author} {\bibfnamefont {M.}~\bibnamefont
  {Tissier}}\ and\ \bibinfo {author} {\bibfnamefont {N.}~\bibnamefont
  {Wschebor}},\ }\href {\doibase 10.1103/PhysRevD.82.101701} {\bibfield
  {journal} {\bibinfo  {journal} {Phys.Rev.}\ }\textbf {\bibinfo {volume}
  {D82}},\ \bibinfo {pages} {101701} (\bibinfo {year} {2010})},\ \Eprint
  {http://arxiv.org/abs/1004.1607} {arXiv:1004.1607 [hep-ph]} \BibitemShut
  {NoStop}%
\bibitem [{\citenamefont {Tissier}\ and\ \citenamefont
  {Wschebor}(2011)}]{Tissier:2011ey}%
  \BibitemOpen
  \bibfield  {author} {\bibinfo {author} {\bibfnamefont {M.}~\bibnamefont
  {Tissier}}\ and\ \bibinfo {author} {\bibfnamefont {N.}~\bibnamefont
  {Wschebor}},\ }\href {\doibase 10.1103/PhysRevD.84.045018} {\bibfield
  {journal} {\bibinfo  {journal} {Phys. Rev.}\ }\textbf {\bibinfo {volume}
  {D84}},\ \bibinfo {pages} {045018} (\bibinfo {year} {2011})},\ \Eprint
  {http://arxiv.org/abs/1105.2475} {arXiv:1105.2475 [hep-th]} \BibitemShut
  {NoStop}%
\bibitem [{\citenamefont {Serreau}\ and\ \citenamefont
  {Tissier}(2012)}]{Serreau:2012cg}%
  \BibitemOpen
  \bibfield  {author} {\bibinfo {author} {\bibfnamefont {J.}~\bibnamefont
  {Serreau}}\ and\ \bibinfo {author} {\bibfnamefont {M.}~\bibnamefont
  {Tissier}},\ }\href {\doibase 10.1016/j.physletb.2012.04.041} {\bibfield
  {journal} {\bibinfo  {journal} {Phys. Lett.}\ }\textbf {\bibinfo {volume}
  {B712}},\ \bibinfo {pages} {97} (\bibinfo {year} {2012})},\ \Eprint
  {http://arxiv.org/abs/1202.3432} {arXiv:1202.3432 [hep-th]} \BibitemShut
  {NoStop}%
\bibitem [{\citenamefont {Siringo}(2015)}]{Siringo:2015aka}%
  \BibitemOpen
  \bibfield  {author} {\bibinfo {author} {\bibfnamefont {F.}~\bibnamefont
  {Siringo}},\ }\href@noop {} {\  (\bibinfo {year} {2015})},\ \Eprint
  {http://arxiv.org/abs/1509.05891} {arXiv:1509.05891 [hep-ph]} \BibitemShut
  {NoStop}%
\bibitem [{\citenamefont {Siringo}(2016)}]{Siringo:2015wtx}%
  \BibitemOpen
  \bibfield  {author} {\bibinfo {author} {\bibfnamefont {F.}~\bibnamefont
  {Siringo}},\ }\href {\doibase 10.1016/j.nuclphysb.2016.04.028} {\bibfield
  {journal} {\bibinfo  {journal} {Nucl. Phys.}\ }\textbf {\bibinfo {volume}
  {B907}},\ \bibinfo {pages} {572} (\bibinfo {year} {2016})},\ \Eprint
  {http://arxiv.org/abs/1511.01015} {arXiv:1511.01015 [hep-ph]} \BibitemShut
  {NoStop}%
\bibitem [{\citenamefont {Siringo}(2017)}]{Siringo:2017svp}%
  \BibitemOpen
  \bibfield  {author} {\bibinfo {author} {\bibfnamefont {F.}~\bibnamefont
  {Siringo}},\ }\href {\doibase 10.1103/PhysRevD.96.114020} {\bibfield
  {journal} {\bibinfo  {journal} {Phys. Rev.}\ }\textbf {\bibinfo {volume}
  {D96}},\ \bibinfo {pages} {114020} (\bibinfo {year} {2017})},\ \Eprint
  {http://arxiv.org/abs/1705.06160} {arXiv:1705.06160 [hep-ph]} \BibitemShut
  {NoStop}%
\bibitem [{\citenamefont {Carpenter}\ \emph {et~al.}(2017)\citenamefont
  {Carpenter}, \citenamefont {Gelman}, \citenamefont {Hoffman}, \citenamefont
  {Lee}, \citenamefont {Goodrich}, \citenamefont {Betancourt}, \citenamefont
  {Brubaker}, \citenamefont {Guo}, \citenamefont {Li},\ and\ \citenamefont
  {Riddell}}]{JSSv076i01}%
  \BibitemOpen
  \bibfield  {author} {\bibinfo {author} {\bibfnamefont {B.}~\bibnamefont
  {Carpenter}}, \bibinfo {author} {\bibfnamefont {A.}~\bibnamefont {Gelman}},
  \bibinfo {author} {\bibfnamefont {M.}~\bibnamefont {Hoffman}}, \bibinfo
  {author} {\bibfnamefont {D.}~\bibnamefont {Lee}}, \bibinfo {author}
  {\bibfnamefont {B.}~\bibnamefont {Goodrich}}, \bibinfo {author}
  {\bibfnamefont {M.}~\bibnamefont {Betancourt}}, \bibinfo {author}
  {\bibfnamefont {M.}~\bibnamefont {Brubaker}}, \bibinfo {author}
  {\bibfnamefont {J.}~\bibnamefont {Guo}}, \bibinfo {author} {\bibfnamefont
  {P.}~\bibnamefont {Li}}, \ and\ \bibinfo {author} {\bibfnamefont
  {A.}~\bibnamefont {Riddell}},\ }\href {\doibase 10.18637/jss.v076.i01}
  {\bibfield  {journal} {\bibinfo  {journal} {Journal of Statistical Software,
  Articles}\ }\textbf {\bibinfo {volume} {76}},\ \bibinfo {pages} {1} (\bibinfo
  {year} {2017})}\BibitemShut {NoStop}%
\bibitem [{\citenamefont {Cyrol}\ \emph
  {et~al.}(2018{\natexlab{a}})\citenamefont {Cyrol}, \citenamefont {Mitter},
  \citenamefont {Pawlowski},\ and\ \citenamefont {Strodthoff}}]{Cyrol:2017qkl}%
  \BibitemOpen
  \bibfield  {author} {\bibinfo {author} {\bibfnamefont {A.~K.}\ \bibnamefont
  {Cyrol}}, \bibinfo {author} {\bibfnamefont {M.}~\bibnamefont {Mitter}},
  \bibinfo {author} {\bibfnamefont {J.~M.}\ \bibnamefont {Pawlowski}}, \ and\
  \bibinfo {author} {\bibfnamefont {N.}~\bibnamefont {Strodthoff}},\ }\href
  {\doibase 10.1103/PhysRevD.97.054015} {\bibfield  {journal} {\bibinfo
  {journal} {Phys. Rev.}\ }\textbf {\bibinfo {volume} {D97}},\ \bibinfo {pages}
  {054015} (\bibinfo {year} {2018}{\natexlab{a}})},\ \Eprint
  {http://arxiv.org/abs/1708.03482} {arXiv:1708.03482 [hep-ph]} \BibitemShut
  {NoStop}%
\bibitem [{\citenamefont {Cyrol}\ \emph
  {et~al.}(2018{\natexlab{b}})\citenamefont {Cyrol}, \citenamefont {Mitter},
  \citenamefont {Pawlowski},\ and\ \citenamefont {Strodthoff}}]{Cyrol:2017ewj}%
  \BibitemOpen
  \bibfield  {author} {\bibinfo {author} {\bibfnamefont {A.~K.}\ \bibnamefont
  {Cyrol}}, \bibinfo {author} {\bibfnamefont {M.}~\bibnamefont {Mitter}},
  \bibinfo {author} {\bibfnamefont {J.~M.}\ \bibnamefont {Pawlowski}}, \ and\
  \bibinfo {author} {\bibfnamefont {N.}~\bibnamefont {Strodthoff}},\ }\href
  {\doibase 10.1103/PhysRevD.97.054006} {\bibfield  {journal} {\bibinfo
  {journal} {Phys. Rev.}\ }\textbf {\bibinfo {volume} {D97}},\ \bibinfo {pages}
  {054006} (\bibinfo {year} {2018}{\natexlab{b}})},\ \Eprint
  {http://arxiv.org/abs/1706.06326} {arXiv:1706.06326 [hep-ph]} \BibitemShut
  {NoStop}%
\bibitem [{\citenamefont {Glimm}\ and\ \citenamefont
  {Jaffe}(1981)}]{glimm1981quantum}%
  \BibitemOpen
  \bibfield  {author} {\bibinfo {author} {\bibfnamefont {J.}~\bibnamefont
  {Glimm}}\ and\ \bibinfo {author} {\bibfnamefont {A.}~\bibnamefont {Jaffe}},\
  }\href {https://books.google.de/books?id=hKTvAAAAMAAJ} {\emph {\bibinfo
  {title} {Quantum physics: a functional integral point of view}}}\ (\bibinfo
  {publisher} {Springer-Verlag},\ \bibinfo {year} {1981})\BibitemShut {NoStop}%
\bibitem [{\citenamefont {Cuniberti}\ \emph {et~al.}(2001)\citenamefont
  {Cuniberti}, \citenamefont {De~Micheli},\ and\ \citenamefont
  {Viano}}]{Cuniberti:2001hm}%
  \BibitemOpen
  \bibfield  {author} {\bibinfo {author} {\bibfnamefont {G.}~\bibnamefont
  {Cuniberti}}, \bibinfo {author} {\bibfnamefont {E.}~\bibnamefont
  {De~Micheli}}, \ and\ \bibinfo {author} {\bibfnamefont {G.~A.}\ \bibnamefont
  {Viano}},\ }\href {\doibase 10.1007/s002200000324} {\bibfield  {journal}
  {\bibinfo  {journal} {Commun. Math. Phys.}\ }\textbf {\bibinfo {volume}
  {216}},\ \bibinfo {pages} {59} (\bibinfo {year} {2001})},\ \Eprint
  {http://arxiv.org/abs/cond-mat/0109175} {arXiv:cond-mat/0109175
  [cond-mat.str-el]} \BibitemShut {NoStop}%
\bibitem [{\citenamefont {Lowdon}(2017)}]{Lowdon:2017uqe}%
  \BibitemOpen
  \bibfield  {author} {\bibinfo {author} {\bibfnamefont {P.}~\bibnamefont
  {Lowdon}},\ }\href {\doibase 10.1103/PhysRevD.96.065013} {\bibfield
  {journal} {\bibinfo  {journal} {Phys. Rev.}\ }\textbf {\bibinfo {volume}
  {D96}},\ \bibinfo {pages} {065013} (\bibinfo {year} {2017})},\ \Eprint
  {http://arxiv.org/abs/1702.02954} {arXiv:1702.02954 [hep-th]} \BibitemShut
  {NoStop}%
\bibitem [{\citenamefont {Lowdon}(2018)}]{Lowdon:2018mbn}%
  \BibitemOpen
  \bibfield  {author} {\bibinfo {author} {\bibfnamefont {P.}~\bibnamefont
  {Lowdon}},\ }\href@noop {} {\  (\bibinfo {year} {2018})},\ \Eprint
  {http://arxiv.org/abs/1801.09337} {arXiv:1801.09337 [hep-th]} \BibitemShut
  {NoStop}%
\bibitem [{\citenamefont {Backus}\ and\ \citenamefont {Gilbert}()}]{BG1}%
  \BibitemOpen
  \bibfield  {author} {\bibinfo {author} {\bibfnamefont {G.}~\bibnamefont
  {Backus}}\ and\ \bibinfo {author} {\bibfnamefont {F.}~\bibnamefont
  {Gilbert}},\ }\href {\doibase 10.1111/j.1365-246X.1968.tb00216.x} {\bibfield
  {journal} {\bibinfo  {journal} {Geophysical Journal of the Royal Astronomical
  Society}\ }\textbf {\bibinfo {volume} {16}},\ \bibinfo {pages} {169}},\
  \Eprint
  {http://arxiv.org/abs/https://onlinelibrary.wiley.com/doi/pdf/10.1111/j.1365-246X.1968.tb00216.x}
  {https://onlinelibrary.wiley.com/doi/pdf/10.1111/j.1365-246X.1968.tb00216.x}
  \BibitemShut {NoStop}%
\bibitem [{\citenamefont {Backus}\ and\ \citenamefont {Gilbert}(1970)}]{BG2}%
  \BibitemOpen
  \bibfield  {author} {\bibinfo {author} {\bibfnamefont {G.}~\bibnamefont
  {Backus}}\ and\ \bibinfo {author} {\bibfnamefont {F.}~\bibnamefont
  {Gilbert}},\ }\href {\doibase 10.1098/rsta.1970.0005} {\bibfield  {journal}
  {\bibinfo  {journal} {Philosophical Transactions of the Royal Society of
  London A: Mathematical, Physical and Engineering Sciences}\ }\textbf
  {\bibinfo {volume} {266}},\ \bibinfo {pages} {123} (\bibinfo {year}
  {1970})},\ \Eprint
  {http://arxiv.org/abs/http://rsta.royalsocietypublishing.org/content/266/1173/123.full.pdf}
  {http://rsta.royalsocietypublishing.org/content/266/1173/123.full.pdf}
  \BibitemShut {NoStop}%
\bibitem [{\citenamefont {Press}\ \emph {et~al.}(1997)\citenamefont {Press},
  \citenamefont {Teukosly}, \citenamefont {Vetterling},\ and\ \citenamefont
  {Flannery}}]{Press:1997nr}%
  \BibitemOpen
  \bibfield  {author} {\bibinfo {author} {\bibfnamefont {W.~H.}\ \bibnamefont
  {Press}}, \bibinfo {author} {\bibfnamefont {S.~A.}\ \bibnamefont {Teukosly}},
  \bibinfo {author} {\bibfnamefont {V.~T.}\ \bibnamefont {Vetterling}}, \ and\
  \bibinfo {author} {\bibfnamefont {B.~P.}\ \bibnamefont {Flannery}},\
  }\href@noop {} {\emph {\bibinfo {title} {Numerical recipes in C}}}\ (\bibinfo
   {publisher} {Cambridge University Press},\ \bibinfo {address} {Cambridge},\
  \bibinfo {year} {1997})\BibitemShut {NoStop}%
\bibitem [{\citenamefont {Tripolt}\ \emph {et~al.}(2018)\citenamefont
  {Tripolt}, \citenamefont {Gubler}, \citenamefont {Ulybyshe},\ and\
  \citenamefont {Von~Smekal}}]{Tripolt:2018xeo}%
  \BibitemOpen
  \bibfield  {author} {\bibinfo {author} {\bibfnamefont {R.-A.}\ \bibnamefont
  {Tripolt}}, \bibinfo {author} {\bibfnamefont {P.}~\bibnamefont {Gubler}},
  \bibinfo {author} {\bibfnamefont {M.}~\bibnamefont {Ulybyshe}}, \ and\
  \bibinfo {author} {\bibfnamefont {L.}~\bibnamefont {Von~Smekal}},\
  }\href@noop {} {\  (\bibinfo {year} {2018})},\ \Eprint
  {http://arxiv.org/abs/1801.10348} {arXiv:1801.10348 [hep-ph]} \BibitemShut
  {NoStop}%
\bibitem [{\citenamefont {Burnier}\ and\ \citenamefont
  {Rothkopf}(2013)}]{Burnier:2013nla}%
  \BibitemOpen
  \bibfield  {author} {\bibinfo {author} {\bibfnamefont {Y.}~\bibnamefont
  {Burnier}}\ and\ \bibinfo {author} {\bibfnamefont {A.}~\bibnamefont
  {Rothkopf}},\ }\href {\doibase 10.1103/PhysRevLett.111.182003} {\bibfield
  {journal} {\bibinfo  {journal} {Phys. Rev. Lett.}\ }\textbf {\bibinfo
  {volume} {111}},\ \bibinfo {pages} {182003} (\bibinfo {year} {2013})},\
  \Eprint {http://arxiv.org/abs/1307.6106} {arXiv:1307.6106 [hep-lat]}
  \BibitemShut {NoStop}%
\bibitem [{\citenamefont {{Stan Development Team. 2017}}()}]{rstan}%
  \BibitemOpen
  \bibfield  {author} {\bibinfo {author} {\bibnamefont {{Stan Development Team.
  2017}}},\ }\href@noop {} {\enquote {\bibinfo {title} {Rstan: the r interface
  to stan. r package version 2.16.2.}}\ }\bibinfo {howpublished}
  {\url{http://mc-stan.org}}\BibitemShut {NoStop}%
\end{thebibliography}%

\end{document}